\documentclass[twoside,12pt,a4paper]{report}
\usepackage{graphicx}
\usepackage{pgf}
\usepackage{fancyhdr}
\usepackage{multirow}
\usepackage{amssymb}
\usepackage{amsbsy}
\usepackage{amsmath}
\usepackage{epsfig,subfigure}
\usepackage{color}
\usepackage{bm}
\usepackage{verbatim} 
\usepackage{rotating} 
\usepackage{hhline}
\usepackage{xspace}
\usepackage{subfigure}
\usepackage{anysize}
\usepackage{fixmath}
\usepackage{color}
\usepackage{cite}
\usepackage[toc,page]{appendix}
\usepackage[utf8]{inputenc}
\usepackage[T1]{fontenc}
\usepackage{cite} 
\usepackage{comment}
\usepackage[acronym]{glossaries}
\usepackage{lineno} 
\marginsize{3,5cm}{3,5cm}{2,5cm}{2,5cm}

\newcommand{\LumReacta}{\ensuremath{dd\rightarrow {}^{3}\hspace{-0.03cm}\mbox{He} n}}
\newcommand{\LumReactb}{\ensuremath{dd\rightarrow p p n_{sp} n_{sp}}}
\newcommand{\MainReact}{dd\rightarrow({}^{4}\hspace{-0.03cm}\mbox{He}$-$\eta)_{bound} \rightarrow$ $^{3}\hspace{-0.03cm}\mbox{He} n \pi{}^{0}}
\newcommand{\BcgReacta}{\ensuremath{dd\rightarrow {}^{3}\hspace{-0.03cm}\mbox{He} n \pi^{0}}}
\newcommand{\BcgReactb}{\ensuremath{dd\rightarrow {}^{3}\hspace{-0.03cm}\mbox{He} N^{*} \rightarrow {}^{3}\hspace{-0.03cm}\mbox{He} n \pi^{0}}}
\newcommand{\BcgReactW}{\ensuremath{dd\rightarrow {}^{3}\hspace{-0.03cm}\mbox{He} p \pi^{-}}}

\newcommand{\Hea}{\ensuremath{^{3}\hspace{-0.03cm}\mbox{He}}} 
\newcommand{\Heb}{\ensuremath{^{4}\hspace{-0.03cm}\mbox{He}}}
\newcommand{\piz}{\ensuremath{\pi^{0}}}
\newcommand{\BSbound}{\ensuremath{(^{4}\hspace{-0.03cm}\mbox{He}$-$\eta)_{bound}}}
\newcommand{\BS}{\ensuremath{^{4}\hspace{-0.03cm}\mbox{He}$-$\eta}}

\usepackage{titlesec}

\titleclass{\subsubsubsection}{straight}[\subsection]

\newcounter{subsubsubsection}[subsubsection]
\renewcommand\thesubsubsubsection{\thesubsubsection.\arabic{subsubsubsection}}

\titleformat{\subsubsubsection}
  {\normalfont\normalsize\bfseries}{\thesubsubsubsection}{1em}{}
\titlespacing*{\subsubsubsection}{0pt}{3.25ex plus 1ex minus .2ex}{1.5ex plus .2ex}

\makeatletter
\renewcommand\paragraph{\@startsection{paragraph}{5}{\z@}%
  {3.25ex \@plus1ex \@minus.2ex}%
  {-1em}%
  {\normalfont\normalsize\bfseries}}
\renewcommand\subparagraph{\@startsection{subparagraph}{6}{\parindent}%
  {3.25ex \@plus1ex \@minus .2ex}%
  {-1em}%
  {\normalfont\normalsize\bfseries}}
\def\toclevel@subsubsubsection{4}
\def\toclevel@paragraph{5}
\def\toclevel@paragraph{6}
\def\l@subsubsubsection{\@dottedtocline{4}{7em}{4em}}
\def\l@paragraph{\@dottedtocline{5}{10em}{5em}}
\def\l@subparagraph{\@dottedtocline{6}{14em}{6em}}
\makeatother

\begin{document}

\begin{titlepage} 

\begin{center}
\large{
A DOCTORAL DISSERTATION\\
PREPARED IN THE INSTITUTE OF PHYSICS\\
OF THE JAGIELLONIAN UNIVERSITY,\\
SUBMITTED TO THE FACULTY OF PHYSICS, ASTRONOMY
AND APPLIED COMPUTER SCIENCE\\
OF THE JAGIELLONIAN UNIVERSITY
}
\end{center}

\begin{figure}[h]
\begin{center}
\includegraphics[width=4.0cm,height=5.0cm]{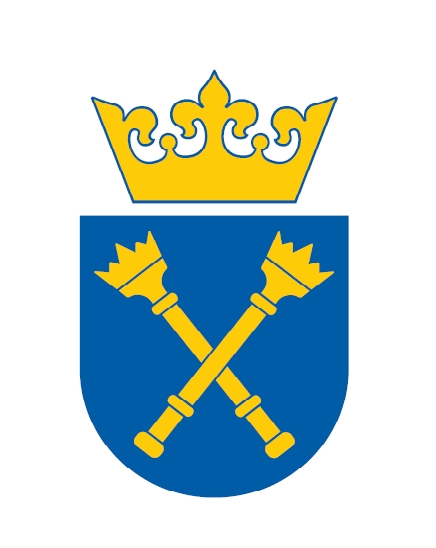}
\end{center}
\end{figure}

\begin{center}
	\Huge \textbf{Search for $\mathbold{\eta}$-mesic helium via} $\mathbold{dd\rightarrow}$ $\mathbold{^{3}}\hspace{-0.03cm}\bm{\mbox{He}} \mathbold{n \pi^{0}}$ \textbf{reaction by means of the WASA-at-COSY facility\\}
\end{center}

\begin{center}
\Large \textbf{Magdalena Skurzok}
\end{center}

\vspace{1.0cm}

\begin{center}
\normalsize{THESIS ADVISOR:\\ PROF. DR HAB. PAWEŁ MOSKAL}\\
\vspace{0.5cm}
\normalsize{CO-ADVISOR:\\ DR WOJCIECH KRZEMIE{\'N}}

\vspace{2.0cm}

Cracow, 2015

\end{center}

\end{titlepage}


\newpage
\thispagestyle{empty}
\begin{center}

\end{center}




\newpage
\thispagestyle{empty}

\begin{center}
\large \textbf{Abstract}
\end{center}


\noindent The existence of $\eta$-mesic nuclei in which the $\eta$ meson is bound with nucleus via the strong interaction was postulated by Haider and Liu over twenty years ago, however till now no experiment confirmed it empirically. \\

\noindent In November 2010, we performed a search for a $^{4}\hspace{-0.03cm}\mbox{He}$-$\eta$ bound state by measuring the excitation function for the $dd\rightarrow$ $^{3}\hspace{-0.03cm}\mbox{He} n \pi{}^{0}$  and $dd\rightarrow$ $^{3}\hspace{-0.03cm}\mbox{He} p \pi{}^{-}$ reactions in the vicinity of the $\eta$ production threshold. The measurement was performed with high statistic and high acceptance with the WASA detector, installed at the cooler synchrotron COSY in the Forschungszentrum J{\"u}lich.~The experiment was carried out using a deuteron COSY beam and deuteron pellet target. The beam momentum varied continuously in each of acceleration cycle from 2.127~GeV/c to 2.422~GeV/c, which corresponds to a range of excess energy \mbox{$Q$~$\in$~(-70,30)~MeV}. \\

\noindent This dissertation is about the search for $^{4}\hspace{-0.03cm}\mbox{He}$-$\eta$ bound state in $dd\rightarrow$ $^{3}\hspace{-0.03cm}\mbox{He} n \pi{}^{0}$ reaction. The excitation function for the process was determined after identification of all outgoing particles and the application of the selection conditions based on Monte Carlo simulations of $\eta$-mesic helium production and its decay via excitation of the $N^{*}$ resonance. The total integrated luminosity was calculated based on the $dd\rightarrow$ $^{3}\hspace{-0.03cm}\mbox{He} n$ and $\LumReactb$ reactions, while the luminosity dependence on the excess energy, used for normalization of the excitation function, was determined based on quasi-elastic proton-proton scattering. 
No narrow structure of the $\eta$-mesic helium was observed in the excitation function. The upper limit of the total cross section for the bound state formation and its decay in $\MainReact$ process was determined on the 90\% confidence level. It varies from 21 to 36~nb for the bound state width ranging from 5~MeV to 50~MeV, respectively. However, an indication for a broad state was observed. The kinematic region, where we expect the evidence of the signal from the bound state, cannot be fully described only by the combination of the considered background processes. In contrast, the experimental excitation function is very well fitted by the background contributions for the region where the signal is not expected.\\

\clearpage



\newpage
\thispagestyle{empty}
\begin{center}

\end{center}


\pagestyle{fancy}
\renewcommand{\chaptermark}[1]{ \markboth{#1}{}}
\renewcommand{\sectionmark}[1]{ \markright{#1}{}}

\fancyhf{}
\fancyhead[LE,RO]{\thepage}
\fancyhead[RE]{\textbf{\nouppercase{\leftmark}}}
\fancyhead[LO]{\textbf{\nouppercase{\rightmark}}}


\tableofcontents
\clearpage


\chapter{Introduction}

For decades, physicists wrestled with basic questions about the surrounding universe:~What kind of objects it consists of and what kind of interactions are responsible for its existence? All matter around us is made of elementary particles, which occur in two basic types called quarks and leptons. Unlike leptons, quarks have color charge, which causes the strong interaction. Quantum chromodynamics (QCD) is the quantum field theory describing the strong interactions between quarks and gluons carrying the color charge. According to this theory, hadrons consist of three quarks \textit{qqq} (baryons) or quark-antiquark pairs \textit{q-$\overline{q}$} (mesons). The most important baryons are the protons and the neutrons, the building blocks of the atomic nuclei.

One of the most fruitful experimental investigations in the field of nuclear physics is the search for new, uncommon objects. Many of them, such as hypernuclei~\cite{Danysz}, tetraquarks~\cite{Tetraq}, pentaquarks~\cite{Pentaquark} or dibaryons~\cite{dibar_jeden,dibar_dwaa,dibar_trzy}, have been already discovered, however still a lot is waiting to be explored. One of those theoretically predicted and till now not discovered object is \textit{mesic nuclei}. This new kind of exotic nuclear matter consists of nucleus bound via strong interaction with neutral meson such as $\eta$, $\eta'$, $K$, $\omega$. One of the most promising candidates for such states are the $\eta$-mesic nuclei, postulated by Haider and Liu in 1986~\cite{HaiderLiu1}. The coupled-channel analysis of the $\pi N \rightarrow \pi N$ , $\pi N \rightarrow \pi \pi N$ and $\pi N \rightarrow \eta N$ reactions showed that in the close-to-threshold region, the $\eta$-nucleon interaction is attractive and strong enough to form an $\eta$-nucleus bound system~\cite{BhaleraoLiu}. However, till now none of experiments confirmed it empirically. The first theoretical predictions indicated that due to the large number of nucleons the $\eta$ meson is more likely to bind to a heavy nucleon, therefore the experimental searches concentrated on the heavy nuclei systems. Nevertheless those experiments have not brought expected effect~\cite{chrien_bnl}. Current researches indicate that $\eta$ nucleon interaction is considerably stronger than it was expected earlier~\cite{Moskal_habil}. A wide range of possible values of the $\eta N$ scattering length $a_{\eta N}$ calculated for hadronic- and photoproduction of the $\eta$ meson has not excluded the formation of $\eta$-nucleus bound states for a light nuclei such as $^{4}\hspace{-0.03cm}\mbox{He}$, $^{3}\hspace{-0.03cm}\mbox{He}$, T~\cite{Wilkin1,WycechGreen} and even for deuteron~\cite{Green}. 


The existence of mesonic bound state would give unique possibility for better understanding the elementary meson-nucleon interaction in nuclear medium for low energies. Moreover it would provide information about $N^{*}(1535)$ resonance~\cite{Jido} and about $\eta$ meson properties in nuclear matter~\cite{InoueOset}. According to Bass and Thomas~\cite{BassTom, BassTomek}, the $\eta$ meson binding inside nuclear matter is very sensitive to the singlet component in the quark-gluon wave function of this meson, therefore the investigation of the $\eta$ mesic bound states is important also in terms of the understanding of $\eta$ and $\eta'$ meson structure.

It is indicated that a good candidate for experimental search of possible binding is \mbox{$^{4}\hspace{-0.03cm}\mbox{He}$-$\eta$} system~\cite{WycechGreen}. An observed steep rise in the cross section for $dd \rightarrow$ $^{4}\hspace{-0.03cm}\mbox{He} \eta$ reaction close to kinematic threshold is a sign of very strong final state interaction (FSI), which could be the evidence for the existence of the bound system.


We developed the experimental method which allows for the search for \mbox{$^{4}\hspace{-0.03cm}\mbox{He}$-$\eta$} bound state in deuteron-deuteron fusion reaction.~The proposal for the experiment was presented at the meeting of the Program Advisory Committee in Research Center J\"ulich in Germany and accepted for the realization in November 2010~\cite{Proposal_old}.~The search was performed with high statistic and high acceptance at the COSY accelerator by means of the WASA detection system~\cite{proc_erice2012, mspmwk_actasuppl2013, wkpmms_acta2014, wkpmms_fbs2014, mswkpm_epj_2014, wkpmjsms_epj2014}.~The measurement was carried out with deuteron COSY beam scattered on internal deuteron pellet target.~During each of acceleration cycle the beam momentum was varied continuously from 2.127~GeV/c to 2.422~GeV/c crossing the kinematic threshold for the $dd \rightarrow$ $^{4}\hspace{-0.03cm}\mbox{He} \eta$ reaction at 2.336~GeV/c. This range of the beam momenta corresponds to an excess energy range from -70~MeV to 30~MeV. The unique ramped beam momentum technique allows to reduce the systematic uncertainties.~The data were effectively taken for about one week whereof the measurement with magnetic field was carried out for only two days because of the failure of cooling system of Superconducting Solenoid. 

~The search for $\eta$-mesic helium was conducted via the measurement of the excitation function for the $dd\rightarrow$ $^{3}\hspace{-0.03cm}\mbox{He} n \pi{}^{0}$  and $dd\rightarrow$ $^{3}\hspace{-0.03cm}\mbox{He} p \pi{}^{-}$ reactions in the vicinity of the $\eta$ production threshold.~The present work is devoted to the investigation of the $dd\rightarrow$ $^{3}\hspace{-0.03cm}\mbox{He} n \pi{}^{0}$ reaction.~The excitation function for the reaction was determined after the detailed analysis of the experimental data. ~The \mbox{existence} of the bound system should manifest itself as a resonance-like structure in the excitation curve for $\MainReact$ reaction below the $dd\rightarrow$ $^{4}\hspace{-0.03cm}\mbox{He} \eta$ reaction threshold. In order to interpret the achieved experimental excitation functions the advanced Monte Carlo simulations of signal $dd\rightarrow$ ($^{4}\hspace{-0.03cm}\mbox{He}$-$\eta)_{bound} \rightarrow$ $^{3}\hspace{-0.03cm}\mbox{He} n \pi{}^{0}$ reaction were carried out. The simulations were prepared based on the kinematic model of bound state production and decay. According to this model $(^{4}\hspace{-0.03cm}\mbox{He}$-$\eta)_{bound}$ nucleus is created in deuteron - deuteron collision, $\eta$ meson is absorbed on one of the nucleons inside helium and may propagate in the nucleus via consecutive excitation of nucleons to the $N^{*}$(1535) state until the resonance decays into the pion-neutron pair. Before the decay, it is assumed that $N^{*}$ resonance moves with Fermi momentum distribution of nucleons inside $\Heb$. The $\Hea$ nucleus, formed from three other nucleons, plays then a role of a spectator. The simulations were carried out under assumption that the bound state has a Breit-Wigner resonance structure with fixed binding energy $B_{s}$ and a width $\Gamma$ and that the beam momentum is ramped around threshold for $\eta$ production.\\

This thesis is divided into ten chapters. The second Chapter presents theoretical aspects of search for $\eta$-mesic nuclei. In Chapter 3 the experimental background of the search for the $\eta$-mesic nuclei is presented. The fourth Chapter includes general informations about the performed experiment: detection facility, the analysis tools, detector calibration and data preselection. The Chapter 5 is devoted to the simulations of the $\MainReact$ reaction. Description of the data analysis is presented in Chapter 6 while the determination of detection efficiency is presented in the subsequent Chapter. Chapter 8 describes the luminosity determination. Chapter 9 presents the final results: the excitation function and the upper limit of the total cross section for considered process. A summary and the outlook are provided in Chapter~10.

\chapter{Phenomenology of mesic nuclei}

This chapter is devoted to the overview of theoretical investigations of $\eta$-mesic nuclei.~The first two sections describe the interaction of $\eta$ meson with nucleon and the bound states in the scattering theory. In the third section we present several predictions for $\eta$-mesic bound states while the fourth section includes the physical motivation of the research presented in this thesis. Theoretical background including description of the bound and virtual states in scattering theory, basic definitions and formulas are presented in Ref.~\cite{Krzemien_PhD}.~Detailed information reader can also find in the cited literature.


\section{$\mathbold{\eta}$--$\mathbold{N}$ interaction~\label{etaN_inter}}

The interaction between $\eta$ meson, which properties are presented in Appendix~\ref{Etaprop}, and nucleons has been studying since many years paying special attention to possibility of the bound states creation. Since, it is impossible to create the $\eta$ beams due to its short lifetime, the $\eta$-nucleon studies are based on the investigation of $\eta N$ scattering amplitude for the processes like $\pi N \rightarrow \eta N$, $\gamma N \rightarrow \eta N$ and also $N N \rightarrow NN \eta$ ($pp \rightarrow pp \eta$~\cite{Moskal2}, $pn \rightarrow pn \eta$~\cite{Moskal3}). In those reactions $\eta$ meson interacts with recoiling nucleon and in the low momentum region the interaction is dominated by broad nucleon $S_{11}$ resonance $N^{*}(1535)$, which is very close to the $\eta$ production threshold (49~MeV above the $\eta N$ threshold) and has width 150~MeV. The resonance is strongly coupled to the $s$-wave $\pi-N$ and the $\eta-N$ channels~\cite{WycechGreen} and causes the steep rise in the pion-nucleon cross section.~Recent and previous experimental data are reviewed in ~\cite{Prakhov, Arndt2} and ~\cite{Clajus, Arndt}, respectively. \\
\indent In order to determine the $\eta$-nucleon scattering amplitude, coupled channel calculations have been performed and their results were fitted to the available data. The first calculation carried out by Bhalerao and Liu~\cite{BhaleraoLiu} including $\eta-N$, $\pi-N$ and $\Delta-\pi$ channels results in the strong and attractive interaction between $\eta$ and nucleon in the low energy ($s$-wave) region. It was confirmed by later calculations~\cite{GreenWycech_99,InoueOset1,Shyam,Durand} and allows to postulate possible existence of $\eta$-mesic bound states.

\section{Bound states in the scattering theory}

\begin{figure}[h!]
\centering
\includegraphics[width=9.0cm,height=5.0cm]{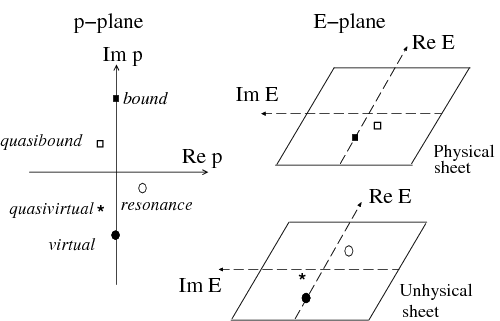} 
\caption{Complex momentum (left) and energy planes (right) with marked poles corresponding to bound, quasi-bound, resonant, virtual and quasi-virtual states. Figure is adopted from~\cite{Kelkar}.~\label{scatt_matrix}}
\end{figure}

The bound state in a usual sense is an object which mass is smaller than the sum of its constituent masses. However, in non relativistic quantum mechanics binding is more complex. The existence of the unstable states is attributed to the occurrence of poles in the scattering matrix in the complex momentum or energy plane. At the low momenta the scattering matrix can be written as~\cite{WycechKrzemien}: 

\begin{equation}
S=\frac{a}{1-ipa},~\label{scatt_1}
\end{equation}

\noindent where $p$ and $a$ are a complex relative $\eta$-nucleus momentum and a scattering length, respectively. The complex energy $E$ can be expressed by the complex momentum $p$ as $E=\frac{p^{2}}{2m_{\mu}}$, where $m_{\mu}$ is reduced mass of $\eta$-nucleus system. Then the real and imaginary parts are related as $Re(E)=\frac{Re^{2}(p)+Im^{2}(p)}{2 m_{\mu}}$ and $Im(E)=\frac{Re(p) Im(p)}{m_{\mu}}$. The pole lying in the physical sheet of momentum and energy plane fulfilling conditions $Im(p)>0$ or $Re(E)<0$ corresponds to the bound state or quasi-bound state, which is schematically presented in Fig.~\ref{scatt_matrix}. \\

\indent The bound state is related to the case when interaction is described only by a real potential ($Im(a)=0$). The pole is then located on the positive imaginary $p$ axis. In case of inelastic interaction which is associated with absorption ($Im(a)\ne 0$) we have the quasi bound state located in the second quadrant of the complex momentum plane. The resonance and virtual/quasi-virtual state poles lie on the unphysical momentum sheet ($Im(p)<0$) in the third and fourth quadrant, respectively.

\section{Theoretical predictions for $\mathbold{\eta}$-mesic nuclei}

The first theoretical predictions concerning the $\eta$-mesic nuclei existence were declared by Haider and Liu in 1986~\cite{HaiderLiu1} based on coupled channel calculations reported by Bhalerao and Liu~\cite{BhaleraoLiu} the year before. Based on the obtained $\eta N$ scattering length ($a_{\eta N}=(0.28+0.19i)$~fm), they postulated the formation of $\eta$-mesic nuclei with masses $A\geq12$. However, later phenomenological and theoretical studies of $\eta$ production in hadronic- and photo- induced reactions brought much wider range of possible scattering length from $a_{\eta N}=(0.18+0.16i)$~fm to $a_{\eta N}=(1.03+0.49i)$~fm~\cite{Kelkar}. The larger scattering lengths do not exclude the formation of a bound states for the helium~\cite{Wilkin1,WycechGreen} and even deuteron~\cite{Green}. \\
\indent The standard theoretical investigations of the possible binding were focused on the construction of the optical potential for the $\eta$-mesic nucleus based on information about $\eta N$ scattering lengths obtained by fitting the different models to experimental data and thus, the solution of wave equation. This method was used especially in theoretical searches of heavy $\eta$-mesic nuclei using two approaches~\cite{Kelkar}. In the first approach $\eta$-nucleon optical potential is constructed using "$T\rho$" approximation ($U_{opt}=V+iW=-\frac{2\pi}{\mu}T(\eta N \rightarrow \eta N)A\rho(r)$~\cite{Machner_2015}, where $\mu$ is reduced $\eta N$ mass, $T(\eta N \rightarrow \eta N)$ is $\eta$-nucleon transition matrix and $\rho$ is nuclear density). The calculations based on this approach provide information about binding energies and widths of $\eta$-mesic nuclei for $A>12$~\cite{HaiderLiu1,HaiderLiu3,Chiang,Garcia}. Another approach is QCD based quark-meson-coupling (QMC), where optical potential is constructed with assumption that $\eta$ is submerged in the nuclear medium and couples to quarks and mixes with $\eta'$~\cite{Tsushima, BassTomek2}. Using this potential and solving the Klein-Gordon equation, authors obtained the single particle energies for the $\eta$ meson for closed shell nuclei as well as $^{6}\hspace{-0.03cm}\mbox{He}$, $^{11}\hspace{-0.03cm}\mbox{B}$ and $^{26}\hspace{-0.03cm}\mbox{Mg}$. Obtained results suggest that one should expect bound states in all of those nuclei.\\
\indent In case of light nuclei, the existence of $\eta$-mesic bound states is manifested as poles in the scattering matrix and the corresponding $\eta$-nucleus scattering lengths $a_{\eta-nucleus}$. The formation of the $\eta$-mesic nucleus can proceed if the $Re(a_{\eta-nucleus})$ is negative, what corresponds to attractive nature of the interaction, and the the following inequality is fulfilled~\cite{HaiderLiu3}: 

\begin{equation}
|Re(a_{\eta-nucleus})|>|Im(a_{\eta-nucleus})|.\label{eq:jeden}
\end{equation}

One of the first predictions concerning light $\eta$-mesic nuclei was carried out using few body equations~\cite{Ueda}. The author considered $\eta NN - \pi NN$ coupled system and observed pole structure
corresponding to a quasibound state with mass 2430~MeV and width 10-20~MeV. The idea was later used to study possible production of $d$-$\eta$, $^{3}\hspace{-0.03cm}\mbox{He}$-$\eta$ and $\BS$ bound states within finite rank approximation (FRA)~\cite{Rakityansky}. The obtained complex poles in the scattering amplitude correspond to the bound states for $Re(a_{\eta N})\in(0.27,0.98)$~fm. \\
\indent The new approach including information about $\eta$ production mechanism and the final state interaction FSI was presented by Neelima Kelkar et al.~\cite{Kelkar,Kelkar_2015}. The authors performed analysis of the $\eta$ production, calculated $\eta$-nucleus amplitudes and locate the $d$-$\eta$, $^{3}\hspace{-0.03cm}\mbox{He}$-$\eta$ and $\BS$ mesic nuclei using the concept of Wigner’s time delay. This analysis shows, that the formation of light $\eta$-mesic bound states is possible for only small values of $a_{\eta N}$ while higher scattering lengths correspond to resonances~\cite{Kelkar_2007}. \\
\indent Recent phenomenological studies of the $\BS$ bound state production in $dd \rightarrow$ $\Hea p \pi^{-}$ reaction were carried out by Wycech and Krzemie{\'n}~\cite{WycechKrzemien} based on approximation of the scattering amplitude for two body process. The authors estimated the cross section for $dd \rightarrow$ $\BSbound \rightarrow$ $\Hea p \pi^{-}$ process to $\sigma \simeq 4.5$~nb. The result is more than two times higher than the value estimated in~\cite{KrzemienSmyrski,Krzemien_PhD} based on the simple assumption for probability of the $\BSbound$ decay in one of possible channel.

\section{Motivation for the research}

The discovery of postulated $\eta$-mesic nuclei would be interesting on its own since till now no experiment provides empirical confirmation of its existence. The observation of such object would allow to determine its properties and thus investigate many important issues in the $\eta$ meson physics.
 
One of them are studies of the $\eta$ meson interaction with nucleons inside the nuclear matter which would lead to determination of the $\eta N$ scattering length which is quite poorly known~\cite{Machner_2015,Kelkar} cause $a_{\eta N}$ cannot be extracted directly from the existing experimental data.\\
\indent Moreover, the existence of $\eta$-mesic nuclei would also provide information about $N^{*}(1535)$ resonance properties in medium~\cite{Jido,Hirenzaki_2010}. As it was mentioned in Sec.~\ref{etaN_inter}, the resonance is coupled to pion and $\eta$ meson in the low energy region and the bound state studies could provide unique chance to study the chiral symmetry of baryons since $N^{*}(1535)$ resonance is a chiral partner of nucleon~\cite{Jido}. The investigations of $\eta$-mesic bound states can also be useful in testing different approaches related to structure of $N^{*}(1535)$ resonance~\cite{Krzemien_PhD}, cause it is very hard to distinguish between theoretical models from the existing data~\cite{Hirenzaki_2015Acta}.\\  
\indent Another aspect which could be studied via the $\eta$-mesic nuclei is the structure of $\eta$ meson. According to ~\cite{BassTomek2, BassTom, Hirenzaki_2014} its binding energy is strongly related to the contribution of the flavour singlet component of the quark-gluon wave function of the $\eta$ meson. The bound states investigation could bring valuable information about the magnitude of the glue content in the $\eta$ wave function. Moreover, the $\eta$ mass shift inside the nucleus allows to study the axial $U_{A}$(1) dynamics~\cite{BassTomek2}.

\chapter{Search for $\mathbold{\eta}$-mesic nuclei in previous experiments}

The issue of the $\eta$-mesic bound states has become popular already over 25 years ago when Haider and Liu postulated their existence~\cite{HaiderLiu1}. Since then many experiments in different laboratories were dedicated to search for this new kind of nuclear matter. The overview of previous measurements carried out in heavy and light nuclei regions was described in~\cite{Krzemien_PhD}. This chapter shows the summary of experiments and presents current results.

\section{Heavy nuclei region}

The first theoretical prediction of the $\eta$-mesic bound states regarded nuclei with atomic masses greater than 12~\cite{HaiderLiu1,BhaleraoLiu}. Therefore, in the beginning, the measurements were performed for the heavy nuclei region. 

First such experiment devoted to the search for $\eta$-mesic nuclei was carried out at BNL (Brookhaven National Lab)~\cite{chrien_bnl} by measurement of $\pi^{+} + A \rightarrow p + (A-1)_{\eta}$ reaction with the lithium, carbon, oxygen and aluminium targets. Obtained proton spectra did not reveal any peak structure which could be interpreted as an indication of the bound state. However, this fact does not necessarily mean that the ($\pi$, $N$) reaction is not a good way to produce $\eta$-mesic nuclei. The new investigations with pion beam are going to be performed at \mbox{J-PARC}~\cite{Fujioka_ActPhysPol2010, Fujioka2} with a new optimal kinematic conditions. It is proposed to study of $(\pi^{-},n)$ reaction on $^{7}\hspace{-0.03cm}\mbox{Li}$ and $^{12}\hspace{-0.03cm}\mbox{C}$. The main advantages over a previous BNL experiment will be: (i) back-to-back proton-pion pair detection and (ii) the recoilless conditions fulfilled with the pion beam momentum in the range between 0.7 and 1.0~GeV/c together with detecting zero-degree neutrons (for the BNL measurement at scattering angle 15$^{\circ}$ the momentum transfer was larger than 200~MeV/c). Moreover, \mbox{PILOT} experiment is planned with deuteron target ($\pi^{+}+d \rightarrow p+p+\eta$) in order to estimate background level for the considered reactions.

Another type experiment based on double charge exchange reaction (DCX) was performed at LAMPF (Los Alamos Meson Physics Facility)~\cite{Lampf} in Los Alamos where the $\eta$-mesic $^{18}\hspace{-0.03cm}\mbox{F}$ was searched in $\pi^{+} + $ $^{18}\hspace{-0.03cm}\mbox{O}\rightarrow \pi^{-} +$ $^{18}\hspace{-0.03cm}\mbox{Ne}$ reaction. In this case the bound state was considered to be produced via collision of $\pi^{+}$ beam with the neutron inside oxygen nucleus and decays via absorption of the $\eta$ meson on the neutron and, consequently, the emission of negatively charged pion. Obtained excitation functions also did not reveal clear structure which could be associated to the $\eta$ mesic nuclei. 
 
The first experiment which claimed an evidence for existence of an $\eta$-mesic bound states was performed at LPI (Lebedev Physical Institute)~\cite{Sokol_1999,Sokol_2001}.~The $\eta$-mesic nuclei were searched in photoproduction process: $\gamma + ^{12}$\hspace{-0.03cm}$\mbox{C}\rightarrow N + (A-{\eta}) \rightarrow N + \pi^{+}+n+X$, where $A$ denotes $^{11}\hspace{-0.03cm}\mbox{C}$ or $^{11}\hspace{-0.03cm}\mbox{B}$ nuclei.~The invariant mass distribution of the correlated $\pi^{+}n$ pairs shows a narrow peak structure below the position of $N^{*}(1535)$ resonance (shifted by about 90~MeV/c$^{2}$). The width and binding energy of the obtained resonance structure were determined to be about 100~MeV and 40~MeV, respectively. Obtained results are in agreement with theoretical prediction according to which the production of $\eta$-mesic nuclei proceeds via $N^{*}(1535)$ resonance excitation and its decay into $\pi$-nucleon pair. 
\indent A similar experiment at LPI was dedicated to search for $\eta$-mesic nuclei through observation of the two-nucleon decay mode arising to the two-nucleon absorption of the captured $\eta$ in the nucleus~\cite{Baskov_2012}. In the experiment~proton-neutron pairs outgoing from carbon target in \mbox{$\gamma + ^{12}$\hspace{-0.03cm}$\mbox{C}\rightarrow N+p+n+X$} reaction were measured in coincidence. The protons velocity obtained for photon energy $E_{\gamma}$=850 MeV (above $\eta$ photoproduction) peaked at about 0.7 what can be interpreted as the result of production of low-energy $\eta$-mesons followed by their two-nucleon annihilation ($\eta NN$ $\rightarrow NN$). In contrast standard photoproduction (for $E_{\gamma}$=650 MeV) does not give the particles with such high momenta. Assuming that the observed events from both of described LPI measurements ($\pi^{+} n$ and $pn$ decays) are related with the formation and decay of $\eta$-mesic nuclei, the upper limit of the photoproduction cross section was determined and is equal to $10~\mu$b.

The search for back-to-back $\pi^{-} p$ pairs related to the $\eta$-mesic bound states was also carried out at JINR (Joint Institute for Nuclear Research) with the internal deuteron beam~\cite{Afanasiev}.~The $d$ + $^{12}\hspace{-0.03cm}\mbox{C} \rightarrow \pi^{-}+p + X$ reaction was measured for the deuteron beam energy 2.1~GeV/nucl. In the experiment the $\pi^{-} p$ back-to-back correlation was clearly observed and resonance like peak was found below the $\eta$ production threshold. The result could be associated with the signature of the two-body $N^{*}$ resonance decay related with formation of an $\eta$-mesic nucleus. However, the investigation need more intensive beam and the higher acceptance of the spectrometer.

At GSI (Centre for Heavy Ion Research in Darmstadt)~\cite{Gillitzer_ActaPhysSlov2006} the search for $\eta$-mesic nuclei was carried out in recoil-free transfer reaction using similar method as in case of measurements of deeply bound pionic states~\cite{Yamazaki_INSRep1996}. In the experiment ($d$,$^{3}\hspace{-0.03cm}\mbox{He}$) reaction was measured on $^{7}\hspace{-0.03cm}\mbox{Li}$ and $^{12}\hspace{-0.03cm}\mbox{C}$ targets at GSI Fragment Separator System (FRS). The analysis of this data is in progress. So far no final result is published.

A very strong claim for the discovery of the resonance like structure corresponding to the $\eta$-mesic magnesium was made by the COSY-GEM group after the analysis of $p(^{27}\hspace{-0.03cm}\mbox{Al},$\Hea$)\pi^{-}p'X$ reaction~\cite{Budzanowski,Machner_ActaPhys}. Similarly like in case of GSI measurement~\cite{Gillitzer_ActaPhysSlov2006}, this experiment fulfilled the recoilless kinematics conditions. The obtained missing mass spectrum of the $^{3}\hspace{-0.03cm}\mbox{He}$ shows enhancement for binding energy of about -13~MeV with the width of about 10~MeV. According to the authors, the peak could be interpreted as a signal from $^{25}\hspace{-0.03cm}\mbox{Mg}$-$\eta$ bound state. However, it is important to confirm the result with higher statistics.

\section{Light nuclei region}

A wide range of possible values of the $\eta$N scattering length $a_{\eta N}$ extracted from hadronic- and photoproduction of the $\eta$ meson (overview in Ref.~\cite{Kelkar, Machner_2015}) has not excluded the formation of $\eta$-nucleus bound states for a light nuclei as $^{4}\hspace{-0.03cm}\mbox{He}$, $^{3}\hspace{-0.03cm}\mbox{He}$, T~\cite{Wilkin1,WycechGreen} and even for deuteron~\cite{Green}. In case of light nuclei $\eta$ absorption is smaller and the bound states are expected to be narrower in comparison to the case of heavy nuclei. Therefore, the light bound states seems to be good candidates for the study of possible binding. 

\begin{figure}[h!]
\centering
\includegraphics[width=6.cm,height=5.0cm]{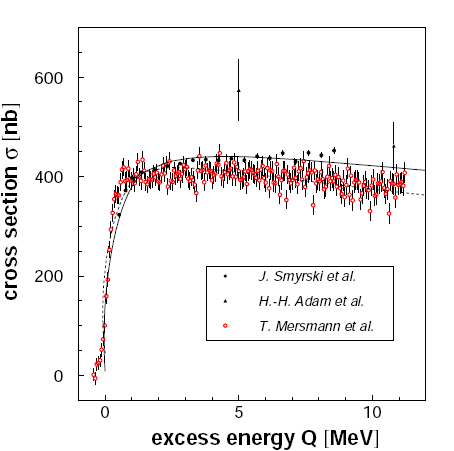} \hspace{0.5cm}
\includegraphics[width=5.8cm,height=4.8cm]{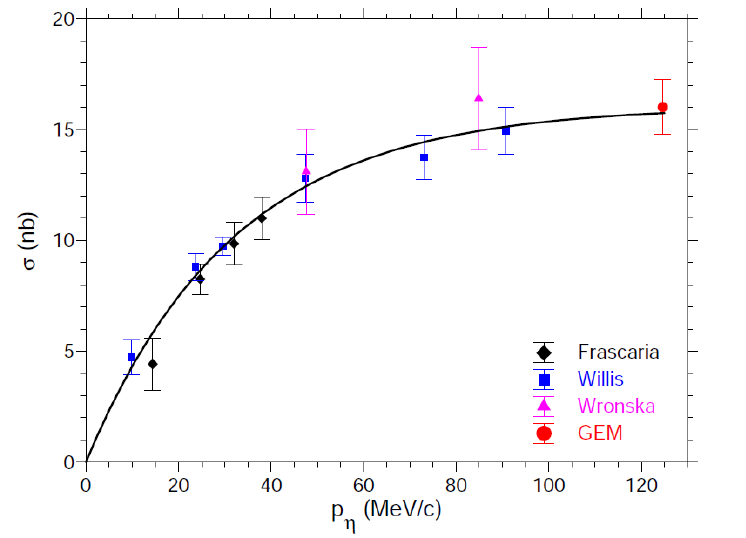}
\vspace{-0.3cm}
\caption{(left) Total cross-section for the $dp\rightarrow$ $^{3}\hspace{-0.03cm}\mbox{He}\eta$ reaction measured with the \mbox{ANKE} (open circles)~\cite{Mersmann} and the \mbox{COSY-11} facilities (closed circles)~\cite{Smyrski1} and (triangles)~\cite{Adam}. Scattering length fit to the ANKE and COSY-11 data is represented with dashed and solid lines, respectively. (right) Total cross-section for the $dd\rightarrow$ $^{4}\hspace{-0.03cm}\mbox{He}\eta$ reaction as a function of CM momentum obtained from the measurements of Frascaria et al.~\cite{Frascaria} (black diamonds), Willis et al.~\cite{Willis} (blue squares), Wro{\'n}ska et al.~\cite{Wronska} (magenta triangles) and Budzanowski et al.~\cite{Budzanowski1} (red circle). The solid line represents a fit in the scattering length approximation. The figure is adopted from~\cite{Machner_ActaPhys}.\label{sigma_3He_4He}}
\end{figure}

\indent
The experimental studies of the final state interactions (FSI) in $^{3}\hspace{-0.03cm}\mbox{He}\eta$ and $^{4}\hspace{-0.03cm}\mbox{He}\eta$ systems result in observations which may suggest the existence of the $\eta$-mesic helium bound states.~The measurements performed by \mbox{SPES-4}~\cite{Berger}, \mbox{SPES-2}~\cite{Mayer}, \mbox{COSY-11}~\cite{Smyrski1} and \mbox{COSY-ANKE}~\cite{Mersmann} as well as in \mbox{SPES-4}~\cite{Frascaria}, \mbox{SPES-3}~\cite{Willis}, GEM~\cite{Machner_ActaPhys} and \mbox{COSY-ANKE}~\cite{Wronska} collaborations revealed a strong enhancement in the cross section of the \mbox{$dp\rightarrow$ $^{3}\hspace{-0.03cm}\mbox{He}\eta$} and $dd\rightarrow$ $^{4}\hspace{-0.03cm}\mbox{He}\eta$ reactions, respectively. This results can be interpreted as a possible indications of the \mbox{$\eta$-mesic} helium. Fig.~\ref{sigma_3He_4He} shows the cross sections measured for both of considered processes. The fits to the experimental data marked in left and right panels of Fig.~\ref{sigma_3He_4He} with solid lines gave the value of the \mbox{$\eta$-$\mbox{helium}$} scattering length \mbox{$a_{\eta^{3}\hspace{-0.03cm}{He}}=[\pm(2.9\pm0.6)+(3.2\pm0.4)i]$}~fm~\cite{Smyrski1} and $a_{\eta^{4}\hspace{-0.03cm}{He}}=[\pm(3.1\pm0.5)+(0.0\pm0.5)i]$~fm~\cite{Budzanowski1}, respectively. However, these values do not allow to check the basic condition for the bound state existence cause it is not possible to verify if the real part of the scattering length is larger than the imaginary part.

The COSY-11 and ANKE groups performed additionally measurement of the differential cross section for $dp\rightarrow$ $^{3}\hspace{-0.03cm}\mbox{He}\eta$ process. The cross section near the threshold has not isotropic form because not only $s$ wave but also $p$ wave contributes to the process. It depends linearly on $cos\theta_{\eta}$ and therefore asymmetry can be defined as: 

\begin{equation}
\frac{d\sigma(\theta_{\eta})}{d\Omega}=\frac{\sigma_{tot}}{4\pi}(1-\alpha cos\theta_{\eta}).\label{eq:alpha}
\end{equation}

\begin{figure}[h!]
\begin{center}
\includegraphics[width=9.0cm,height=6.5cm]{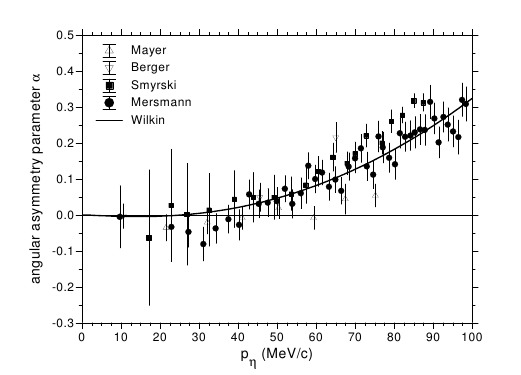} 
\vspace{-0.4cm}
\caption{The asymmetry parameter $\alpha$ as a function of the $\eta$ meson momentum. The data are from~\cite{Mayer} (open triangles up), ~\cite{Berger} (open triangles down), ~\cite{Smyrski1} (full squares) and ~\cite{Mersmann} (full dots). The solid curve represents the fit according to~\cite{Wilkin2}. Figure is adopted from~\cite{Machner_2015}.\label{ANKE_tensor20}}  
\end{center}
\end{figure}

\noindent Asymmetry parameter $\alpha$ as a function of $\eta$ momentum is presented in Fig.~\ref{ANKE_tensor20}. 
The experimental data were fitted with assumption of very strong variation of the $s$-wave amplitude and not to fast changes of $p$-wave amplitude~\cite{Wilkin2}. The fit is in agreement with measured data and implies the small and constant value of the tensor analysing power $t_{20}$ for deuteron. The tensor analysing power was recently measured with ANKE group~\cite{Papenbrock} in $\vec{d}p\rightarrow$ $^{3}\hspace{-0.03cm}\mbox{He}\eta$ process. The measurement was carried out for the excess energy range $Q$~$\in$~(0,11)~MeV with the polarized deuteron beam. 
The angle averaged tensor analysing power was determined in this region and it varies around -0.2. However, the variation is smaller than the error bars what suggest the constant behaviour of $t_{20}$. The obtained result supports strongly the FSI interpretation in the near-threshold region.  

\vspace{-0.5cm}

\begin{figure}[h!]
\centering
\includegraphics[width=14.0cm,height=4.0cm]{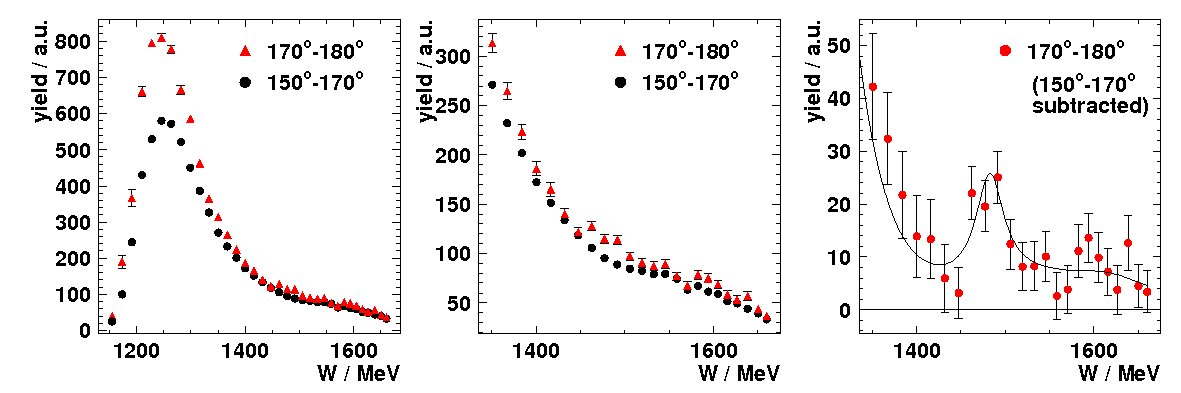}
\includegraphics[width=9.0cm,height=6.5cm]{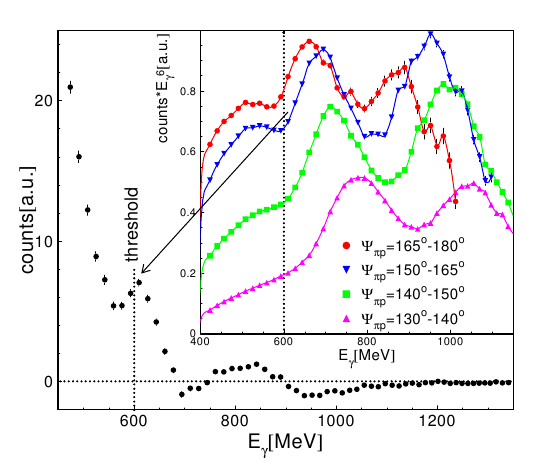} 
\vspace{-0.3cm}
\caption{(upper panel) Excitation functions of the $\pi^{0}$-proton production for relative angles of $170^{0}-180^{0}$ (red triangles) and $150^{0}-170^{0}$ (black circles) in the $\gamma^{3}\hspace{-0.03cm}\mbox{He}$ center-of-mass sytem are shown in the left and center panels. In the right panel the difference between both distributions with superimposed line denoting the results of the fit of the \mbox{Breit-Wigner} distribution plus background are presented. The figure is adopted from~\cite{Pfeiffer}. (lower panel) Difference between excitation functions for the opening angle $165^{\circ}-180^{\circ}$ and $150^{\circ}-165^{\circ}$ obtained by~\cite{Pheron}. Vertical line corresponds to the $\eta$ production threshold.\label{mami}}
\end{figure}

The first direct experimental indication of a light $\eta$-nucleus bound states was reported by the TAPS collaboration~\cite{Pfeiffer} for the $\eta$ photoproduction process $\gamma^{3}$\hspace{-0.03cm}$\mbox{He}\rightarrow \pi^{0}pX$. The reaction was measured with the TAPS calorimeter at the electron accelerator facility Mainz Microtron (MAMI).~The~measurements of~the excitation functions of~the~$\pi^{0}$-proton production for two ranges of~the~relative angle between those particles were carried~out (upper panel of Fig.~\ref{mami}).~It appeared that a~difference between excitation curves for opening angles of~$170^{\circ}-180^{\circ}$ and $150^{\circ}-170^{\circ}$ in~the~center-of-mass frame revealed an~enhancement just below the threshold of the $\gamma^{3}\hspace{-0.03cm}\mbox{He}\rightarrow$ $^{3}\hspace{-0.03cm}\mbox{He}\eta$ reaction. It was interpreted as a~possible signature of a~{$^{3}\hspace{-0.03cm}\mbox{He}$-$\eta$} bound state where $\eta$ meson captured by one of~nucleons inside helium forms an intermediate $N^{*}(1535)$ resonance which decays into pion-nucleon pair.~A~binding energy and width for~the anticipated $\eta$-mesic bound state were deduced from the~fit~of~the~\mbox{Breit-Wigner} distribution function~\cite{Pfeiffer} to the~experimental points and are equal to $(-4.4\pm4.2)$~MeV and $(25.6\pm6.1)$~MeV, respectively. Those values are consistent with expectations for $\eta$-mesic nuclei. However, the later measurement performed by the TAPS collaboration using upgraded detection setup~\cite{Pheron} with much higher statistics allows to ascertain that the structure observed in the $\pi^{0}$-$p$ excitation function is an artefact of the complicated behaviour of the background. Obtained results are presented in lower panel of~Fig.~\ref{mami}. The excitation functions were measured for the higher photon energies what allowed to observe the structures corresponding to second and third resonance regions of the nucleon. The subtraction of the excitation functions for opening angles $165^{\circ}-180^{\circ}$ and $150^{\circ}-165^{\circ}$ reveal narrow peak located at the $\eta$ production threshold which appears due to the shifting of the low energy tails of the second resonance region.

\vspace{-0.5cm}

\begin{figure}[h!]
\centering
\includegraphics[width=9.0cm,height=7.0cm]{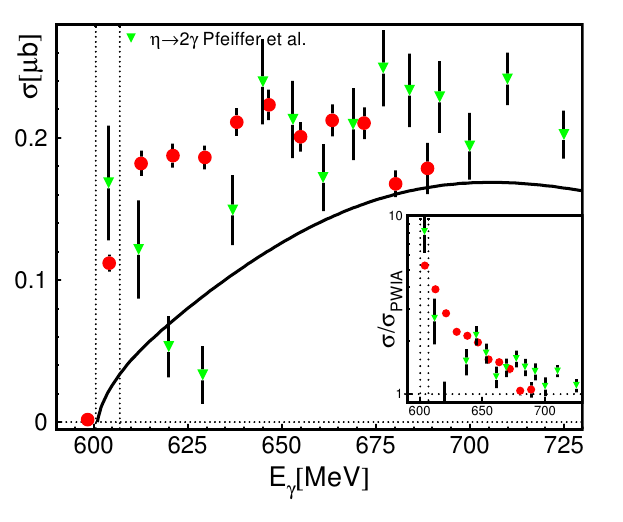}
\vspace{-0.3cm}
\caption{Total cross section for the $\gamma^{3}\hspace{-0.03cm}\mbox{He}\rightarrow$ $^{3}\hspace{-0.03cm}\mbox{He}\eta$ reaction. The green triangles are from~\cite{Pfeiffer} while red dots from~\cite{Pheron}. The two vertical lines indicate the coherent and the break up thresholds. The inserts show the ratio of data and PWIA prediction. Figure is adopted from~\cite{Krusche_Wilkin}.\label{mami_xs}}
\end{figure}

The cross sections obtained in both analyses~\cite{Pfeiffer,Pheron}, presented in Fig.~\ref{mami_xs}, rises very steeply from the production threshold and then stays almost constant. The result is similar to those observed for $\Hea \eta$ hadrono-production at COSY~\cite{Mersmann,Smyrski1}. It suggests that the rise of the cross section above threshold is independent of the initial channel and is therefore a strong argument for the existence of the pole in the scattering matrix which could be associated with $\Hea$-$\eta$ bound state.

The search for the $\eta$-mesic $^{3}\hspace{-0.03cm}\mbox{He}$ was also performed by COSY-11~\mbox{~\cite{Moskal_2010,SmyMosKrze1,Krzemien1,Smyrski2,Smyrski3}} and COSY-TOF~\cite{Gillitzer_ActaPhysSlov2006} groups via measurement of excitation function of the $dp\rightarrow ppp\pi^{-}$ and $dp\rightarrow$ $\Hea\pi^{0}$ reactions around the $\eta$ production threshold. For the first experiment the upper limit of total cross section for $dp\rightarrow$
\mbox{$(^{3}\hspace{-0.03cm}\mbox{He}$-$\eta)_{bound} \rightarrow ppp\pi^{-}$} process was estimated to the value of 270~nb and for $dp\rightarrow$ $(^{3}\hspace{-0.03cm}\mbox{He}$-$\eta)_{bound} \rightarrow$ $\Hea\pi^{0}$ to the value 70~nb. The analysis of COSY-TOF measurement is still in progress. 

\begin{figure}[h!]
\begin{center}
\includegraphics[width=9.0cm,height=6.0cm]{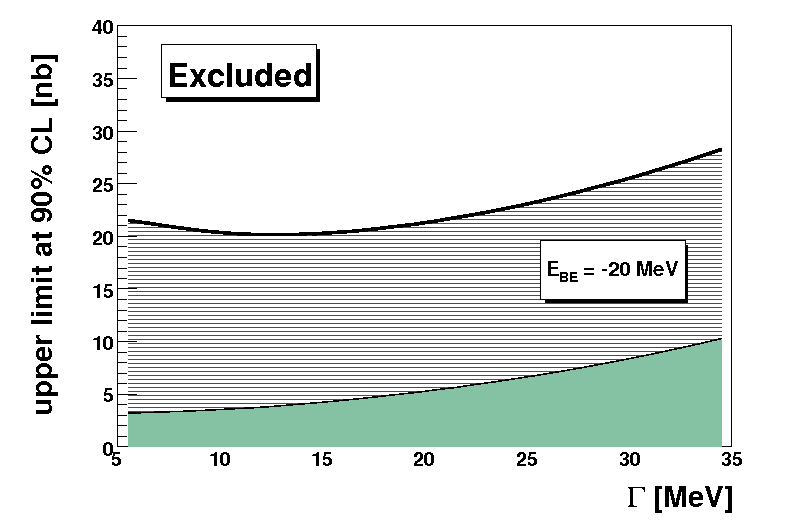} 
\vspace{-0.4cm}
\caption{Upper limit at 90\% confidence level of the cross section for formation of the $^{4}\hspace{-0.03cm}\mbox{He}$-$\eta$ bound state and its decay via the $dd\rightarrow$ $(^{4}\hspace{-0.03cm}\mbox{He}$-$\eta)_{bound}$ $\rightarrow$ $^{3}\hspace{-0.03cm}\mbox{He} p \pi{}^{-}$ reaction as a function of the width of the bound state. The binding energy was set to 20~MeV. The green area at the bottom represents the systematic uncertainties. Figure is adopted from~\cite{BS_WASA}.\label{Wojtek}}  
\end{center}
\end{figure}

In June 2008 WASA-at-COSY collaboration performed the exclusive measurement dedicated to search for the $^{4}\hspace{-0.03cm}\mbox{He}$-$\eta$ bound state in deuteron-deuteron fusion reaction. The $\eta$-mesic nuclei was searched via studying of excitation function for the $dd\rightarrow$ $^{3}\hspace{-0.03cm}\mbox{He} p \pi{}^{-}$ reaction in the vicinity of $^{4}\hspace{-0.03cm}\mbox{He}$-$\eta$ threshold. The measurement was carried out for the beam momentum slowly ramped around the $\eta$ production threshold corresponding to the range of excess energy $Q$ from about -51~MeV to 22~MeV. Excitation function obtained for the $dd\rightarrow$ $^{3}\hspace{-0.03cm}\mbox{He} p \pi{}^{-}$ reaction does not show the resonance like structure which could be interpreted as a signature of \mbox{$\eta$-mesic} $^{4}\hspace{-0.03cm}\mbox{He}$ bound state~\cite{Krzemien_PhD,BS_WASA}. Therefore, an upper limit for the cross-section for the bound state formation and decay in the process $dd\rightarrow$ $(^{4}\hspace{-0.03cm}\mbox{He}$-$\eta)_{bound}$ $\rightarrow$ $^{3}\hspace{-0.03cm}\mbox{He} p \pi{}^{-}$ was determined at the 90\% confidence level.~For this purpose the excitation function was fitted with Breit-Wigner function with fixed binding energy and width combined with second order polynomial. Obtained upper limit presented in Fig.~\ref{Wojtek} for binding energy 20~MeV varies from 20~nb to 27~nb as the width of the bound state varies from 5~MeV to 35~MeV. The upper limits depend mainly on the width of the bound state and only slightly on the binding energy.

The new data set collected in 2010 with much higher statistics allowed to achieve a sensitivity of the cross section of the order of few nb for the bound state production in $^{3}\hspace{-0.03cm}\mbox{He} p \pi{}^{-}$ reaction. The data analysis for this channel is in progress.

\chapter{Experiment}

This chapter includes general information about the experiment dedicated for the search of $\eta$-mesic $\Heb$ which was carried out in 2010. In the first section the brief description of the WASA-at-COSY detection system is presented. The second section contains information about the tools used in data analysis. The last three sections are devoted to accelerator beam settings, calibration of appropriate parts of the detector and the data preselection, respectively.  

\section{Detector Setup}

The experiment described in this thesis was carried out in the Forschungszentrum Jülich, Germany with the WASA (Wide Angle Shower Apparatus) detector installed at COSY accelerator.~In this section the characteristics of the facility is briefly described.~More detailed description can be found in Ref.~\cite{Adam1,Prasuhn}.

\subsection{COoler SYnchrotron COSY}

\begin{figure}[h!]
\centering
\includegraphics[width=10.0cm,height=12.0cm]{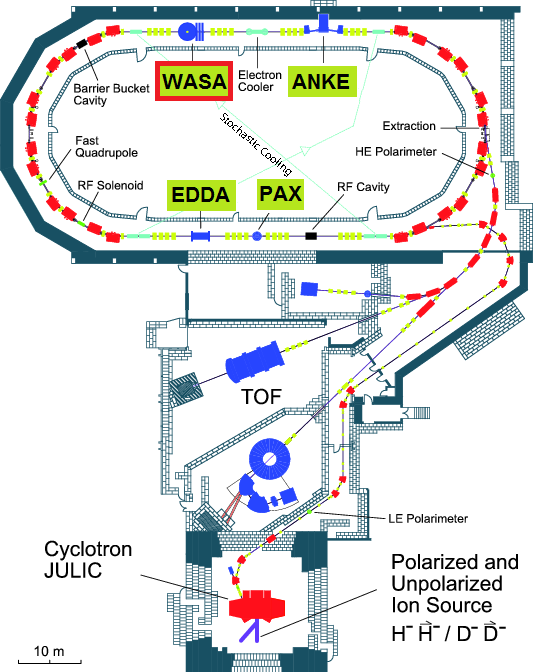} 
\caption{The COSY accelerator facility with highlighted internal and external experimental setups.~\label{fig_COSY}}
\end{figure}

The COSY accelerator complex~\cite{Maier_COSY} presented in Fig.~\ref{fig_COSY} consists of a 184~m circumference cooler synchrotron ring, the JULIC cyclotron and the internal and external experimental targets. In the COSY ring, protons and deuterons (also polarized), pre-accelerated before by JULIC cyclotron, might be accelerated in the momentum range between 0.3~GeV/c and 3.7~GeV/c. The ring can be filled with up to $10^{11}$ unpolarized particles leading to luminosities of $10^{31}$~cm$^{-2}$s$^{-1}$ in case of internal cluster target (ANKE, COSY-11)~\cite{Brauksiepe_COSY, Smyrski_COSY} and $10^{32}$~cm$^{-2}$s$^{-1}$ in case of pellet target applied at WASA~\cite{Adam1}. The beam preparation includes injection, accumulation and acceleration and takes about few seconds, while its lifetime with the pellet target (see Sec.~\ref{Pellet}) is of the order of several minutes. Beams are cooled by means of electron cooling as well as stochastic cooling~\cite{Prasuhn_cooling} at injection and high energies, respectively. It allows to reach the high beam momentum resolution and decrease the luminosity losses during the beam interaction with targets in case of internal experiments.
The greatest advantage of COSY, in point of view of this work, is the ramped beam technique, which permits to perform measurement in slow acceleration mode for a given momentum interval within each acceleration cycle (see Sec.~\ref{Sec_Beam}). This method allows to reduce the systematic uncertainties which appear in case of separate set for each momentum value.\\


\subsection{The WASA Facility}

\begin{figure}[h!]
\centering
\includegraphics[width=13.0cm,height=7.0cm]{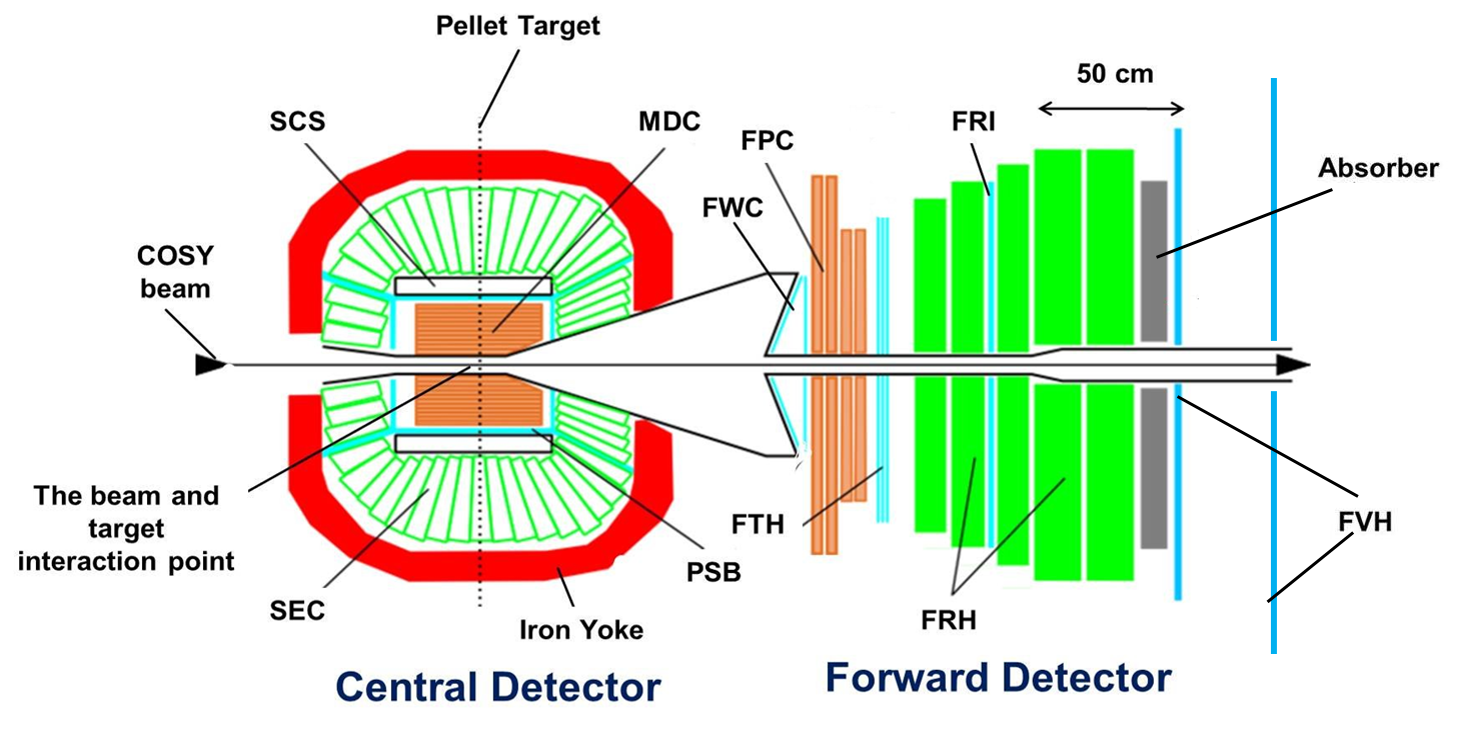} 
\caption{Scheme of WASA-at-COSY detection system. The reaction takes place in the centre of the detector at the crossing of the COSY beam and pellet beam. Gamma quanta, electrons and charged pions being products of mesons decays are registered in the Central Detector. Scattered projectiles and charged recoil particles like $\Hea$, deuterons and protons are registered in the Forward Detector. The abbreviations of the detectors names are explained in the text.\label{fig_wasa}}
\end{figure}

The WASA facility~\cite{Adam1} is an internal detection system installed at COSY since 2007. Before, up to 2005, it was operating at the CELSIUS storage ring at The Svedberg Laboratory in Uppsala, Sweden~\cite{Celsjus}. The physics program of the WASA-at-COSY facility is dedicated mainly to study of $\eta$ and $\omega$ rare decays~\cite{eta_jeden,eta_dwa}, to the study of dibaryon production~\cite{dibar_jeden,dibar_dwaa} and the search for \mbox{$\eta$-mesic} nuclei~\cite{BS_WASA,Krzemien_PhD}. The WASA detector vertical cross section is schematically presented in Fig.~\ref{fig_wasa}. 

The 4$\pi$ WASA detector consists of two main parts: Forward Detector (FD) and Central Detector (CD) optimized for tagging the recoil particles and registering the meson decay products, respectively. The internal target of the pellet-type is installed in the central part of the detection system (its position is marked with dotted line in Fig.~\ref{fig_wasa}). All individual components of the WASA facility are described briefly in the next subsections. 

\subsubsection{Pellet Target System~\label{Pellet}}

\begin{figure}[h!]
\centering
\includegraphics[width=12.0cm,height=7.0cm]{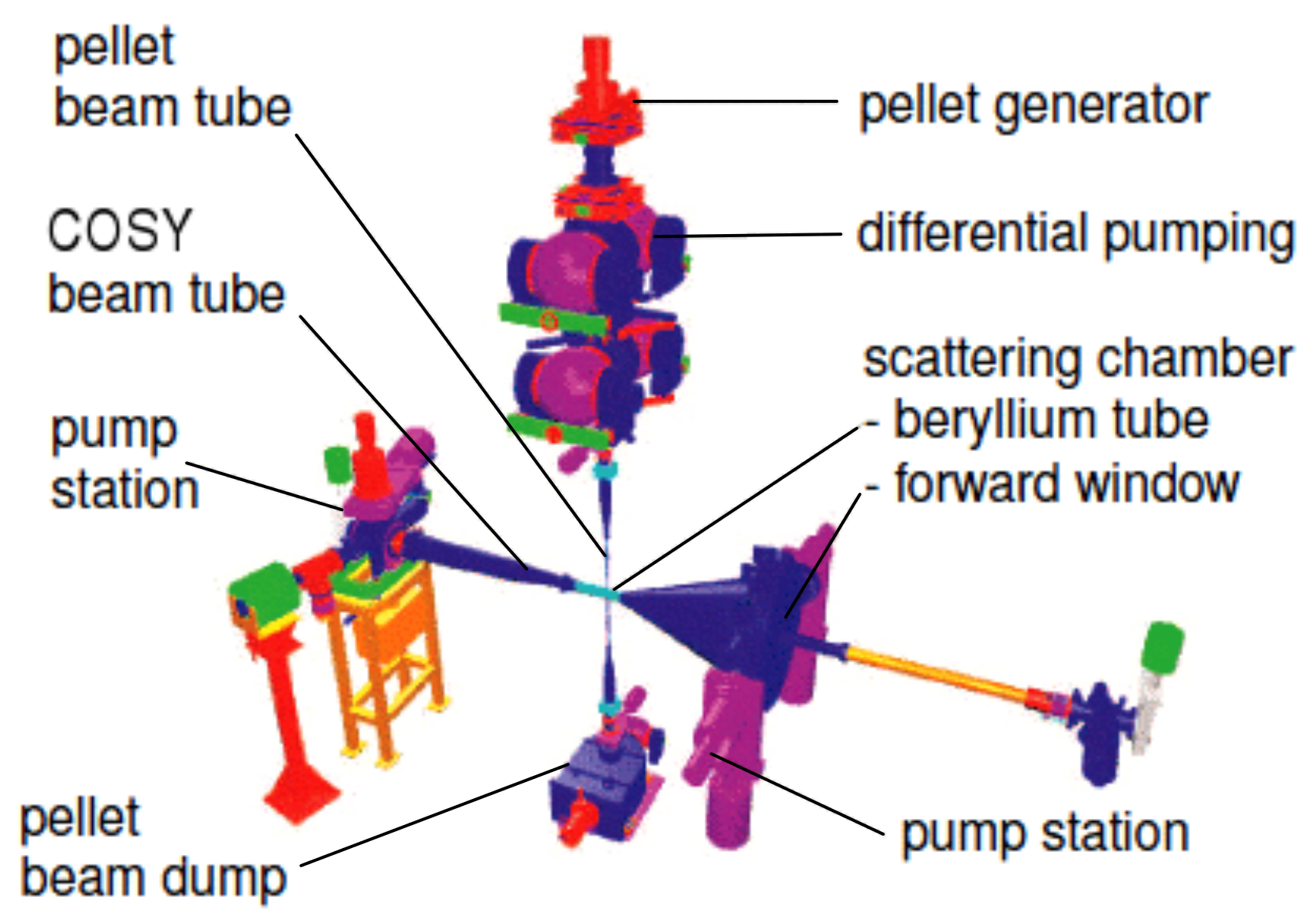} 
\caption{The WASA-at-COSY Pellet Target system.\label{fig_pellet}}
\end{figure}

\noindent The Pellet Target system~\cite{pellet_WASA} has been developed for the WASA facility to fulfil the conditions required for the studies of the rare processes. The main components of the system are shown in Fig.~\ref{fig_pellet}.\\

\vspace{0.5cm}

The Pellet Target setup provides a stream of pellets (frozen droplets) of hydrogen (H$_{2}$) and deuterium (D$_{2}$). They are produced in the pellet generator, located above the Central Detector, where the droplets from the high purity liquid jet (H$_{2}$ or D$_{2}$) are formed with the vibrating nozzle. The nozzle vibrations frequency is typically 70~kHz. The droplets freezes by evaporation while passing through the chamber becoming the pellets. Then, after the entering the 7~cm vacuum-injection capillary the pellets are accelerated up to velocities of 60-80~m/s. Finally a skimmer collimates the pellet beam before it enters the 2~m long pellet tube of 7~mm diameter which is used to guide the pellet beam to the interaction region. The average rates of pellets passing the interaction point is about few thousand per second. The main pellets characteristics are summarized in Table.~\ref{table_pellets}.


\newpage

\begin{table}[h!]
\begin{normalsize}
\begin{center}
\begin{tabular}{l l}\hline
\hline
\vspace{0.2cm}
pellet size &$\approx 35~\mu$m\\
\vspace{0.2cm}
pellet frequency  &5-12~kHz  \\
\vspace{0.2cm}
pellet velocity &60-80~m/s\\
\vspace{0.2cm}
pellet stream divergence & $\sim$0.04$^{\circ}$ \\
\vspace{0.2cm}
pellet stream diameter at beam & 2-4~mm \\
\vspace{0.2cm}
areal target thickness & $>10^{15}$atoms$\cdot$cm$^{-2}$ \\
\hline 
\hline
\end{tabular}
\end{center}
\vspace{-0.8cm}
\begin{center}
\caption{Pellet Target properties.\label{table_pellets}}
\end{center}
\end{normalsize}
\end{table}

\subsubsection{Forward Detector (FD)}

The detection and identification of forward scattered projectiles and target-recoil particles such as protons, deuterons and helium nuclei and also of neutrons and charged pions are carried out with the Forward Detector which covers the range of polar angles from 3$^{\circ}$ to 18$^{\circ}$. It consists of fourteen planes of plastic scintillators forming Forward Window Counter (FWC), Forward Trigger Hodoscope (FTH), Forward Range Hodoscope (FRH), Forward Range Interleaving Hodoscope (FRI) and Forward Veto Hodoscope (FVH), respectively and proportional counter drift tubes called Forward Proportional Chamber (FPC). Particles are identified based on measurement of energy loss in the detection layers of FWC, FTH and FRH while their trajectories are reconstructed from the signals registered successively in FWC, FPC, FTH and FVH detectors. The registered energy loss permits to determine total particle momentum which direction is reconstructed from the measurement of particles tracks by means of straw detectors constituting FPC. Components of the Forward Detector are presented in Fig.~\ref{fig_wasa} and described in text below.

\vspace{0.6cm}

\noindent \textbf{Forward Window Counter}

\vspace{0.2cm}

\begin{figure}[h!]
\centering
\subfigure[The Forward Window Counter.]{\includegraphics[width=6.0cm,height=5.0cm]{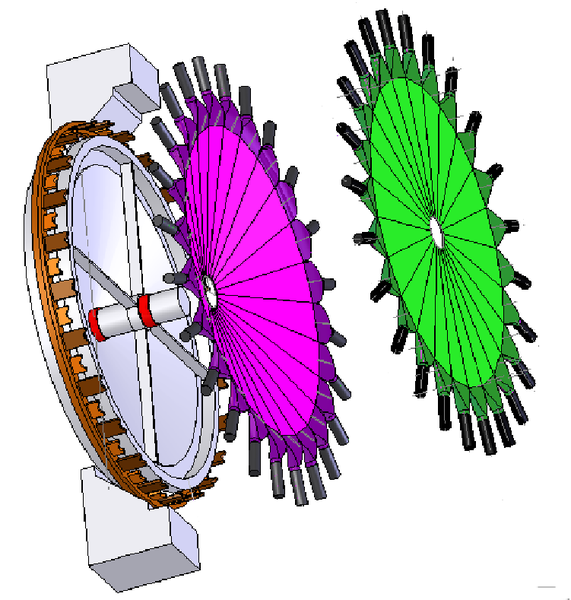}~\label{fig_FDa}} 
\subfigure[3D view of the Forward Proportional Chamber.]{\includegraphics[width=5.0cm,height=5.0cm]{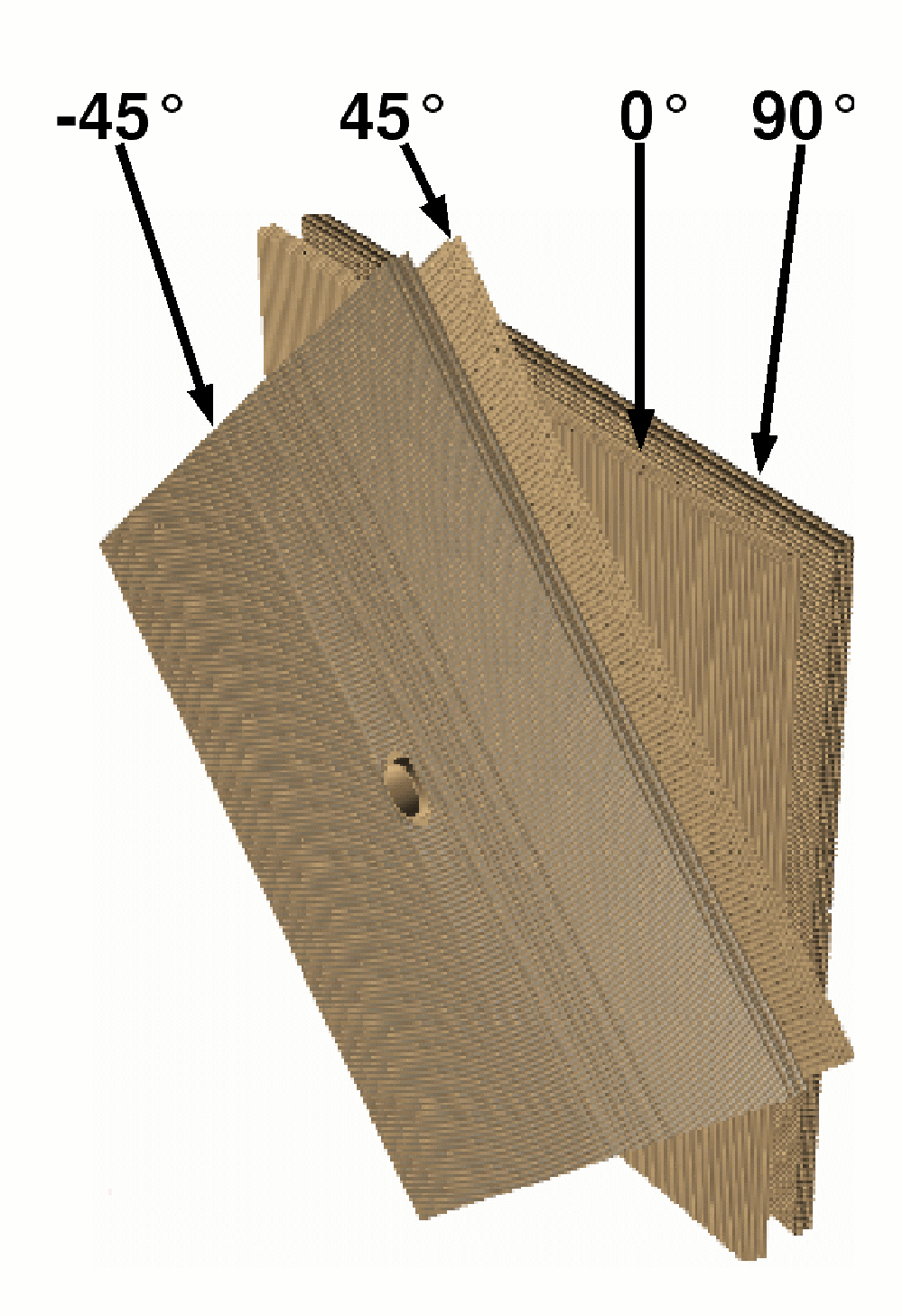}~\label{fig_FDb}}\\
\subfigure[The three layers of the Forward Trigger Hodoscope (left). The intersections of elements define pixels as indicated in the projection of the planes after hit by two particles (right)]{\includegraphics[width=10.0cm,height=5.0cm]{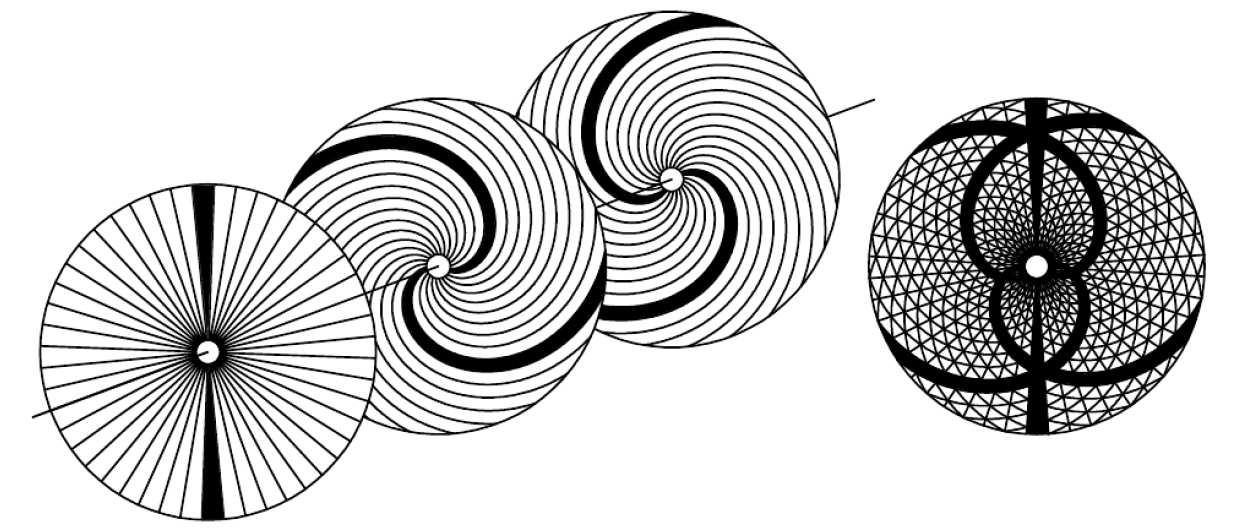}~\label{fig_FDc}} \\
\subfigure[The Forward Range Hodoscope.]{\includegraphics[width=7.0cm,height=6.0cm]{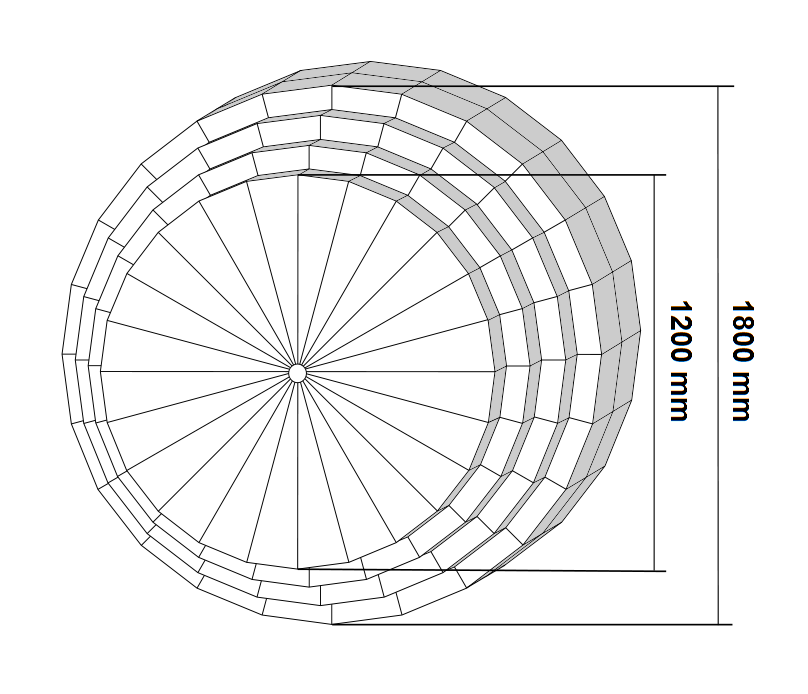}~\label{fig_FDd}}
\caption{Components of the Forward Detector.\label{fig_FD}}
\end{figure}

\noindent The Forward Window Counter (FWC) is the first detector of the Forward Part along the beam direction. It consists of two 3~mm layers, each of 24 plastic scintillators connected to the photomultipliers (PM) via light guides. The FWC is mounted on the paraboloidal stainless steel vacuum window. The first layer is shifted by half an element with respect to second one which is mounted perpendicularly to the beam direction (see Fig.~\ref{fig_FDa}). The Window Counter is used as a first level of the trigger logic which allows to reduce the background coming from the particles scattered downstream beam pipe. It is also one of the detector which can be employed in the $\Hea$ identification via the $\Delta E$--$\Delta E$ method. 

\vspace{0.6cm}

\noindent \textbf{Forward Proportional Chamber}

\vspace{0.2cm}

\noindent The Forward Proportional Chamber (FPC) located directly after FWC is a tracking device providing precise angular information about the particles outgoing from the target region (scattering angle resolution about 0.2$^{\circ}$). It is also used for the accurate reconstruction of the track coordinates of charged particles crossing through. The Chamber is composed of 4 modules, each with 488 proportional drift tubes (straws) of 8~mm diameter made of thin mylar foil and filled with argon-ethan gas mixture. The modules are rotated by 45$^{\circ}$ with respect to each other and their orientation is -45$^{\circ}$, +45$^{\circ}$, 0$^{\circ}$ and 90$^{\circ}$ with respect to the $x$ direction (see Fig.~\ref{fig_FDb}). 

\vspace{0.6cm}

\noindent \textbf{Forward Trigger Hodoscope}

\vspace{0.2cm}

\noindent The Forward Trigger Hodoscope (FTH) is third in the order sub-detector consisting of three layers of plastic sintillators. There are 48 radial elements in the first layer, closest to the FPC, and 24 elements in the form of archimedian spirals oriented clockwise and counter-clockwise in the last two planes (see Fig.~\ref{fig_FDc}). The FTH  provides for the trigger system angular information about the track based on the overlap of hit elements in three consecutive layers. Moreover, it gives information about the track multiplicity and is used for identification of the charged particles in the FD via energy loss.  

\vspace{0.6cm}

\noindent \textbf{Forward Range Hodoscope}

\vspace{0.2cm}

\noindent The five planes of Forward Range Hodoscope are positioned behind the FTH (see Fig.~\ref{fig_FDd}). Each of them consists of 24 plastic scintillator modules with thickness of 11~cm and 15~cm for first three and the last two layers, respectively. ~The energy resolution for particles stopped in the detector is about 3\%. The FRH together with FWC and FTH allows to determine the energy of charged particles stopped in detector or passing through.~The initial kinetic energy reconstruction and identification of charged particle is based on the energy deposited in the different detector planes ($\Delta E$--$\Delta E$ and $\Delta E$--$E$ methods).

\vspace{0.6cm}



\newpage

\newpage

\noindent \textbf{Forward Range Intermediate Hodoscope}

\vspace{0.2cm}

\noindent The Forward Range Intermediate Hodoscope (FRI) can be situated between the second and third layer of the FRH. This two-layer scintillator hodoscope, with modules rotated by 90$^{\circ}$ with respect to each other, delivers precise time and position information in two dimensions. This sub-detector was not used during the experiment dedicated to this thesis.

\vspace{0.6cm}

\noindent \textbf{Forward Veto Hodoscope}

\vspace{0.2cm}

\noindent The Forward Veto Hodoscope (FVH), being the last subdetector in FD, consists of two layers: one of 12 horizontal and second of 12 vertical plastic scintillator bars. Each bar is equipped with the photomultipliers on both sides. The distance between layers is 77~cm. The main goal of FRH is detection of high-energetic particles going through the FRH.

\vspace{0.6cm}

\noindent \textbf{Forward Absorber}

\vspace{0.2cm}

\noindent The Forward Absorber (FRA) can be located between the last layer of the FRH and the FVH. It is iron plane with thickness of usually 5-10~cm. The FRA is used for stopping the slower protons (for example from the $pp \rightarrow pp \eta$ reaction). The fast protons associated with the elastic scattering penetrate the Absorber and induce signals in the FVH which are used for veto purposes in the trigger. The absorber was also not used during the described experiment.

\subsubsection{Central Detector (CD)}

\noindent The Central Detector is built around the interaction point and designed mainly for measurements of photons and charged particles originating from $\pi^{0}$ and $\eta$ mesons decays.~The charged particles momenta and reaction vertex are determined by means of Mini Drift Chamber (MDC). Charged particles are here bending in the magnetic field provided by surrounding Superconducting Solenoid (SCS). First their trajectories are reconstructed, and then knowing the magnetic field, the momentum vector is reconstructed. The identification of charged particles is based on information about energy deposited by particles in Plastic Scintillator Barrel (PSB) and in Scintillator Electromagnetic Calorimeter (SEC). The calorimeter is also used for the photon identification.

\vspace{0.6cm}

\noindent \textbf{Mini Drift Chamber}

\vspace{0.2cm}

\begin{figure}[h!]
\centering
\subfigure[MDC inside the Al-Be cylinder.]{\includegraphics[width=6.0cm,height=5.0cm]{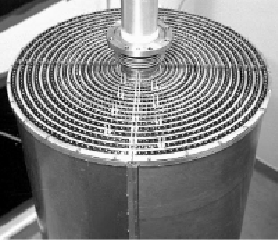}~\label{fig_CDa}}
\subfigure[The Plastic Scintillator Barrel. Endcaps are marked with yellow and red colours.]{\includegraphics[width=6.0cm,height=5.0cm]{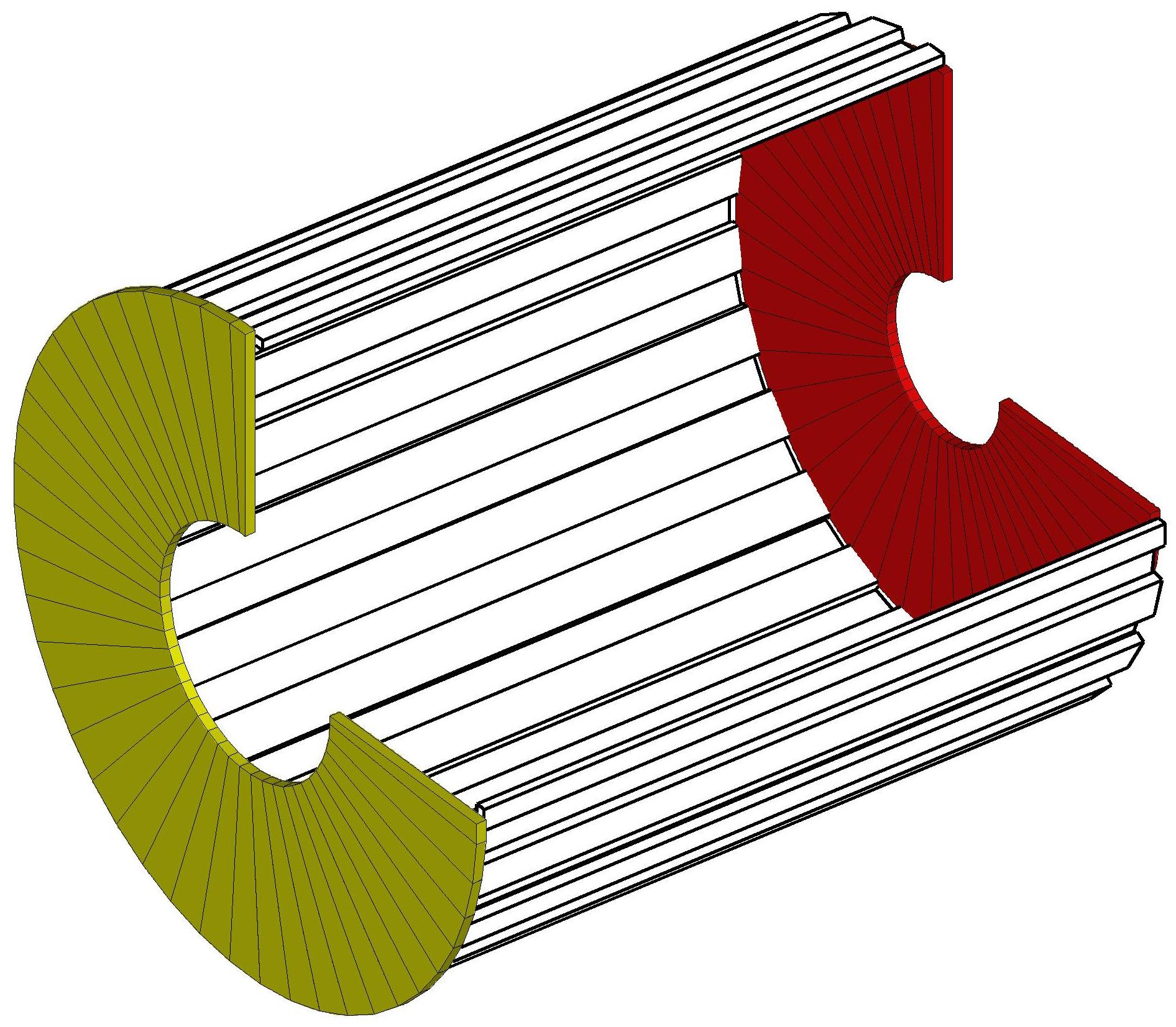}~\label{fig_CDb}} \\
\subfigure[Cross section of the Scintillating Electromagnetic Calorimeter.]{\includegraphics[width=6.0cm,height=5.0cm]{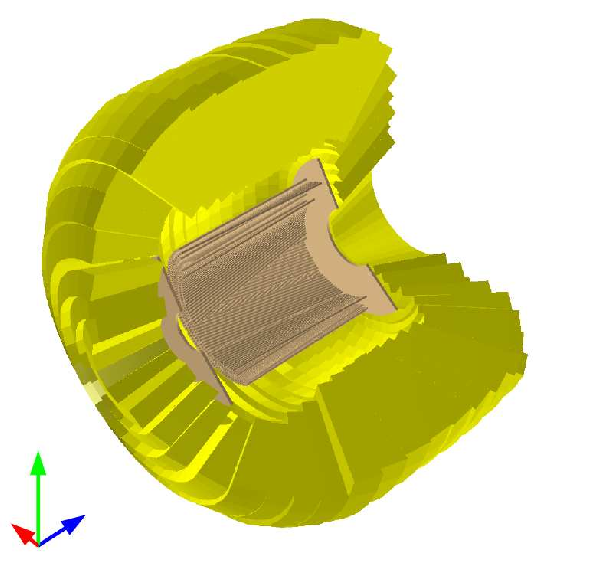}~\label{fig_CDc}}
\subfigure[Angular coverage of the SEC. The numbers above the picture indicate the numbers of crystals while their size is marked on the vertical axis.]{\includegraphics[width=6.0cm,height=4.0cm]{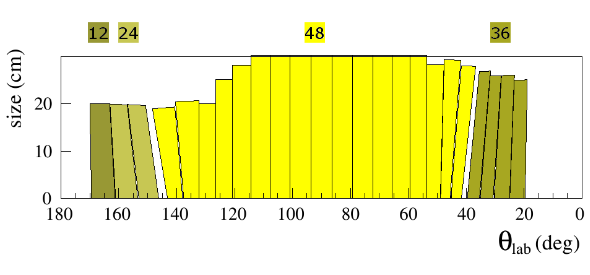}~\label{fig_CDd}} 

\vspace{0.3cm}

\caption{Components of the Central Detector.\label{fig_CD}}
\end{figure}

\noindent The Mini Drift Chamber (MDC) is a sub-detector placed around the 60~mm diameter beryllium beam pipe, close to the interaction region and is covered by 1~mm thick Al-Be cylinder (see Fig.~\ref{fig_CDa}). The chamber consists of 1738 straw tubes arranged in 17 layers covering scattering angles from 24$^{\circ}$ to 159$^{\circ}$. The straw diameter is 4~mm, 6~mm and 8~mm in first five inner layers, in the 6 middle layers and in 6 outer layers, respectively. The straws are made of 25~$\mu$m thin aluminized mylar foil and surround the \mbox{20~$\mu$m} diameter gold plated tungsten anode wire. The first nine inner layers are parallel with respect to the beam axis while the next layers are situated with small skew angles \mbox{(6$^{\circ}$ to 9$^{\circ}$)}. The straws are filled with gas mixture containing argon - ethane in ratio 80\%-20\%. The MDC is immersed in the magnetic field provided by the Superconducting Solenoid which causes the bending of charged particles trajectories. The main purpose of MDC is determination of particle momenta and reaction vertex position. Detailed information about the MDC can be found in Ref.~\cite{Jacewicz_MDC}.

\vspace{0.6cm}

\noindent \textbf{Plastic Scintillator Barrel}

\vspace{0.2cm}

\noindent The Plastic Scintillator Barrel (PSB) is mounted inside the Solenoid Coil and surrounds the Drift Chamber (see Fig.~\ref{fig_CDb}). It consists of three parts: cylindrical central part (48 scintillator bars) and two endcaps (48 "cake-piece" shaped elements each) covering almost 4$\pi$ solid angle.~The main aim of PSB is distinction between neutral and charged tracks as well as identification of charged particles via $\Delta E$--$E$ method using total energy information in Calorimeter and $\Delta E$--$p$ method based on momentum information from MDC.

\vspace{0.6cm}

\noindent \textbf{Superconducting Solenoid}

\vspace{0.2cm}

\noindent The Superconducting Solenoid (SCS) installed inside the calorimeter provides the magnetic field for the momentum reconstruction of the tracks measured by the MDC. The SCS is cooled with the liquid helium and produces the magnetic field up to 1.3~T. A detailed description of the solenoid is presented in~\cite{Ruber_SCS}.

\vspace{0.6cm}

\noindent \textbf{Scintillation Electromagnetic Calorimeter}

\vspace{0.2cm}

\noindent The Scintillation Electromagnetic Calorimeter (SEC) is situated between the SCS and the iron yoke which covers the Central Detector. It is composed of 1012 sodium-doped CsI scintillating crystals and covers the scattering angles from 20$^{\circ}$ to 169$^{\circ}$. The crystals have shape of a truncated pyramid and are organized in 24 layers. One can distinguish the three main parts of the calorimeter: the central with the longest crystals (30~cm), forward build of crystals having 25~cm length and the backward consisting of the shortest crystals (20~cm). The cross-section of SEC and its angular coverage are presented in Fig.~\ref{fig_CDc} and Fig.~\ref{fig_CDd}, respectively. The calorimeter is used for the energy measurement of charged and neutral particles with resolution of 3\% for stopped charged particles and  8\% for 0.1~GeV photons. Together with MDC and PSB, SEC is used for the charged particles identification based on information about their deposited energy. Detailed description of the SEC is presented in~\cite{Koch}.


\subsection{Data Acquisition System (DAQ)}

\begin{figure}[h!]
\centering
\includegraphics[width=13.0cm,height=8.5cm]{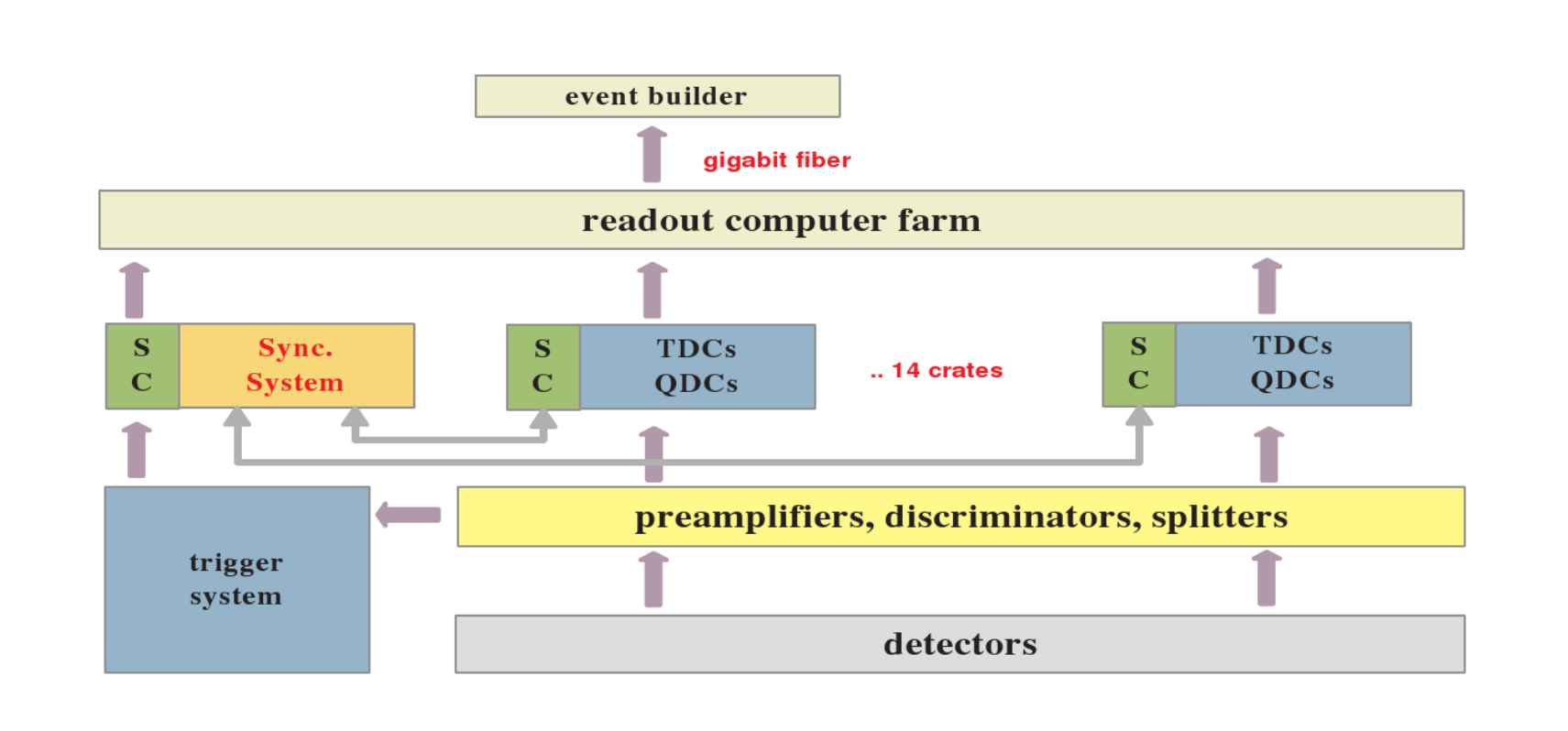} 
\caption{A scheme of the Data Acquisition system (DAQ) for WASA-at-COSY. The figure is taken from~\cite{Kleines2}. \label{fig_DAQ}}
\end{figure}

The main goal of Data Acquisition system is proper processing of the signals from the detector components in order to make them accessible for the data analysis. The DAQ system is based on the third generation of the DAQ systems at COSY and is optimized to cope with the high luminosities~\cite{Kleines1}. The structure of WASA DAQ is schematically presented in Fig.~\ref{fig_DAQ}.\\

The readout electronic based on  Field Programmable Gate Array FPGAs used for digitization and buffering of data  allows to reach event rates of 10~kHz at a live time of~80\% of the system~\cite{Redmer_PhD}. Signals from straws and photomultipliers connected with detectors, are distributed and adapted by front-end electronics (preamplifiers, splitters, discriminators).~Subsequently, the analogue signals from front-end cards are digitized by means of QDC (Charge-to-Digital Converter) and TDC (Time-to-Digital Converter) read out modules located in 14 crates. Next, the digitized signals are marked with a time stamps and put in FIFO ("First In First Out") queue. Synchronization System (SC), called by trigger, computes the event number and send it together with its time stamp to all QDC and TDC boards. Signals with a matching time stamp are marked with the same event number and passed to the computer readout and to the event builder. Finally, the event are written on the discs. The technical details of the DAQ architecture are presented in Refs.~\cite{Kleines1,Kleines2}. 

\section{Analysis tools}

For the purpose of investigations made in this thesis, the events generators for each of considered reactions were prepared based on proper kinematic models. The simulations of the WASA detector response have been carried out with the WASA Monte Carlo (WMC) software based on GEANT package \cite{Geant}. The data analysis was performed within the RootSorter framework \cite{RootSorter} based on the data analysis package ROOT \cite{Root} developed at CERN (the European Organization for Nuclear Research).~The ROOT was used for calculations, fitting and preparing the histograms shown in this thesis.


\section{Beam settings~\label{Sec_Beam}}

The experimental proposal~\cite{Proposal_old} dedicated for the search of $\BSbound$ in $\BcgReacta$ and $\BcgReactW$ reactions with WASA-at-COSY facility was accepted for realisation by Programme Advisory Committee in Forschungszentrum J{\"u}lich in Germany. The two weeks experiment was carried out at the turn of November and December 2010. The data were effectively taken about one week, while the rest of time was spent for the accelerator cycle preparation, beam and experimental triggers adjustments and pellet target regenerations.   

\indent During the experimental run the momentum of the deuteron beam was varied continuously within each acceleration cycle from 2.127~GeV/c to 2.422~GeV/c, crossing the kinematic threshold for the $\eta$ production in the $\Heb\eta$ reaction at 2.336~GeV/c. This beam momentum range corresponds to the excess energy range of interests $Q$ $\in$ (-70,30) MeV ($Q$~=~0 MeV denotes the threshold). For the purpose of data analysis this range was divided into 20 intervals. The settings of the beam cycle are summarized in Table~\ref{Beam_sett}.

\indent The total time of each acceleration cycle in the experimental run has a length of 70.3~s.~At the beginning of the cycle, the beam is accelerated in 5.7~s to the momentum of 2.127~GeV/c via fast ramping. Subsequently, the beam momentum is increased slowly from the value of 2.127~GeV/c to 2.422~GeV/c and this ramping phase takes 57~s. At time $t_{cycle}=5.5$ s the vacuum shutters of the Pellet Target are opened and the acquisition system starts recording data. In the 63.1'th second of the cycle duration Pellet stream is blocked (shutters are closed) while the data taking continues until 66.1~s. Then the DAQ is stopped and the detector voltages are switched off before the beam is decreased.

\begin{table}[h!]
\begin{normalsize}
\begin{center}
\begin{tabular}{l l}\hline
\hline
\vspace{0.2cm}
beam particles &deuterons\\
\vspace{0.2cm}
beam momentum range  &2.127-2.422 GeV/c  \\
\vspace{0.2cm}
beam cycle length &70.3 s\\
\vspace{0.2cm}
start slow ramping & 5.7 s \\
\vspace{0.2cm}
slow ramping time & 57 s \\
\vspace{0.2cm}
start DAQ  & 5.5 s \\
\vspace{0.2cm}
stop DAQ  & 66.1 s \\

data taking within cycle   &60.6 s (86\%) \\

\hline 
\hline
\end{tabular}
\end{center}
\vspace{-0.8cm}
\begin{center}
\caption{Overview of the accelerator cycle parameters for the experimental run.\label{Beam_sett}}
\end{center}
\end{normalsize}
\end{table}

\vspace{-1.0cm}  


\section{Detector Calibration}

The crucial point in the data analysis of the main considered reactions $\BcgReacta$ (Chapter~\ref{Analysis}) and $\LumReacta$ (Sec.~\ref{Lum_integrated}) is an identification of $\Hea$ ion registered in the Forward Detector and the determination of its four-momentum. The kinetic energy of helium is calculated based on the energy losses in the consecutive Forward Range Hodoscope layers. Therefore, it is very important to use precise energy calibration of the FRH. The calibration of the FRH (layer 3 and 4) is described in details in the first subsection. The description of the Electromagnetic Calorimeter calibration optimized for the reconstruction of photons and hence of $\pi^{0}$ mesons is presented in second subsection.

\subsection{Forward Range Hodoscope}
 
\begin{figure}[h!]
\centering
\includegraphics[width=6.8cm,height=5.0cm]{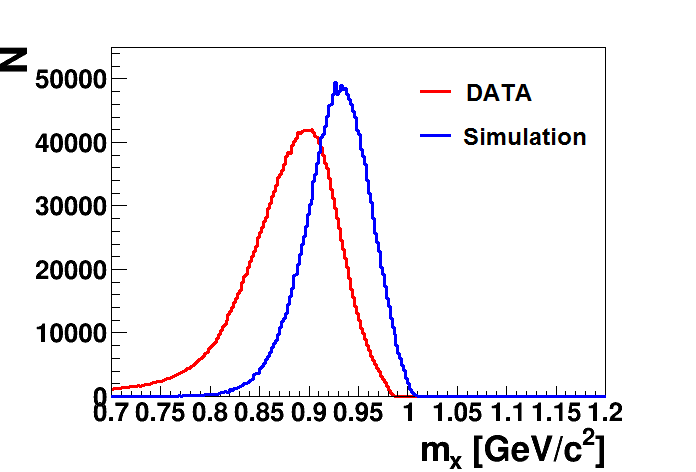}  \includegraphics[width=6.8cm,height=5.0cm]{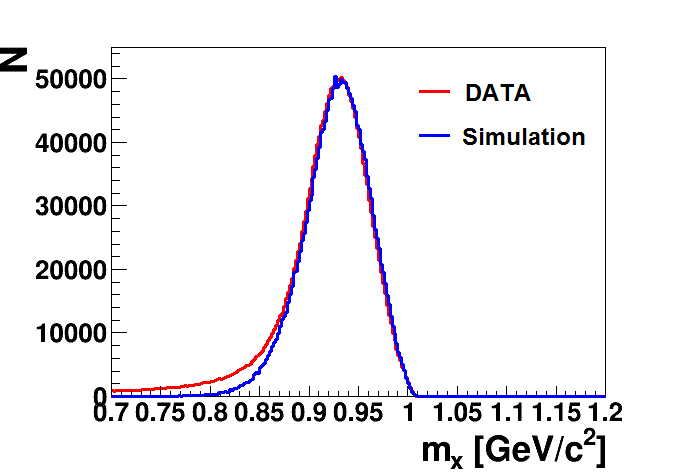}\\
\caption{The missing mass $m_{x}$ spectrum for $dd\rightarrow$ $^{3}\hspace{-0.03cm}\mbox{He} X$ reaction before (left panel) and after calibration tuning (right panel). Data is marked with red line while Monte Carlo simulation is marked with blue line.\label{fig_mm}}
\end{figure}

During the data analysis the calibration of plastic scintillator detectors, based on the conversion at ADC channels into deposited energy~\cite{FD_calib} was taken into account. However, it was noticed that the calibration is incorrect for Forward Range Hodoscope layer 3 and 4 in which high energetic helium outgoing from $\LumReacta$ reaction is stopped. It is shown in left panel of Fig.~\ref{fig_mm} as disagreement in the missing mass spectra of $dd\rightarrow$ $^{3}\hspace{-0.03cm}\mbox{He} X$ reaction obtained from Monte Carlo simulations and from experimental data.   

For the purpose of this analysis, the proper correction for the FRH3 and FRH4 calibration was applied. The spectra of energy deposited Edep(FRH3) and Edep(FRH4) obtained for data were compared with the spectra obtained for WASA Monte Carlo simulations for $\LumReacta$ reaction. The comparison was carried out for 5 intervals of polar angle in Forward Detector $\theta_{FD}$ in range between 3$^{\circ}$ and 10.5$^{\circ}$ and 20 intervals of excess energy $Q$ between -70~MeV and 30~MeV. Edep(FRH3) and Edep(FRH4) spectra for Monte Carlo simulations and data, for each interval, were fitted with gaussian functions in order to find the maxima positions -- $x_{MC}$ and $x_{D}$, respectively. Subsequently, peak position for data was shifted by offset $A$ to fit the peak position obtained from simulations:


\begin{equation}
x_{D}-A=x_{MC}.~\label{trzy}
\end{equation} 

\vspace{0.5cm}

\begin{figure}[h!]
\centering
\includegraphics[width=7.0cm,height=4.8cm]{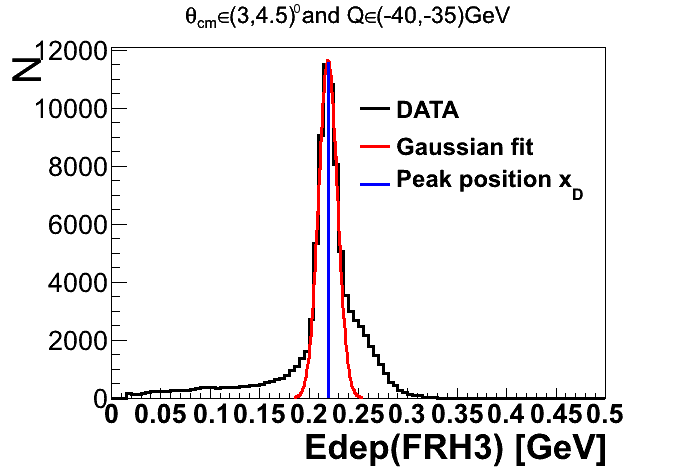}
\hspace{-0.5cm} \includegraphics[width=7.0cm,height=4.8cm]{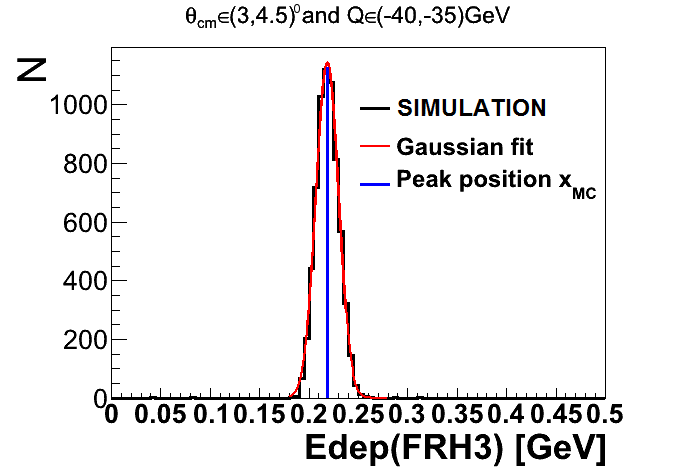}\\ 
\vspace{0.3cm}
\includegraphics[width=7.0cm,height=4.8cm]{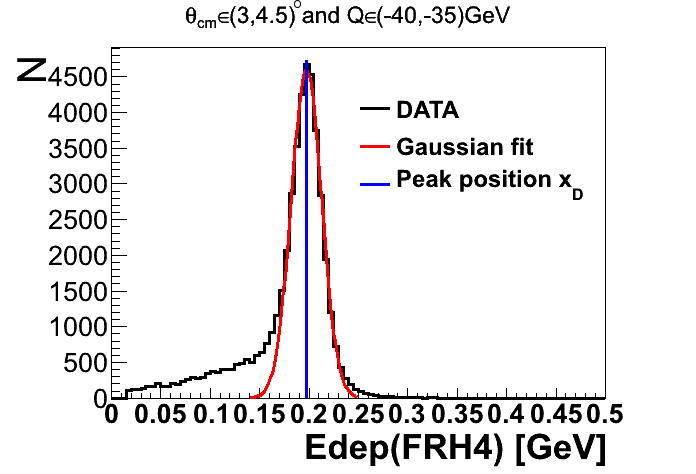}
\hspace{-0.5cm} \includegraphics[width=7.0cm,height=4.8cm]{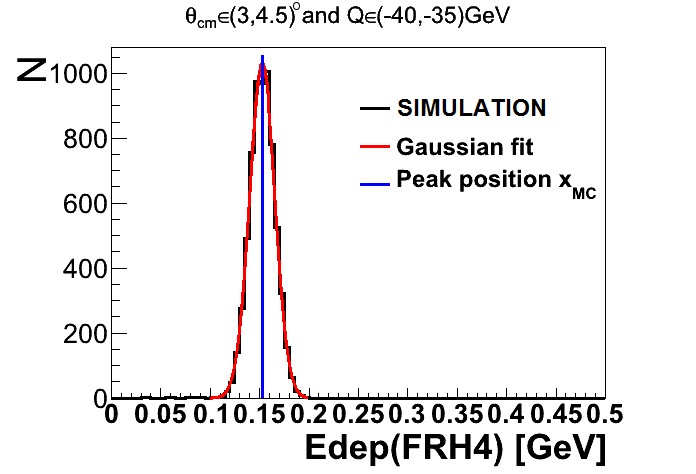}
\caption{Spectra of Edep(FRH3) (upper panel) and of Edep(FRH4) (lower panel) for data and Monte Carlo simulation, respectively. Fitted gaussian functions are marked with red line, while peak positions are marked with blue line.~The spectra are obtained for $\theta_{FD}\in$ (3,4.5)$^{\circ}$ and $Q$ $\in$ (-40,-35)~MeV.~\label{fig_edep}}
\end{figure}

\indent Exemplary distributions of Edep(FRH3) and Edep(FRH4) for one of the chosen $\theta_{FD}$ and $Q$ interval with applied fit are presented for Monte Carlo simulation and data in Fig.~\ref{fig_edep}. Missing mass spectra for Monte Carlo simulation and data after all cuts described in Sec.~\ref{Lum_integrated} with applied calibration tuning fit very well, what is shown in right panel of Fig.~\ref{fig_mm}.

\newpage

\subsection{Electromagnetic Calorimeter~\label{SEC_calib}}

The electromagnetic calorimeter is used for measurement of photons emission angles and energies. The preliminary SEC calibration was performed based on the measurement of cosmic muons and radioactive sources before the WASA installation at COSY~\cite{Koch,Jany}. In order to optimize the photons four-momenta reconstruction, the energy calibration was carried out based on determination of the invariant mass of neutral pions $\pi^{0}$. For this purpose events with exactly two "neutral" clusters in the Central Detector are selected and considered as originating from gamma quanta. Their invariant mass is calculated according to below formula: 

\begin{equation}
m_{\gamma_{1}\gamma_{2}}=\sqrt{2E_{\gamma_{1}}E_{\gamma_{2}}(1 -cos\theta_{1,2})}.
\end{equation} \label{eq:MM_pi0}

\noindent where $E_{\gamma_{1},\gamma_{2}}$ are the measured energies of the photons based on the initial calibration while $\theta_{1,2}$ is the opening angle between the photons momenta. For each gamma quanta pair, being $\piz$ candidate, the invariant mass is assigned to the crystals with the largest energy deposit in the cluster.

\begin{figure}[h!]
\centering
\includegraphics[width=12.0cm,height=8.0cm]{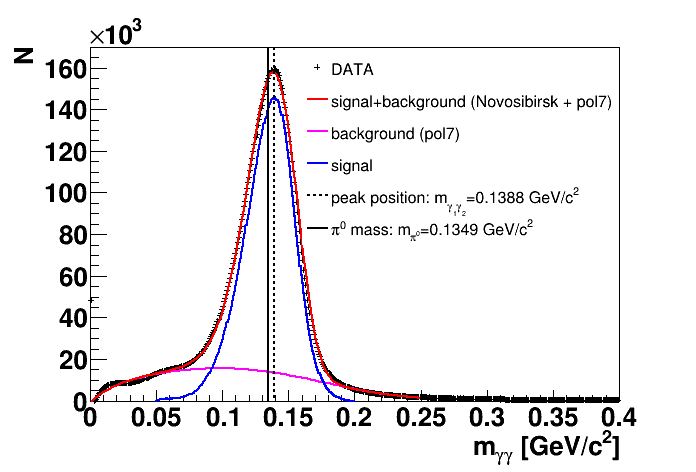} 
\vspace{-0.5cm}
\caption{Invariant mass spectrum for two gamma quanta in Central Detector. The red line shows fit to the signal and background while magenta line shows polynomial fit to the background. Signal peak after background subtraction  is marked as a blue line. The vertical dotted line shows subtracted peak position while the vertical solid line shows the $\pi^{0}$ mass.~\label{fig_fit}}
\end{figure}

\indent In order to apply a global correction for initial calibration, the distribution of invariant mass of two gamma quanta for whole data sample was reconstructed.~The peak position was determined via fitting the sum of signal and background function to the spectrum, what is presented in Fig.~\ref{fig_fit}. 

The signal was fitted with a Novosibirsk function which is described by formula~(\ref{eq_novv1}) and~(\ref{eq_novv2}).

\begin{equation}
P(x)=e^{-0.5(lnq_{y})^{2}/{\Lambda^{2}}+\Lambda^{2}}~\label{eq_novv1},
\end{equation}

\begin{equation}
q_{y}=1+\Lambda(x-x_{0})/\sigma \times \frac{sinh(\Lambda \sqrt{ln4})}{\Lambda \sqrt{ln4}} ~\label{eq_novv2},
\end{equation}
\noindent
where:\\

\noindent
$x_{0}$-peak position,\\
$\sigma$- width of the peak,\\
$\Lambda$-tail.\\

\indent The background was fitted as a seven degree polynomial (magenta line). The total fit (signal + background) is marked in Fig.~\ref{fig_fit} as a red line, while the signal after background subtraction is shown as a blue line. The subtracted peak position $m_{\gamma_{1}\gamma_{2}}$ is depicted as a dotted line and equals 0.1388~GeV/c$^{2}$. The deviation from the actual invariant mass of $\pi^{0}$ is used to set the values of the calibration factor $k$, being the ratio of energy for $\gamma_{1}$ and $\gamma_{2}$ after applied correction to uncorrected energy  $\left(\frac{E^{corr}_{\gamma_{1},\gamma_{2}}}{E_{\gamma_{1},\gamma_{2}}}\right)$, using the formula:

\begin{equation}
k=\frac{m_{\pi^{0}}}{m_{\gamma_{1}\gamma_{2}}}.
\end{equation}

\noindent
The calibration correction factor is applied for each crystal of the calorimeter.


\section{Data Preselection}

Data preselection was carried out in two levels: hardware trigger level and the selection of the raw data with conditions customized for the present analysis. The both levels are discussed in the corresponding sections of this chapter.

\subsection{Trigger settings\label{Sec_Triggers}} 

The main aim of the hardware trigger system is the reduction of the initial event rate to the level that make it possible to be stored on disks, while selecting the events corresponding to the reaction of interest. The trigger conditions are related to multiplicities, coincidences, track matching and energy losses in the plastic as well as to the cluster multiplicities and energy deposition in the electromagnetic calorimeter. 
In present experiment several triggers were set. The main trigger \textit{fHedwr1} selected events with at least one charged particle in the Forward Detector, which corresponds to the track matching between FWC, FTH and FRH and in addition with a high energy loss in the first layer of the Forward Window Counter. The trigger was dedicated for the study of all processes with helium ions in the final state, especially the \mbox{$\MainReact$} (Chapter~\ref{Analysis}) and $\LumReacta$ (Sec.~\ref{Lum_integrated}) reactions. The selection of charged particle with a high energy losses, allowed to suppress significantly the background coming from fast protons and deuterons in FD, for which the deposited energy is small and close to minimum ionizing particle energy loss.

Additional trigger \textit{frhb1|psc1} was used for the luminosity studies with quasi-free $pp$ scattering reaction (Sec.~\ref{Lum_qf}). It required at least one charged particle detected in FD, as well as at least one charged particle in the Central Detector. The prescaling factor for this trigger was equal to 1/4000.


\subsection{Preselection conditions} \label{Sec_Presel}

\begin{figure}[h!]
\centering
\includegraphics[width=7.0cm,height=4.8cm]{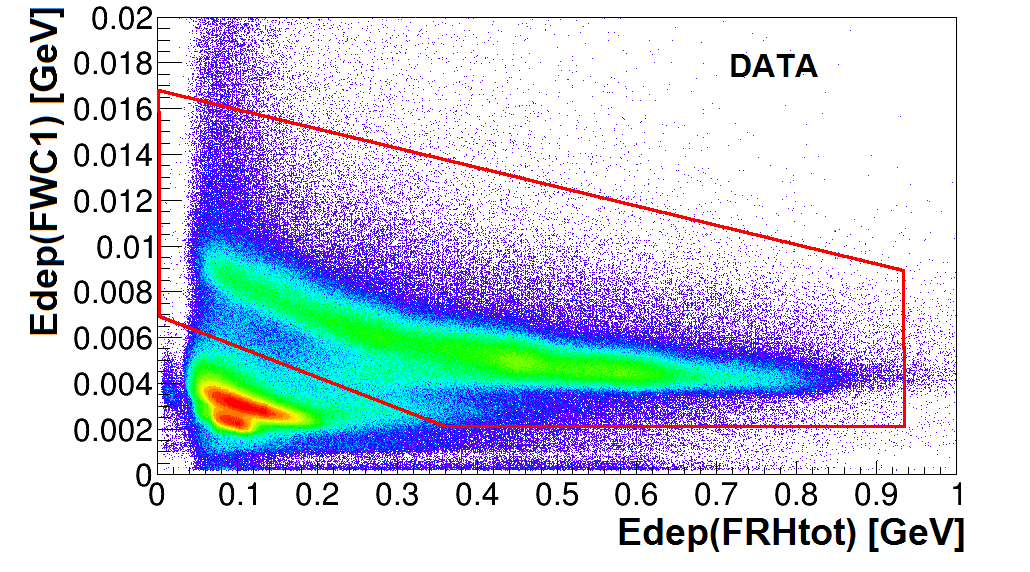}\\
\vspace{0.3cm}
\includegraphics[width=7.0cm,height=4.8cm]{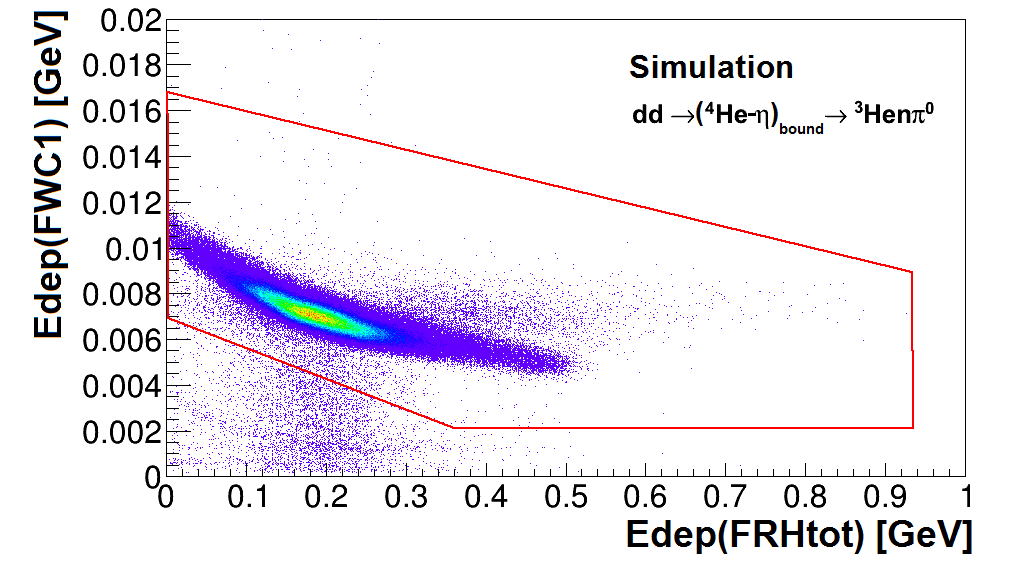} \hspace{-0.3cm}
\includegraphics[width=7.0cm,height=4.8cm]{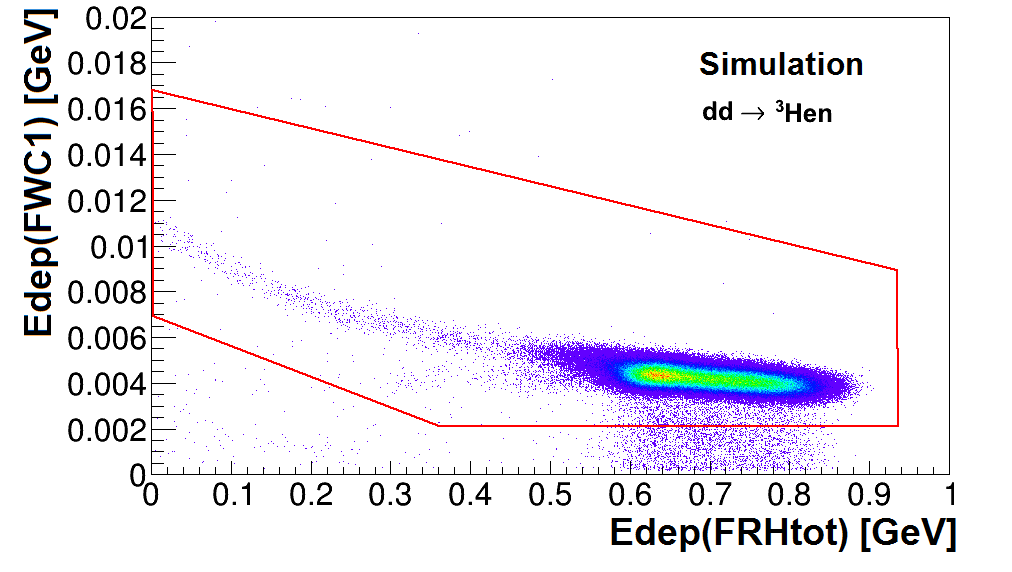}\\
 
\caption{Cut applied in Edep(FWC1) vs. Edep(FRH$_{tot}$) spectrum for: experimental data (upper panel) and WASA Monte Carlo simulations of the main reaction $dd\rightarrow(^{4}\hspace{-0.03cm}\mbox{He}$-$\eta)_{bound}\rightarrow$ $^{3}\hspace{-0.03cm}\mbox{He} n \pi{}^{0}$ (left lower panel), and the normalization reaction $dd\rightarrow$ $^{3}\hspace{-0.03cm}\mbox{He} n$ (right lower panel).~\label{fig_presel}}
\end{figure}

\noindent The data selected by the hardware trigger still includes a large sample of background events. In order to reduce them and also to limit the computation time a preselection of the raw data was carried out. It was performed to select only events corresponding to reactions with $\Hea$ stopped in Forward Detector e.~g. $\MainReact$ and $\LumReacta$. The conditions applied in the preselection are following:

\begin{itemize}

\item Exactly one track corresponding to charge particle in Forward Detector,

\item Polar angle of the track $\theta_{FD}\in(3,18)^{\circ}$ corresponding to the FD acceptance,

\item Graphical cut in Edep(FWC1) vs. Edep(FRH$_{tot}$) spectrum (energy loss in the first layer of Forward Window Counter (FWC1) versus total energy deposited in Forward Range Hodoscope (FRH)) in order to reduce the background associated with protons and charged pions (Fig.~\ref{fig_presel}), 

\item Based on the Monte Carlo simulations, thresholds for the energy deposited in the following layers of Forward Detector were set to the values presented in Table~\ref{table_thr_pres}.

\end{itemize}

\vspace{0.5cm}

\begin{table}[h!]
\begin{normalsize}
\begin{center}
\begin{tabular}{|c|c|c|c|}\hline
FD layer  &threshold [MeV] &FD layer &threshold [MeV]\\
\hline 
FWC1 &0.18 &FRH1 &4.0 \\
FWC2 &0.18 &FRH2 &2.5 \\
FTH1 &1.5  &FRH3 &2.5 \\
FTH2 &0.32 &FRH4 &3.5 \\
FTH3 &0.3  &FRH5 &4.0 \\
\hline
\end{tabular}
\end{center}
\vspace{-0.3cm}
\caption{Thresholds for the energy deposited in the layers of Forward Detector.\label{table_thr_pres}}
\end{normalsize}
\end{table}

\newpage

\chapter{Simulation of the $\mathbold{dd\rightarrow(^{4}}\hspace{-0.03cm}\mathbold{\mbox{He}}$-$\mathbold{\eta)_{bound}\rightarrow}$ $\mathbold{^{3}}\hspace{-0.03cm}\mathbold{\mbox{He} n \pi{}^{0}}$ reaction~\label{Sim_MainReact}} 

Present \hspace{0.01cm} chapter \hspace{0.01cm} is \hspace{0.01cm} devoted \hspace{0.01cm} to \hspace{0.01cm} the \hspace{0.01cm} Monte \hspace{0.01cm} Carlo \hspace{0.01cm} simulations \hspace{0.01cm} of \hspace{0.01cm} the \hspace{0.01cm} $dd\rightarrow$\\
\mbox{$({}^{4}\hspace{-0.03cm}\mbox{He}$-$\eta)_{bound} \rightarrow$ $^{3}\hspace{-0.03cm}\mbox{He} n \pi{}^{0}$} process performed based on the kinematic model of the $\eta$-mesic helium production and decay. It also includes description of the nucleon momentum distribution inside $\Heb$ applied in simulations as well as the comparison of these distributions determined for different models.\\

\section{Kinematics of the $\mathbold{\eta}$-mesic bound state formation and decay~\label{Sym_Kin}}

\begin{figure}[h!]
\centering
\includegraphics[width=14.5cm,height=8.0cm]{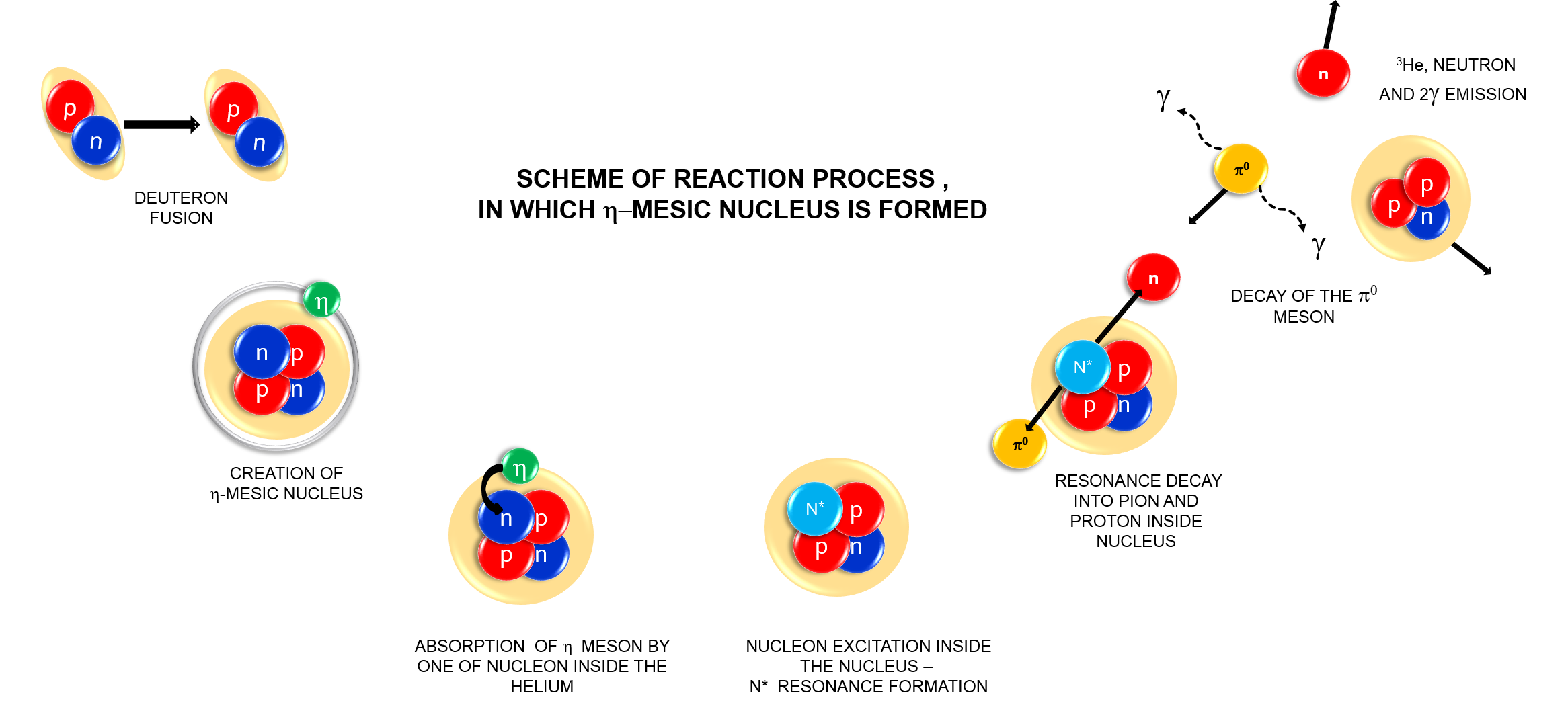}
\caption{Scheme of the $^{4}\hspace{-0.03cm}\mbox{He}$-$\eta$ bound state production and decay in \mbox{$dd\rightarrow$ $^{3}\hspace{-0.03cm}\mbox{He} n \pi{}^{0}$} reaction.\label{free_reaction}}
\end{figure}

\noindent We consider the production of the $^{4}\hspace{-0.03cm}\mbox{He}$-$\eta$ bound state in deuteron-deuteron fusion process.~The mechanism of the reaction is presented schematically in Fig.~\ref{free_reaction}.~According to the scheme, the deuteron from the beam hits the deuteron in the target.~The collision leads to the formation of $^{4}\hspace{-0.03cm}\mbox{He}$ nucleus bound with the $\eta$ meson via strong interaction. The mass of a created bound state is a sum of $\eta$ and $^{4}\hspace{-0.03cm}\mbox{He}$ masses reduced by binding energy $B_{s}$:

\begin{equation}
m_{(^{4}\hspace{-0.05cm}He-\eta)_{bound}}=m_{\eta}+m_{^{4}\hspace{-0.05cm}He}-B_{s}.
\end{equation}\\

\newpage

\vspace{0.5cm}

\noindent
Then, the $\eta$ meson can be absorbed by one of the nucleons inside helium and may propagate in the nucleus via consecutive excitation of nucleons to the $N^{*}(1525)$ state~\cite{Sokol} until the resonance decays into the neutron-$\pi^{0}$ pair, and subsequently $\pi^{0}$ meson decays into two $\gamma$ quanta. It is assumed, that, just before the decay, $N^{*}$ resonance momentum distribution can be well approximated by the Fermi momentum distribution for nucleons inside $\Heb$.~The $\Hea$ plays the role of a spectator which according to the momentum conservation in the $\Heb$ system moves with the Fermi momentum in the opposite direction to the $N^{*}$ resonance. The spectator is considered as a real particle registered in the experiment and in the analysis it is assumed that it is on its mass-shell during the reaction ($\left|\mathbb{P}_{^{3}\hspace{-0.03cm}He}\right|^2 = m_{^{3}\hspace{-0.03cm}He}^2$)~\cite{JKlaja,Moskal1}. A very accurate description of the $\eta$-mesic helium production and decay kinematics including appropriate calculations is presented in Ref.~\cite{Skurzok_Master}.

\section{Simulation scheme~\label{Sim_scheme}}

\noindent The simulation of the $\MainReact$ reaction based on kinematics presented in previous section, can be schematically described in following points:

\begin{enumerate}

\item The deuteron beam momentum $p_{beam}$ is generated with uniform probability density distribution in the range of $p_{beam}\in$ (2.127,2.422)~GeV/c, which corresponds the experimental beam ramping, and then the square of invariant mass of the colliding deuterons $s_{dd}$ is calculated from the beam and target four-momenta ($\mathbb{P}_{d}^{\,b}$, $\mathbb{P}_{d}^{\,t}$) using the formula:

\begin{equation}
s_{dd}=|\mathbb{P}_{d}^{\,b}+\mathbb{P}_{d}^{\,t}|^{2}=2m_{d}\left(m_{d}+\sqrt{m^{2}_{d}+|\vec{p_{b}}|^{2}}\right),
\label{eq_3}
\end{equation}

\noindent where: $m_{d}$ denotes the deuteron mass.

\item It is assumed that the considered bound state has a resonance-like structure with fixed binding energy $B_{s}$ and width $\Gamma$.~Therefore, the invariant mass of the whole system $\sqrt{s_{dd}}$ is distributed randomly according to the Breit-Wigner distribution which is given by formula~(\ref{eq:BW_jeden}) and shown in Fig.~\ref{BW_jeden}:

\begin{equation}
N\left(\sqrt{s_{dd}}\right)=\frac{\Gamma^{2}/4}{\left(\sqrt{s_{dd}}-m_{(^{4}\hspace{-0.05cm}He-\eta)_{bound}}\right)^{2}+\Gamma^{2}/4}~\label{eq:BW_jeden}.
\end{equation}

\vspace{0.5cm}

\begin{figure}[h!]
\centering
\includegraphics[width=13.0cm,height=8.0cm]{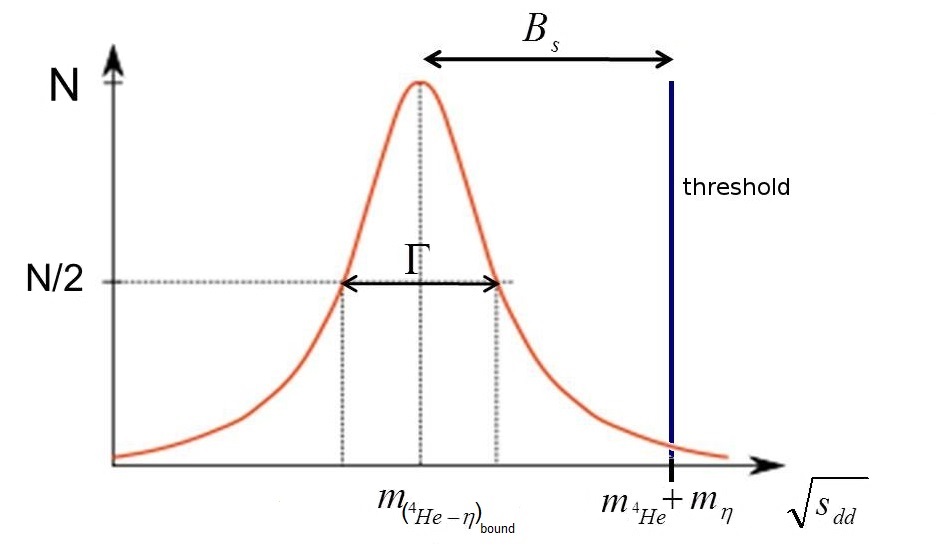}
\caption{Breit-Wigner distribution of the invariant mass $\sqrt{s_{dd}}$ of the bound state system.\label{BW_jeden}}
\end{figure}

\newpage

\item The $N^{*}$ resonance momentum is distributed isotropically in spherical coordinates of $\eta$-mesic nucleus \mbox{(${p}^{\,\,*}_{F}$, $\theta^{*}$, $\phi^{*}$)} with Fermi momentum distribution of nucleons inside $^{4}\hspace{-0.03cm}\mbox{He}$ which is presented for three different models in Fig.~\ref{hel_4} and described in details in next section. 

\item The $^{3}\hspace{-0.03cm}\mbox{He}$ four-momentum vector is calculated in the CM frame based on the momentum conservation and spectator model assumption.

\item The resonance mass $m_{{N}^*}$ is calculated based on invariant mass $\sqrt{s_{dd}}$ and Fermi momentum ${\vec{p}}^{\,\,*}_{F}$ values according to equation~(\ref{eq:102}):

\begin{equation}
m_{{N}^*}=\left(s_{dd}+m^{2}_{^{3}\hspace{-0.05cm}He}-2\sqrt{s_{dd}}\sqrt{m^{2}_{^{3}\hspace{-0.05cm}He}+|\vec{p}^{{\,\,*}}_{F}}|^{2}\right)^{\frac{1}{2}}.
\label{eq:102}
\end{equation}

\item The neutron and pion momentum vectors are simulated isotropically in the $\mbox{N}^{*}$ frame in spherical coordinates. The absolute value of $\vec{p}^{\,\,**}_{n,\pi^{0}}$ is fixed by the equation~(\ref{eq:101}): 

\begin{equation}
|\vec{p}^{\,\,**}_{n,\pi^{0}}|=\frac{\sqrt{\lambda(m^{2}_{{N}^*},m^{2}_{{\pi}^{0}},m^{2}_{{n}})}}{2m_{{N}^*}},
\label{eq:101}
\end{equation}

\noindent
where $\lambda(x,y,z)=(x-y-z)^{2}-4yz$~\cite{Byckling}.

\item The $\gamma$ quanta are simulated isotropically in the $\pi^{0}$ frame in spherical coordinates. The absolute value of $\gamma$s momentum is equal to \mbox{$\vec{p}^{\,\,***}_{\gamma}=m_{\pi^{0}}/2$}.

\item The four-momentum vectors of all ejectiles are transformed into the laboratory frame using the Lorentz transformation. 

\item Simulation of the detection system response is carried out for generated events using a GEANT (WASA Monte Carlo) simulation package.

\end{enumerate}


\vspace{0.5cm}

\begin{figure}[h!]
\centering
\includegraphics[width=6.8cm,height=5.0cm]{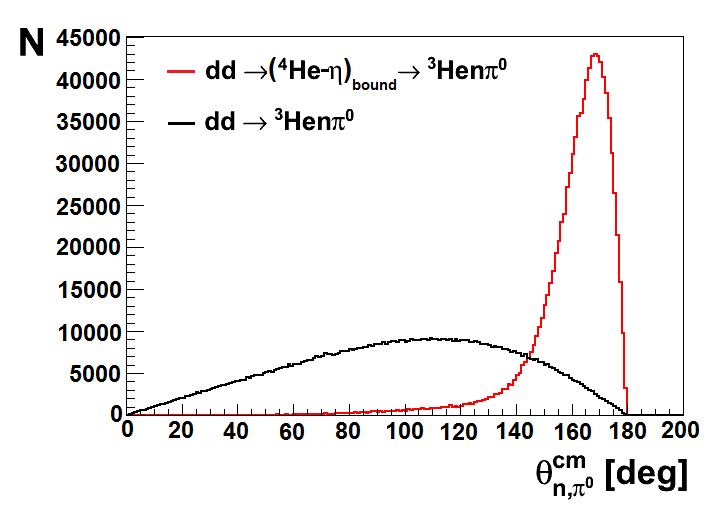}
\includegraphics[width=6.8cm,height=5.0cm]{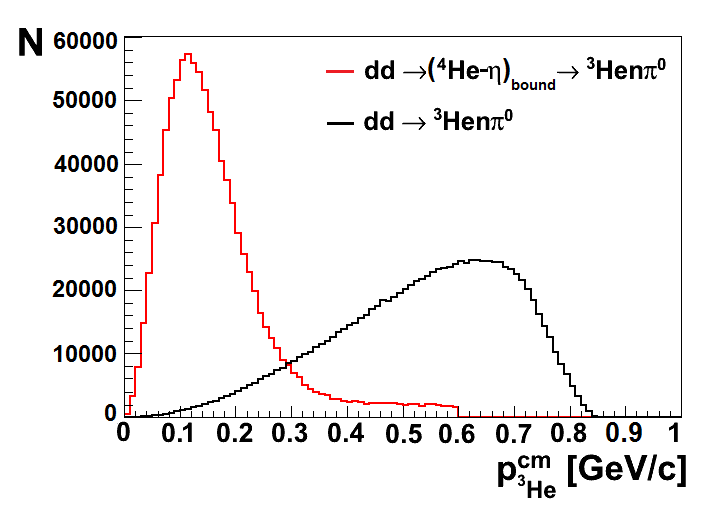}
\vspace{-0.5cm}
\caption{The distribution of the $\pi^{0}$-neutron opening angle (left panel) and the distribution of $\Hea$ momentum (right panel), both in CM system obtained in simulation of the $\MainReact$ reaction (red line) and of the direct $\BcgReacta$ production (black line).~The simulation was carried out for the beam momentum range $p_{beam}\in$ (2.127,2.422)~GeV/c. \label{fig_Simulation}}
\end{figure}

\noindent Fig.~\ref{fig_Simulation} shows spectra obtained for the simulation of the $\BS$ production and decay, carried out according to above description, compared with spectra related to direct production: $\BcgReacta$, which is considered as a one of the main background contribution in the present studies. 

\newpage

For this background process the simulation is performed with an assumption of uniform distribution of the ejectiles over the available phase space. In case of $\MainReact$ reaction, the relative angle $\theta^{cm}_{n,\pi^{0}}$ between neutron and $\pi^{0}$ is equal to $180^\circ$ in the $N^{*}$ reference frame and it is smeared due to the Fermi motion by about $30^\circ$ in the CM frame, while the direct production $\BcgReacta$ distribution covers the full angular range (see left panel of Fig.~\ref{fig_Simulation}). The Fermi motion also determines the $\Hea$ momentum $p^{cm}_{^{3}\hspace{-0.03cm}He}$ in the CM system ($\theta^{cm}_{n,\pi^{0}}$ and $p^{cm}_{^{3}\hspace{-0.03cm}He}$ are strongly correlated). 

\indent In the right panel of Fig.~\ref{fig_Simulation} one can see that the $\Hea$ momentum distribution in the CM system for the direct reaction is much more wider than for the reaction via bound state creation. The cut in the $p^{cm}_{^{3}\hspace{-0.03cm}He}$ spectrum is used as a main criteria in the selection of events corresponding to $\eta$-mesic helium (see next chapter). \\
\indent The simulation of $\MainReact$ process was carried out to check the feasibility of the measurement of $\eta$-mesic helium production with WASA-at-COSY detector setup.~These simulations were also used to set appropriate triggering conditions and the beam momentum range during the experiment, but most of all to compare with experimental data and choose the most optimal analysis conditions and cuts (Chapter~\ref{Analysis}). The Monte Carlo simulations have been finally used to estimate the efficiency including all cuts applied in the analysis that is presented in Chapter~\ref{Efficiency}.\\


\section{Nucleon momentum distribution inside $\mathbold{^{4}}\hspace{-0.03cm}\mbox{He}$~\label{Nucl_mom_4He}}

\noindent As it was shown in the previous section, in the simulation of $\MainReact$ reaction we assume that $N^{*}$ resonance moves with a Fermi momentum given by the distribution for nucleons inside $^{4}\hspace{-0.03cm}\mbox{He}$. Unfortunately, till now no rigorous calculations for $N^{*}$ momentum distribution inside the nucleus are available\footnote{A first theoretical calculations are at present carried out by N. Kelkar~\cite{Kelkar_2015}.}.~The momentum distribution of a particular particle depends significantly on the energy which is required to separate this particle from the bound system. In $^{4}\hspace{-0.03cm}\mbox{He}$ there are only nucleons and each nucleon has the same separation energy and also the same momentum distribution. In case of $^{4}\hspace{-0.03cm}\mbox{He}$-$\eta$ bound system we can assume that it has binding energy about 2~MeV, while the cost of nucleon separation is about 20~MeV. If the $\eta$ forms an $N^{*}$ resonance with one of the nucleons then the separation energy of the $N^{*}$ would be around 22~MeV, and of course the distribution of the $N^{*}$ would be very similar to that of a nucleon inside $^{4}\hspace{-0.03cm}\mbox{He}$. It works if we assume that the mass of the $N^{*}$ is equal to the mass of nucleon and $\eta$ meson. In fact the average mass of the $N^{*}$ in vacuum is much higher which would imply a rather different distribution peaked at higher values. However, in reality we don't know the mass of the $N^{*}$ inside a nucleus and it is difficult to find good solution~\cite{Haidenbauer}. According to ~\cite{Nogga3, Hirenzaki_priv} $N^{*}$ resonance momentum distribution inside the $^{4}\hspace{-0.03cm}\mbox{He}$-$\eta$ bound state can be in a good approximation described by the momentum distribution of neutron inside $^{4}\hspace{-0.03cm}\mbox{He}$ and therefore we apply it in our simulations.

\begin{figure}[h!]
\centering
\includegraphics[width=13.0cm,height=9.0cm]{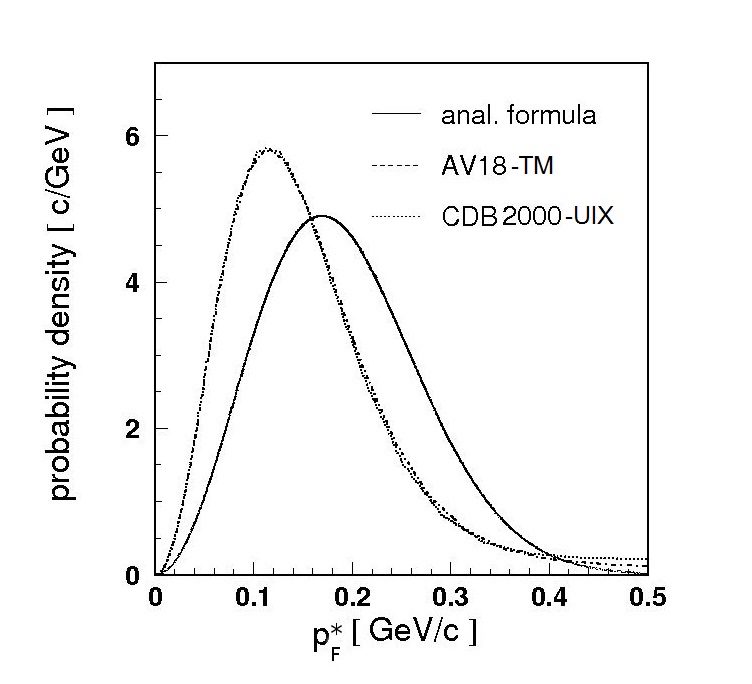}
\vspace{-0.3cm}
\caption{Fermi momentum distribution of nucleons inside $^{4}\hspace{-0.03cm}\mbox{He}$ given by analytic formula (thick solid) and estimated for the AV18-TM model (dashed) and the CDB2000-UIX model (dotted). The distributions were normalized to unity in the momentum range from 0 to 0.5 GeV/c~\cite{Skurzok_Master}.~\label{hel_4}}
\end{figure}

\indent The Fermi momentum distributions for nucleons inside atomic nuclei are calculated based on different interaction models.~For nucleons inside $^{4}\hspace{-0.03cm}\mbox{He}$ the momentum distributions predicted by three independent models are shown in Fig.~\ref{hel_4}. The distribution represented by a thick line is calculated from helium wave function derived based on Fermi three parameter charge distribution of nucleus~\cite{Hejny,Krzemien_PhD}. The momentum distribution is described by the formula (\ref{eq:4.2}): 

\begin{equation}
f{(p)}=\frac{p^2}{a}exp\left(\frac{-p^2}{b}\right), 
\label{eq:4.2}
\end{equation}\\

\noindent
where $a=0.0001989184519$~(GeV/c)$^{3}$, $b=0.0028615450879$~(GeV/c)$^{2}$. Fermi momentum is given in units of GeV/c. 

\newpage


\noindent
The formula is determined based on 3-parameter charge distribution of nucleus~\cite{Hejny}:

\begin{equation}
\rho(r)=\rho_{0} \left(1+\frac{wr^{2}}{c}\right)/ \left[1+exp \left(\frac{r-c}{z}\right)\right], 
\label{eq:4.3}
\end{equation}

\noindent
with parameters: \\
$w$=0.445$\pm$0.020\\
$c$=(1.008$\pm$0.013)~fm\\
$z$=(0.327$\pm$0.002)~fm.\\

\indent The dashed and dotted lines depict distributions obtained from AV18 and the \mbox{CDB2000} nucleon-nucleon interaction models in conjunction with Urbana~IX (UIX) and Tucson-Melbourne (TM) three nucleon interaction (TNI)~\cite{Nogga2}. Due to the fact that $^{4}\hspace{-0.03cm}\mbox{He}$ is symmetrical nucleus, proton and neutron momentum distributions are in good approximation equal. 

\indent In the low momentum region ${p}^{\,\,*}_{F}$ up to 0.4 GeV/c it is no difference between distributions determined using 3 Nuclear Force's 3NF's like AV18-TM and CDB2000-UIX (Fig.~\ref{hel_4}).~A slight discrepancy is visible for ${p}^{\,\,*}_{F}>$0.4~GeV/c and results from different interaction Hamiltonian forms defined for above-cited models.~However, the difference between the distributions derived from AV18 and \mbox{CDB2000} models and given by analytic formula is significant and the maxima of these distributions are shifted by about 45~MeV/c~\cite{Skurzok_Master}.
The discrepancy can result from the fact that the formula (\ref{eq:4.2}) was derived from nucleus charge distribution smeared out by the charge distribution of protons, whereas the AV18 and the CDB2000 models allow for the finite size of nucleus charge distributions and are related to the momentum of the point like protons in the alpha particle~\cite{Nogga3}.

\begin{figure}[h!]
\centering
\includegraphics[width=11.0cm,height=8.5cm]{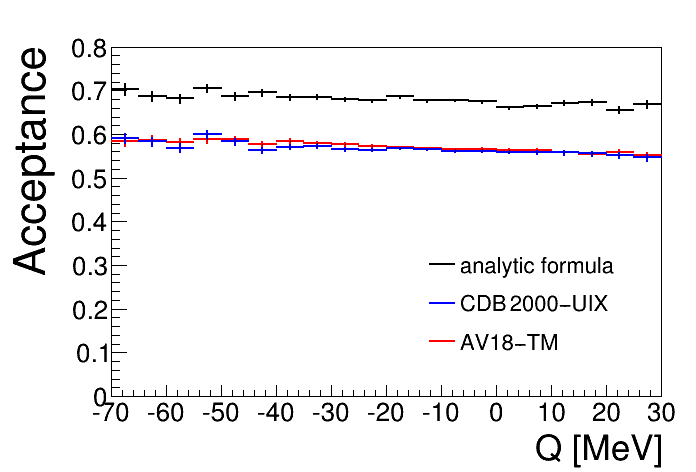}
\vspace{-0.2cm}
\caption{Geometrical acceptance of the WASA-at-COSY detector in case of $dd\rightarrow$ ($^{4}\hspace{-0.03cm}\mbox{He}$-$\eta)_{bound}  \rightarrow$ $^{3}\hspace{-0.03cm}\mbox{He} n \pi{}^{0}\rightarrow$ $^{3}\hspace{-0.03cm}\mbox{He} n \gamma \gamma$ reaction for the three different models of nucleon Fermi momentum distribution inside $\Heb$: analytic formula (black line) AV18-TM (red line) and CDB2000-UIX (blue line). The simulation was carried out for the bound state width $\Gamma$=25~MeV and binding energy $B_{s}$=10~MeV.~\label{acceptance}}
\end{figure}

\indent The simulation of $dd\rightarrow$ ($^{4}\hspace{-0.03cm}\mbox{He}$-$\eta)_{bound} \rightarrow$ $^{3}\hspace{-0.03cm}\mbox{He} n \pi{}^{0}\rightarrow$ $^{3}\hspace{-0.03cm}\mbox{He} n \gamma \gamma$ reaction was carried out for each of the above mentioned models according to description in Sec.~\ref{Sim_scheme}. The geometrical acceptance of WASA detector as a function of excess energy $Q$ for the considered models is presented in Fig.~\ref{acceptance}. The acceptance was determined for simultaneous registration of $\Hea$ in Forward Detector and two $\gamma$ quanta in Central Detector.

\newpage
\noindent
We can see significant difference between result obtained for analytic formula~(\ref{eq:4.2}) and the models AV18-TM and CDB2000-UIX. The acceptance determined from the simulation with the analytic formula is higher of about 15\% than the acceptances for the other two models. It follows that usage the different models describing the nucleon momentum inside $^{4}\hspace{-0.03cm}\mbox{He}$ gives one of the highest contribution to the systematic errors in the data analysis. However, it is important to stress that the shape of the acceptance is model independent and therefore the condition about possible existence of the $\BS$ mesic nuclei does not depend on the model of the Fermi momentum distribution.
In our studies we will consider more realistic nucleon momentum distributions calculated from AV18 and the CDB2000 potentials. \\

\chapter{Analysis of the $\mathbold{^{3}}\hspace{-0.03cm}\mathbold{\mbox{He} n \pi{}^{0}}$ events~\label{Analysis}} 

The analysed data set consists of 810 runs (run numbers: 22758-22889, 22994-23025, 23254-27399) and corresponds to an effective measurement time of about 155 hours. Most (about 84\%) of the measurement was carried out without magnetic field in CD due to the failure of the Solenoid cooling system. The main part of the analysis was devoted to selection and reconstruction of events corresponding to the $\MainReact$ process. This analysis is described step by step in the following sections.

\section{Events Selection~\label{ev_select}} 

\begin{figure}[h!]
\centering
\includegraphics[width=6.7cm,height=4.9cm]{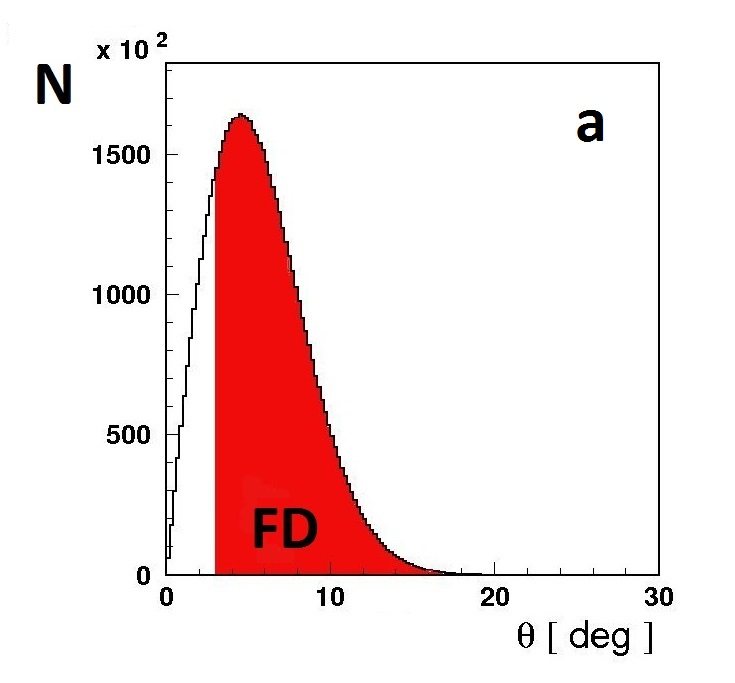} \hspace{-0.5cm} \includegraphics[width=6.7cm,height=4.7cm]{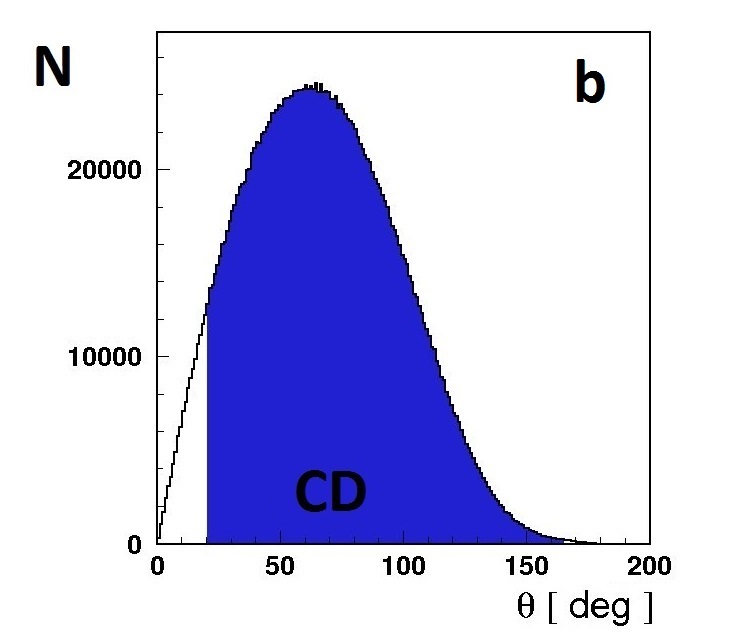}

\caption{Simulated angular distributions of $^{3}\hspace{-0.05cm}\mbox{He}$ (a), and gamma quanta (b) outgoing from the \mbox{$dd\rightarrow(^{4}\hspace{-0.03cm}\mbox{He}$-$\eta)_{bs}\rightarrow$ $^{3}\hspace{-0.03cm}\mbox{He} n \pi{}^{0} \rightarrow$ $^{3}\hspace{-0.03cm}\mbox{He} n \gamma \gamma$} reaction. Figure shows results for Monte Carlo simulation generated using the AV18 potential model describing momentum distribution of nucleons inside $^{4}\hspace{-0.03cm}\mbox{He}$.\label{angular_2}}
\end{figure}

\begin{figure}[h!]
\centering
\includegraphics[width=13.0cm,height=6.4cm]{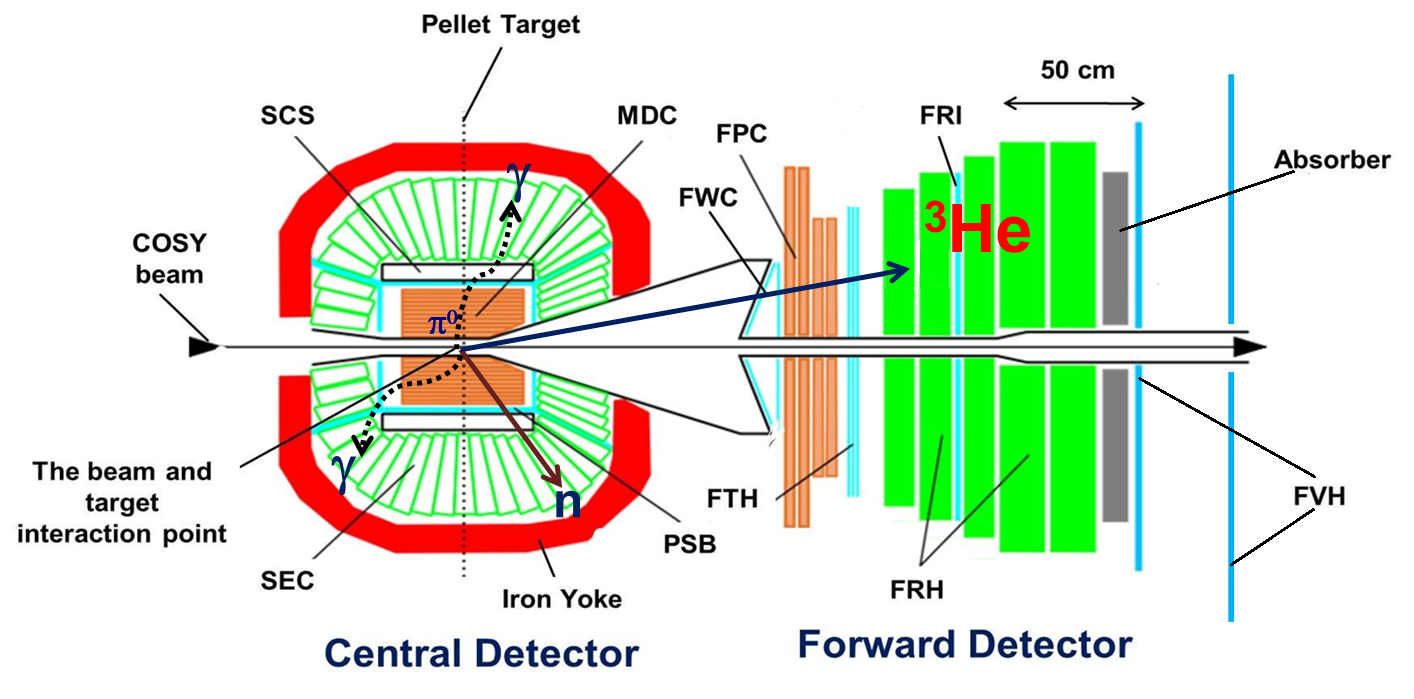} 
\caption{Scheme of the WASA-at-COSY detection system with tagged $\MainReact$ reaction. Helium is registered in Forward Detector whereas gamma quanta are detected in the Central Detector.\label{wasa_myreact}}
\end{figure}

The initial events selection for the considered reaction is carried out on the hardware trigger level which requested at least one charged particle in the FD and a high energy deposition in the FWC detector (Sec.~\ref{Sec_Triggers}). Subsequently, the preselection dedicated for $\Hea$ is performed in order to speed up the analysis, as it was described in Sec.~\ref{Sec_Presel}. In the next step of the analysis, all ejectiles are identified and the events, which may correspond to the production of bound states, are selected with appropriate cuts based on the Monte Carlo simulations. $\Hea$ is registered in the Forward Detector, while gamma quanta from the $\pi^{0}$ decay in the Central Detector. Angular distributions for the outgoing  $\Hea$ and $\gamma$'s are shown in Fig.~\ref{angular_2}.~An angular ranges covered by respective parts of WASA-at-COSY detection setup are marked with shaded areas. The scheme of WASA detection setup with marked $\MainReact$ process is presented in Fig.~\ref{wasa_myreact}. The methods of particles identification are described in next subsections.

\subsection{$\mathbold{\Hea}$ identification in the Forward Detector~\label{hel_select}}

\begin{figure}[h!]
\centering
\includegraphics[width=7.0cm,height=6.0cm]{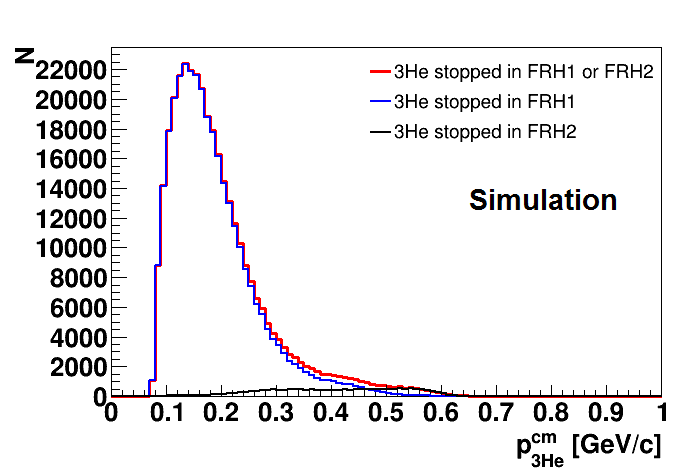}
\hspace{-0.3cm}
\includegraphics[width=7.0cm,height=6.0cm]{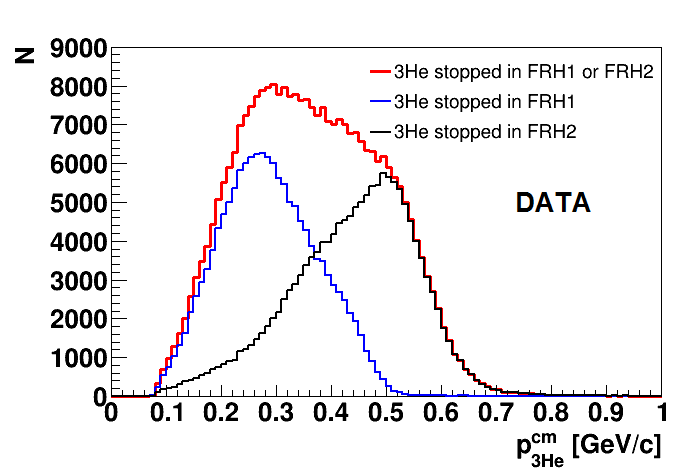}
\caption{Momentum distribution of $\Hea$ in the CM system obtained from simulations (left panel) and experimental data (right panel). The red, blue and black lines denote momentum distribution of all registered $\Hea$ and stopped in FRH1, FRH2, respectively.\label{p3He_comp}}
\end{figure}

According to performed simulations of $\MainReact$ reaction, $\Hea$ is mostly (95\%) stopped in first layer of Range Hodoscope in Forward Detector and just in 5\% in the second layer. It is presented in left panel of Fig.~\ref{p3He_comp} where the spectrum of $\Hea$ momentum in the CM is plotted for helium stopped in FRH1 and FRH2 layers.

\begin{figure}[h!]
\centering
\includegraphics[width=10.0cm,height=7.0cm]{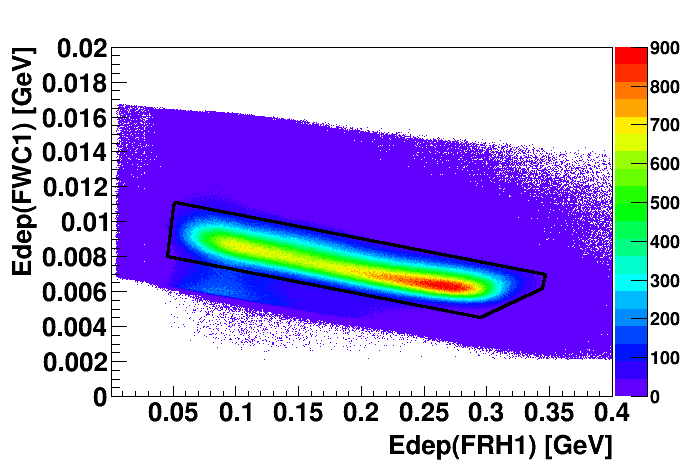} 
\caption{Experimental spectrum of energy deposited in FWC1 and FRH1. The selected area for $\Hea$ is marked with black line. The empty area below comes from the preselection cut (see Sec.~\ref{Sec_Presel}).\label{Hel_ident}}
\end{figure}

In order to reduce significantly background originating from higher energetic helium (right panel of Fig.~\ref{p3He_comp}) with just small signal reduction, the veto condition was set on the second and further FRH layers (deposited energy Edep(FRH2,3,4,5)$<$0.015~GeV). The $\Hea$ was identified with $\Delta E$--$\Delta E$ method based on energy losses in the FWC1 and FRH1. In Fig.~\ref{Hel_ident} one can see the spectrum of the  Edep(FWC1) vs. Edep(FRH1) with marked graphical cut applied for $\Hea$ ions selection.

\subsection{$\mathbold{\piz}$ and neutron identification in the Central Detector}

\begin{figure}[h!]
\centering
\includegraphics[width=10.0cm,height=6.5cm]{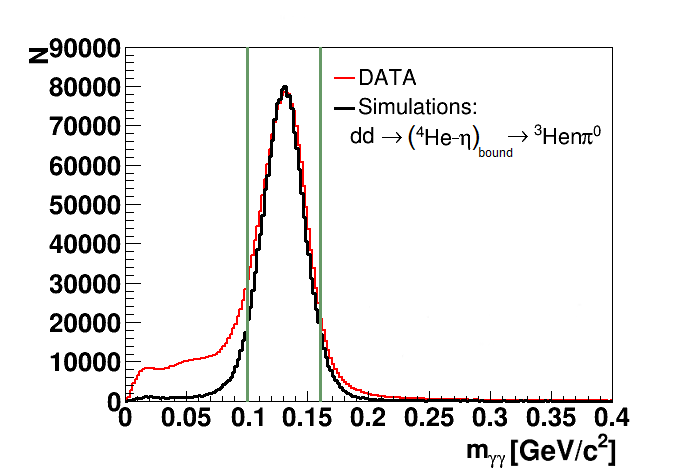}
\vspace{-0.5cm}
\caption{$\pi^{0}$ identification via cut in invariant mass spectrum. Applied cuts are marked with green lines.\label{pion_ident}}  
\end{figure}

As it was mentioned in Sec.~\ref{ev_select}, $\piz$ mesons from $\MainReact$ reaction are registered in the Central Detector. The neutral pions $\pi{}^{0}$ are reconstructed from the invariant mass of two gamma quanta originating from its decay. In the analysis, first all events with at least two neutral clusters in electromagnetic calorimeter are selected. Next, all gamma pair  combinations are considered, and for each pair the invariant mass $m_{\gamma\gamma}$ is calculated. In case of more than two clusters, we take into account only this combination of clusters for which the difference between $\piz$ mass $m_{\pi^{0}}$ and invariant mass of two gamma quanta $m_{\gamma\gamma}$ is minimal. The cut applied in invariant mass spectrum, based on Monte Carlo simulations, is presented in Fig.~\ref{pion_ident}. Experimental data is marked with the red line, while the Monte Carlo simulations with the black line.

Neutron was identified via the missing mass technique. Knowing a four-momenta of deuteron beam ($E_{d}^{\,b}$, $\vec{p}_{b}$), deuteron target ($E_{d}^{\,t}=m_d$, $\vec{p}_{t}=0$), helium ($E_{^{3}\hspace{-0.05cm}He}$, $\vec{p}_{^{3}\hspace{-0.05cm}He}$) and $\piz$ ($E_{\piz}$, $\vec{p_{\piz}}$) and employing the principle of momentum and energy conservation we can calculate the missing mass as follows:

\begin{equation}
m_X^2=E_X^2-\vec{p}\hspace{0.1cm}^{2}_X=(E_{d}^{\,b}+E_{d}^{\,t}-E_{^{3}\hspace{-0.05cm}He}-E_{\piz})^2 -(\vec{p}_{b}+\vec{p}_{t}-\vec{p}_{^{3}\hspace{-0.05cm}He}-\vec{p_{\piz}})^2.
\end{equation}

\begin{figure}[h!]
\centering
\includegraphics[width=7.1cm,height=5.0cm]{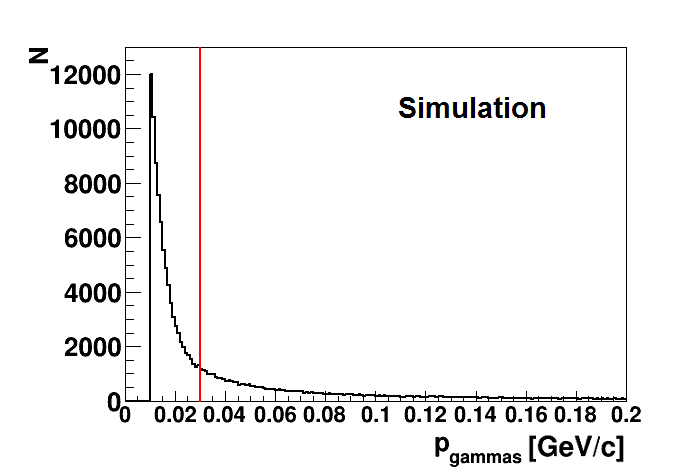}
\hspace{-0.5cm}
\includegraphics[width=7.1cm,height=5.0cm]{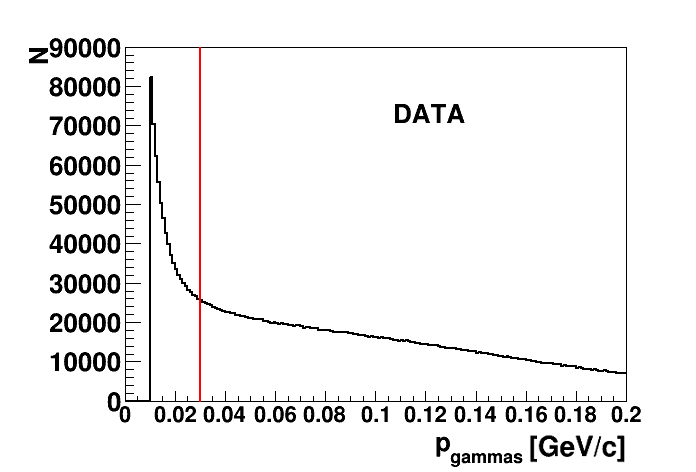}
\caption{Momentum distribution for additional neutral clusters for Monte Carlo simulation (left panel) and data (right panel). The applied cut is marked by red line.\label{p_gammas}}
\end{figure}

Unfortunately, the missing mass spectrum contains a lot of background from reactions with more than two gamma quanta in the decay channel (most probably with the $dd \rightarrow$ $\Hea n \piz \piz$ reaction). This background has been significantly reduced by applying the momentum cut on the calorimeter clusters (third or more), which were not selected as $\gamma$ coming from $\pi{}^{0}$ decay, by the reconstruction procedure.~The cut is based on Monte Carlo simulations, and for the further analysis only these events are accepted for which momentum corresponds to additional cluster is less than 0.03~GeV/c as shown in Fig.~\ref{p_gammas}.

\begin{figure}[h!]
\centering
\includegraphics[width=6.7cm,height=5.0cm]{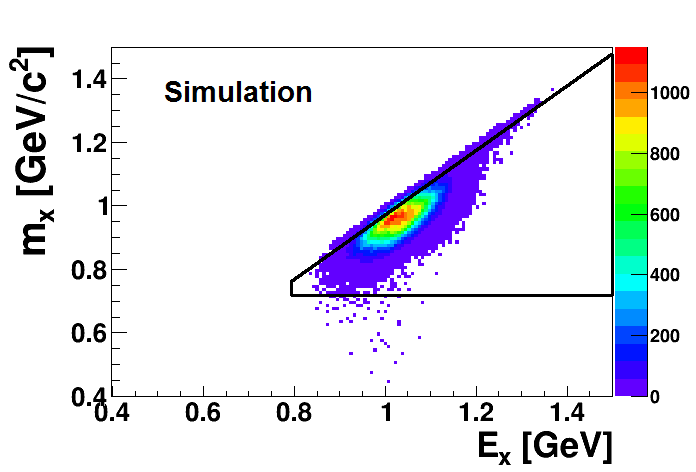}
\includegraphics[width=6.7cm,height=5.0cm]{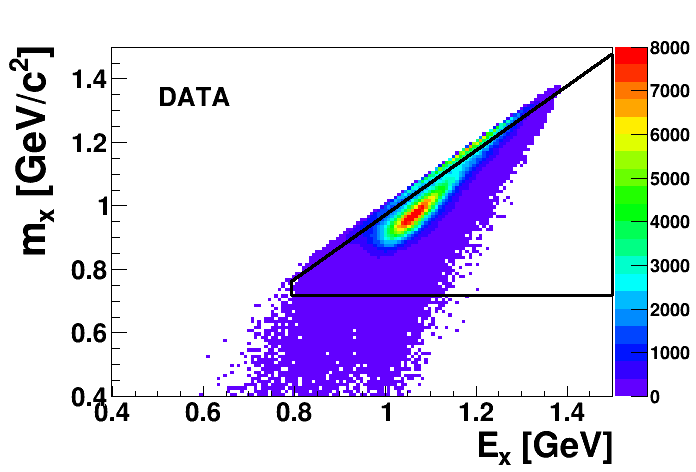}
\caption{Missing mass $m_{x}$ vs. missing energy $E_{x}$ for Monte Carlo simulation (left panel) and experimental data (right panel). The applied cut is marked in black. \label{mx_Ex_main}}
\end{figure}

\noindent Additionally, the cut on the $m_{x}(E_{x})$ spectrum was applied as shown in Fig.~\ref{mx_Ex_main} to reduce the background coming from $dd \rightarrow$ $\Hea n \piz \piz$ reaction. \\

\begin{figure}[h!]
\centering
\includegraphics[width=7.1cm,height=5.0cm]{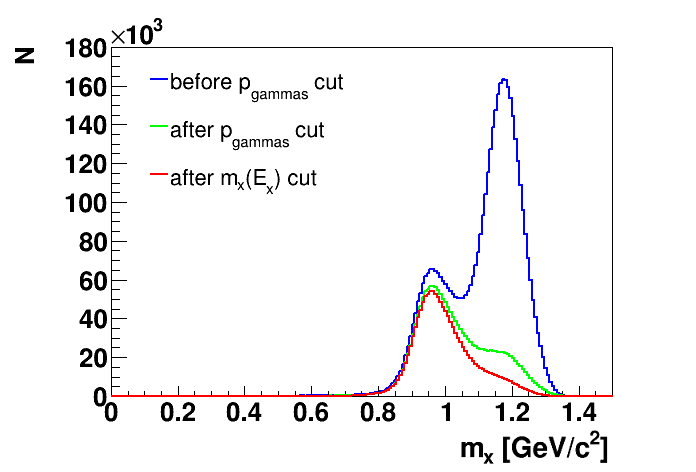}
\hspace{-0.5cm}
\includegraphics[width=7.1cm,height=5.0cm]{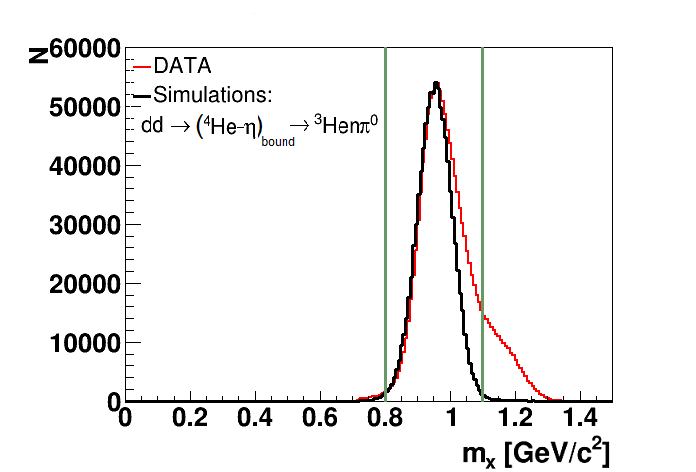}
\caption{The missing mass spectra for $dd \rightarrow$ $\Hea X$ reaction. (left) Comparison of the missing mass $m_{x}$ before and after applied cuts for experimental data. (right) The final missing mass spectrum. Data is marked with red line, the Monte Carlo simulations of the signal is marked with the black line. The region accepted for further analysis is marked with vertical lines.\label{mx_comparison}}
\end{figure}

\newpage
The plot comparing missing mass spectra for data before and after background reduction cuts is presented in the left panel of Fig.~\ref{mx_comparison}, while the neutron spectrum after final cuts is shown in the right panel. Most of the remaining background at the right side of the spectrum was rejected applying cuts marked with vertical lines. 

\subsection{Kinematic cuts for $\mathbold{(\Heb}$-$\mathbold{\eta)_{bound}}$ events selection~\label{kinem_cuts_Main}}

\begin{figure}[h!]
\centering
\includegraphics[width=7.0cm,height=5.0cm]{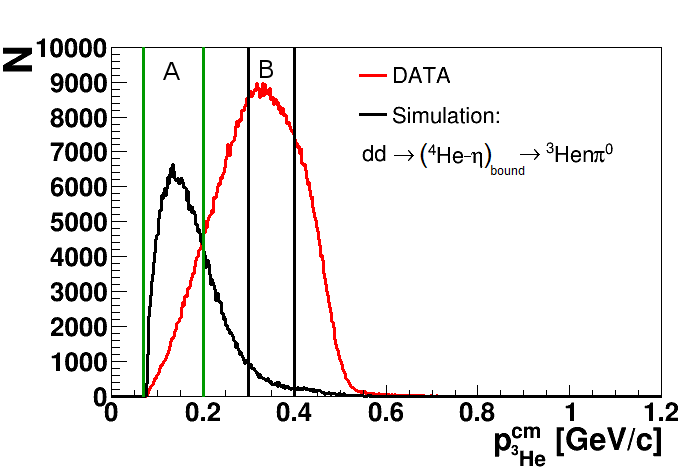}
\hspace{-0.3cm}
\includegraphics[width=7.0cm,height=5.0cm]{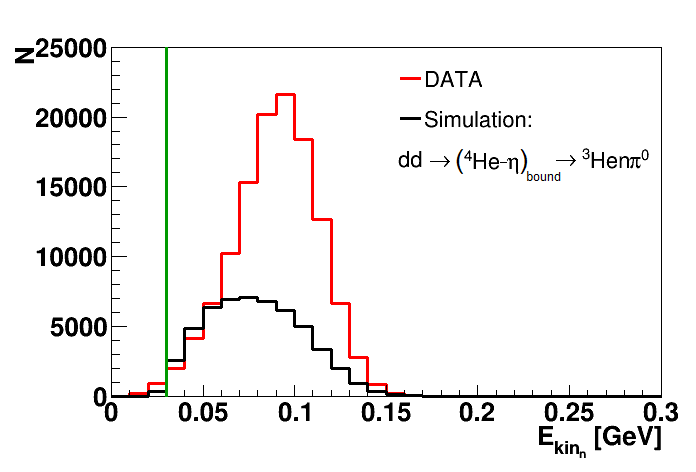}
\includegraphics[width=7.0cm,height=5.0cm]{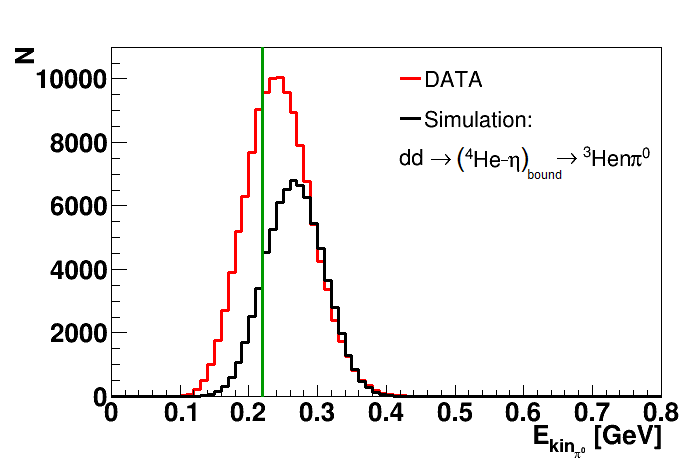}
\hspace{-0.3cm}
\includegraphics[width=7.0cm,height=5.0cm]{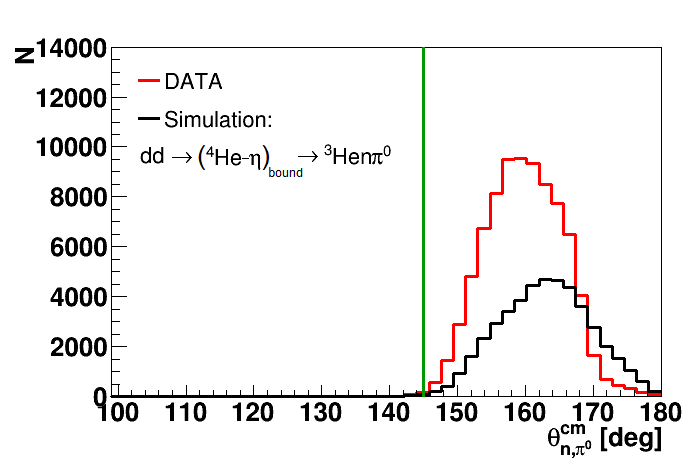}

\caption{Spectrum of $p^{cm}_{^{3}\hspace{-0.05cm}He}$ (left upper panel), $E^{cm}_{kin_{n}}$ distribution in region~A (right upper panel), $E^{cm}_{kin_{\piz}}$ distribution in region~A (left lower panel) and $\theta^{cm}_{n,\piz}$ distribution in region~A (right lower panel). Data are shown in red. Monte Carlo simulations of signal are shown in black, while the applied cuts are marked with the green lines.\label{kinem_cuts}}
\end{figure}

As it was mentioned in Sec.~\ref{Sym_Kin}, the $\Hea$ in the $\MainReact$ reaction plays a role of the spectator which is moving with the low momentum corresponding to the Fermi momentum of the nucleons inside $\Heb$. Therefore, in the $p^{cm}_{^{3}\hspace{-0.05cm}He}$ spectrum we selected two regions: region where we expect a significant contribution form the bound state signal for $p^{cm}_{^{3}\hspace{-0.05cm}He}\in$~(0.07,0.2)~GeV/c and the region poor in signal where the background $\BcgReacta$ and $\BcgReactb$ processes are dominating \mbox{(region $p^{cm}_{^{3}\hspace{-0.05cm}He}\in$~(0.3,0.4)~GeV/c)}. These regions referred to as "Signal Rich" and "Signal Poor" are marked in the left upper panel of Fig.~\ref{kinem_cuts} as a region A and B, respectively.

In order to improve the selection of events corresponding to the $\BSbound$, an additional cuts reducing the background contributions, were applied in the neutron and pion kinetic energies in the CM system. The energy spectra with marked cuts are presented in the upper right and lower left panels of Fig.~\ref{kinem_cuts}. 

Based on the spectrum obtained from simulations of $\MainReact$ process and presented in the left panel of Fig.~\ref{fig_Simulation}, we applied also a cut in the neutron-$\piz$ opening angle in the CM frame corresponding to the range between 145$^{\circ}$ and 180$^{\circ}$. Since the opening angle is strongly correlated with the $\Hea$ momentum the cut removes only a small amount of events below 145$^{\circ}$ what is visible in the right lower panel of Fig.~\ref{kinem_cuts}.

\chapter{Detection efficiency~\label{Efficiency}} 

\begin{figure}[h!]
\centering
\includegraphics[width=9.0cm,height=7.0cm]{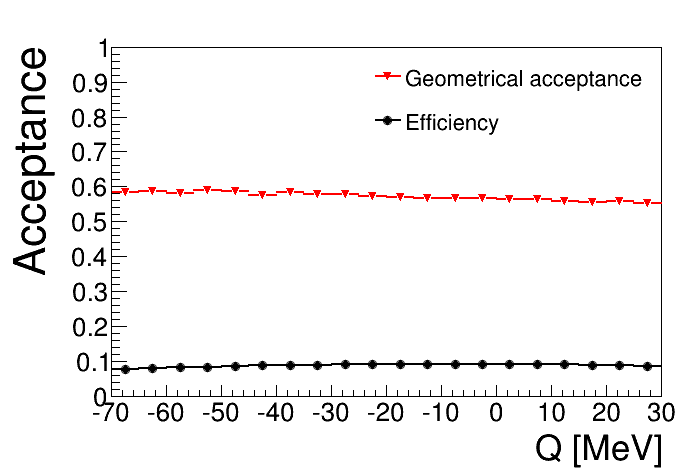} 
\vspace{-0.4cm}
\caption{The acceptance and efficiency for the registration and reconstruction of $\MainReact$ reaction as a function of excess energy $Q$. The geometrical acceptance of the WASA detector is shown with red triangles while  the full efficiency including detection and reconstruction efficiency for the region rich in signal is shown with black circles.\label{acc_eff}}
\end{figure}

\begin{figure}[h!]
\centering
\includegraphics[width=9.0cm,height=7.0cm]{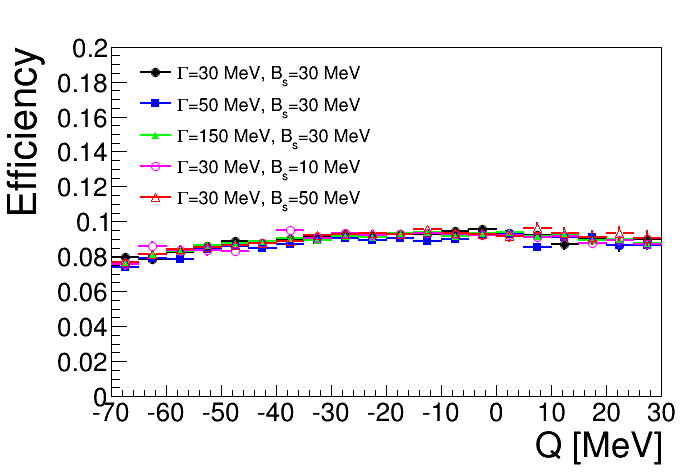} 
\vspace{-0.4cm}
\caption{The efficiency for the registration and reconstruction of $\MainReact$ reaction for the region rich in signal as a function of excess energy $Q$. Results obtained for different bound state widths $\Gamma$ and the binding energies $B_{s}$ are marked with different colours.~\label{eff_gamma_bs}}
\end{figure}

The experimental data are collected with non perfect geometrical acceptance as well as non-perfect detection and reconstruction efficiency. In order to correct the obtained results for those detector effects, one should carefully study the behaviour of acceptance and efficiency distributions. \\
\indent The overall detection and reconstruction efficiency, was determined based on the Monte Carlo simulation for the $\MainReact$ process carried out taking into account detection system response and all selection cuts described in Chapter~\ref{Analysis}. The efficiency was calculated as a ratio of the number of events accepted by detection system to the number of generated events. The correction of the experimental data is applied by dividing the determined distributions of observables of interests by the full efficiency. The efficiency for the "Signal Rich" region A (see Sec.~\ref{kinem_cuts_Main}), together with the detector acceptance are presented as a function of the excess energy in Fig.~\ref{acc_eff}. It is worth to emphasize that the efficiency does not depend on the bound state width $\Gamma$ and the binding energy $B_{s}$ as it is shown in Fig.~\ref{eff_gamma_bs}.\\
\indent The geometrical acceptance of the WASA-at-COSY detector for the $\MainReact$ reaction is equal to about 57\% while the full efficiency including all cuts applied in the analysis is about 9\% and is smooth in the whole excess energy range.

\newpage

\vspace{0.5cm}

\chapter{Luminosity Determination~\label{Lum_determ}}

\noindent In this chapter two methods of the luminosity determination are presented\footnote{The description of luminosity determination has been already published by the author in a form of conference proceedings (Acta Phys. Polon. B46 (2015) 1, 133)}. As it was described in Sec.~\ref{Sec_Beam}, the technique of continuous change of the beam momentum in one accelerator cycle was applied in the experiment.~During an acceleration process the luminosity could vary due to beam losses caused by the interaction with the target and with the rest gas in the accelerator beam line, as well as due to the changes in the beam-target overlap correlated with momentum variation and adiabatic shrinking of the beam size~\cite{Lorentz}.~Therefore, it is necessary to determine not only the total integrated luminosity but also its dependence on the excess energy.
\noindent The total integrated luminosity is determined based on the $dd\rightarrow$ $^{3}\hspace{-0.03cm}\mbox{He} n$ and quasi free $pp\rightarrow p p$ reactions for which the cross sections were already experimentally established. 
Because of the acceptance variation for the beam momentum range for which $^{3}\hspace{-0.03cm}\mbox{He}$ ions are stopped between two Forward Detector layers, the excess energy dependence of the luminosity is determined based on quasi-free $pp\rightarrow p p$ reaction for which the WASA acceptance is a smooth function of the beam momentum.\\ 
\indent The precise luminosity determination as a function of excess energy $Q$ is important for the normalization of the obtained excitation function for $\BcgReacta$ reaction and hence for the interpretation of the result in view of the hypothesis of the $\BS$ bound state production.

\section{Integrated luminosity -- $\mathbold{dd\rightarrow}$ $\mathbold{^{3}}\hspace{-0.03cm}\mathbold{\mbox{He} n}$ reaction analysis~\label{Lum_integrated}}

\noindent \textbf{Cross section determination}

\noindent The absolute value of the integrated luminosity was determined using the experimental data on the $dd\rightarrow$ $^{3}\hspace{-0.03cm}\mbox{He} n$ cross-sections measured by SATURNE collaboration for four beam momenta in the range between 1.65 and 2.49~GeV/c~\cite{Bizard_SATURNE}. The cross section $\sigma_{dd\rightarrow ^{3}\hspace{-0.03cm}{He} n}$ dependence on the square of the momentum transfer $t=(\mathbb{P}_{^{3}\hspace{-0.03cm}{He}}-\mathbb{P}_{beam})^{2}$ may be parametrized as follows~\cite{Bizard_SATURNE,Pricking_PhD}:

\begin{equation}
\frac{d\sigma(t-t_{max})}{dt}=\sum_{i=1}^{3} a_{i} e^{b_{i}(t-t_{max})}~\label{eq_10},
\end{equation}

\noindent
where $t_{max}$ denotes maximal value of $t$ measured for a given beam momentum at SATURNE. Parameters $a_{i}$ and $b_{i}$  are described as a function of the total energy $\sqrt{s_{dd}}$:

\begin{equation}
par_{i}(\sqrt{s_{dd}})=\frac{p_{i}}{\sqrt{s_{dd}}-q_{i}}+r_{i}~\label{eq_11},
\end{equation}

\noindent where the values of $p_{i}$, $q_{i}$ and $r_{i}$ were determined~\cite{Pricking_PhD} by the fit of the above formula to the cross sections measured at SATURNE~\cite{Bizard_SATURNE}. The cross section parametrization was described in details in~\cite{Pricking_PhD}. The parameters obtained from the fit to SATURNE data~\cite{Bizard_SATURNE} are shown in the Table~\ref{tab_SATURNE}.\\

\begin{table}[h]
\begin{normalsize}
\begin{center}
\begin{tabular}{|c|ccc|c|}\hline
 &$p_{i}$ &$q_{i}$ &$r_{i}$\\
\hline 
$a_{1}$  &11.64 &4.05 &-14.49\\
$b_{1}$  &0.78 &3.92 &9.04 \\
$a_{2}$  &2327.04 &-1.44 &-399.27 \\
$b_{2}$  &0.78 &3.92 &9.04\\
$a_{3}$  &0.22 &4.08 &1.24 \\
$b_{3}$  &0.78 &3.92 &9.04\\
\hline
\end{tabular}
\end{center}
\caption{Parameters for the $t$ and $\sqrt{s_{dd}}$ dependence of the total cross section of the \mbox{$dd\rightarrow$ $^{3}\hspace{-0.03cm}\mbox{He} n$} reaction.  In the applied parametrization, the $t$ and $\sqrt{s_{dd}}$ values are expressed in (GeV/c)$^{2}$, GeV and $\mu$b/(GeV/c)$^{2}$, respectively.\label{tab_SATURNE}}
\end{normalsize}
\end{table} 

\vspace{0.5cm}  
  
\begin{figure}[h]
\centering
\includegraphics[width=6.6cm,height=5.0cm]{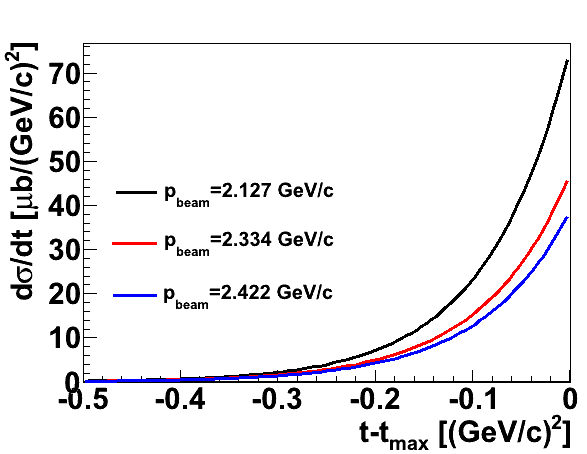} 
\includegraphics[width=6.8cm,height=4.8cm]{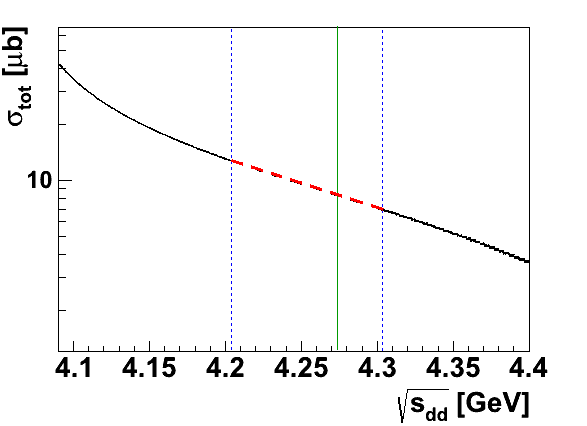}
\caption{(left) Differential cross section for $p_{beam}$ = {2.127, 2.334, 2.422}~GeV/c and (right) total cross section as a function of the  $\sqrt{s_{dd}}$ -- dashed red line covers the experimental beam momentum range \mbox{$p_{beam}\in$~(2.127,2.422)~GeV/c} while green line shows the threshold for $\eta$ meson production.~\label{fig_paramAnnette}}
\end{figure}

\indent The differential cross section as a function of $t-t_{max}$ for the three different beam momentum values from our experimental range $p_{beam}\in$~(2.127,2.422)~GeV/c and the total cross section as a function of the invariant mass are presented in the left and right panel of Fig.~\ref{fig_paramAnnette}, respectively.\\

\begin{figure}[h!]
\centering
\includegraphics[width=9.0cm,height=5.7cm]{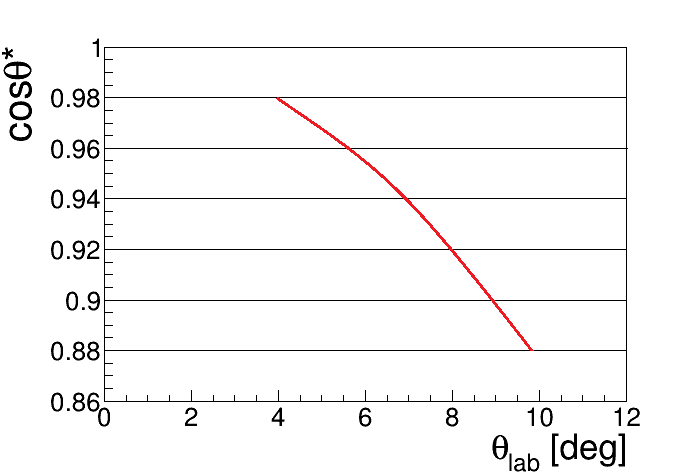} 
\caption{$cos\theta^{*}$ dependence on the $\theta_{lab}$ angle for $\LumReacta$ reaction (red line). The horizontal lines mark the selected intervals of $cos\theta^{*}$.~\label{fig_costheta}} 
\end{figure}

\indent We may determine angular dependence of the cross section using a following relation:

\begin{equation}
\frac{d\sigma}{d(cos\theta^{*})}=\frac{d\sigma}{dt} \cdot \frac{dt}{d(cos\theta^{*})}~\label{jeden}
\end{equation}

\noindent
with the Jacobian term $\frac{dt}{d(cos\theta^{*})}=2\cdot |\vec{p}^{\,\,*}_{beam}|\cdot |\vec{p}^{\,\,*}_{^{3}\hspace{-0.05cm}He}|$ calculated based on the momentum transfer squared in the CM system: 

\vspace{-0.5cm}

\begin{equation}
t=(\mathbb{P}_{^{3}\hspace{-0.03cm}{He}}-\mathbb{P}_{beam})^{2}=m_d^2 + m^2_{^{3}\hspace{-0.05cm}He}-2\cdot E^{\,\,*}_{beam}\cdot E^{\,\,*}_{^{3}\hspace{-0.05cm}He}+2\cdot |\vec{p}^{\,\,*}_{beam}|\cdot |\vec{p}^{\,\,*}_{^{3}\hspace{-0.05cm}He}|\cdot cos\theta^{*},
\end{equation}

\noindent where $E^{\,\,*}_{beam}$, $E^{\,\,*}_{^{3}\hspace{-0.05cm}He}$, $\vec{p}^{\,\,*}_{beam}$, $\vec{p}^{\,\,*}_{^{3}\hspace{-0.05cm}He}$ and $\theta^{*}$ are beam and $\Hea$ energy, momenta and the $^{3}\hspace{-0.03cm}\mbox{He}$ emission angle in the CM frame, respectively.\\
\indent The relation between the $\Hea$ scattering angle $\theta_{lab}$ and $cos\theta^{*}$ is presented in~Fig.~\ref{fig_costheta}. The $^{3}\hspace{-0.03cm}\mbox{He}$ angular range from about 4$^{\circ}$ to 10$^{\circ}$ corresponds to the $cos\theta^{*}\in$~(0.88,0.98).

\begin{figure}[h!]
\centering
\includegraphics[width=6.8cm,height=5.0cm]{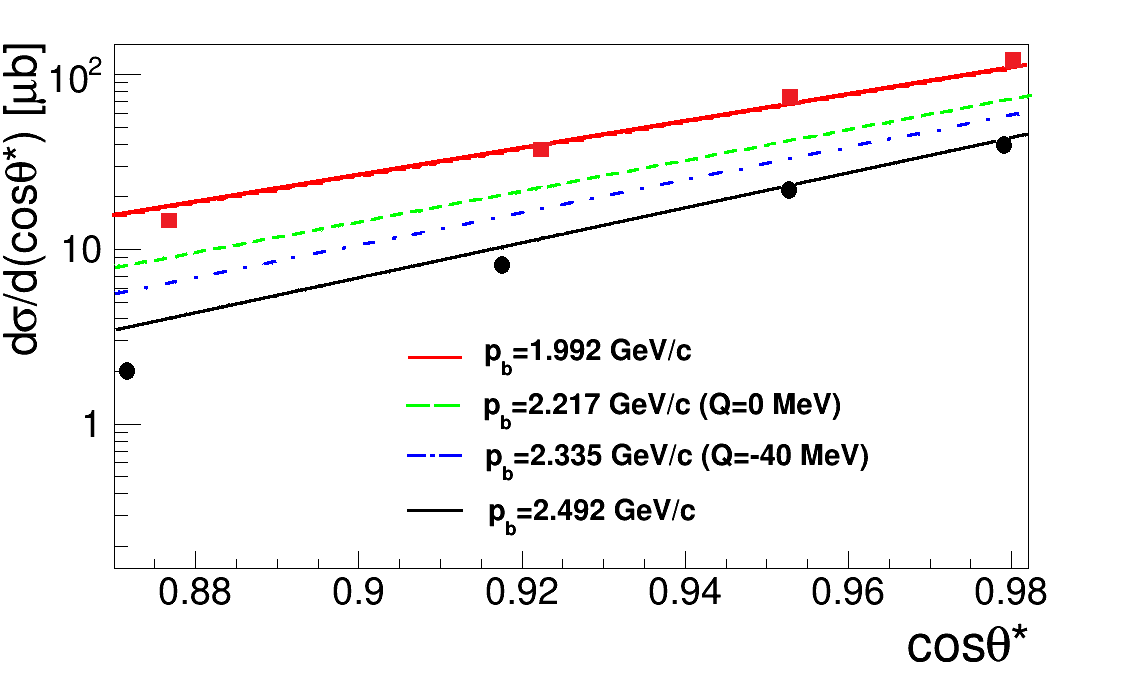} 
\includegraphics[width=7.0cm,height=5.2cm]{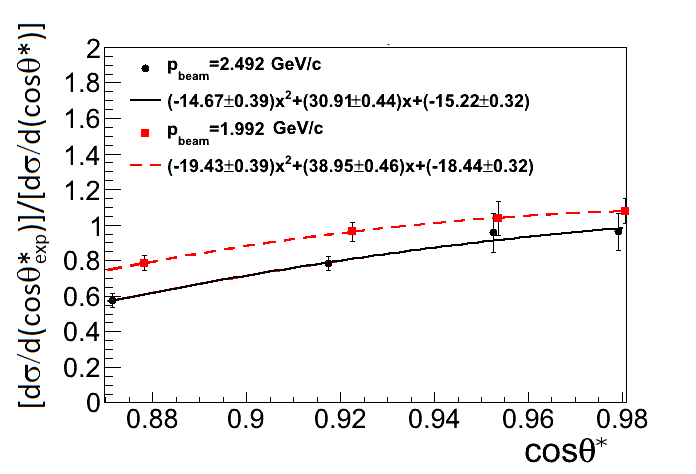} 
\caption{(left) Differential cross section as a function of $cos\theta^{*}$ for SATURNE experimental data (squares/red and dots/black points for fixed beam momentum $p_{beam}$=1.992~GeV/c and $p_{beam}$=2.492~GeV/c, respectively) and obtained from parametrization (top solid/red, dashed/green, dash-dotted/blue, and bottom
solid/black lines for $p_{beam}$ equal to 1.992~GeV/c, 2.217~GeV/c, 2.335~GeV/c and 2.492~GeV/c, respectively). (right) The ratio of experimental and parametrized cross section $\frac{d\sigma}{d(cos\theta^{*})_{exp}}/\frac{d\sigma}{d(cos\theta^{*})}$ for $p_{beam}$=1.992~GeV/c (squares/red) and $p_{beam}$=2.492~GeV/c (dots/black) fitted with second degree polynomial functions (dashed/red and solid/black lines, respectively). The marked errors result from the statistical experimental uncertainties.~\label{fig_Bizzard}}
\end{figure}

\indent The available SATURNE experimental data closest to the range of beam momentum used in the experiment for the angular range relevant for our analysis are shown in the left panel of Fig.~\ref{fig_Bizzard}. Superimposed lines present results of the above described parametrisation for beam momenta corresponding to the experimental points: 1.992~GeV/c and 2.492~GeV/c (red and black, respectively) and for two exemplary momenta corresponding to $Q$ = 0 and \mbox{$Q$ = -40~MeV.}

\indent In the angular region of interest the experimental points lie below the curves obtained based on the parametrization defined in Eq.~(\ref{eq_10}) and (\ref{eq_11}). The discrepancy significantly affects the luminosity determination, therefore correction for the parametrization was necessary and was applied for the $cos\theta^{*}\in$~(0.88,0.98). The ratio between experimental and parametrized cross section $\frac{d\sigma}{d(cos\theta^{*})_{exp}}/\frac{d\sigma}{d(cos\theta^{*})}$ was fitted with a second degree polynomial function for both experimental beam momentum values: 1.992~GeV/c and 2.492~GeV/c.~Obtained result is presented in the right panel of Fig.~\ref{fig_Bizzard}. The cross section correction is calculated for fixed $cos\theta^{*}$ using the fitted functions and linearly interpolated for the proper beam momentum value from range $p_{beam}\in$~(2.127,2.422)~GeV/c.\\ 
 
\noindent \textbf{Selection of $\mathbold{dd\rightarrow}$ $\mathbold{^{3}}\hspace{-0.03cm}\mathbold{\mbox{He} n}$ events} 

\noindent The measurement of the \mbox{$\LumReacta$} reaction was based on the registration of the outgoing helium in the Forward Detector. In the first step of analysis at least one charged particle in the FD and a high energy deposition in the FWC Detector (Sec.~\ref{Sec_Triggers}) were required to reduce the background especially from protons and pions. Then the preselected data (see Sec.~\ref{Sec_Presel}) were taken into account. 

\begin{figure}[h!]
\centering
\includegraphics[width=6.5cm,height=5.0cm]{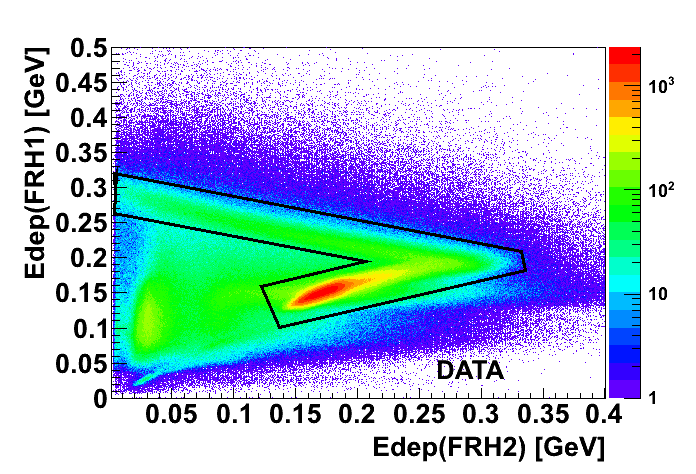}
\includegraphics[width=6.5cm,height=5.0cm]{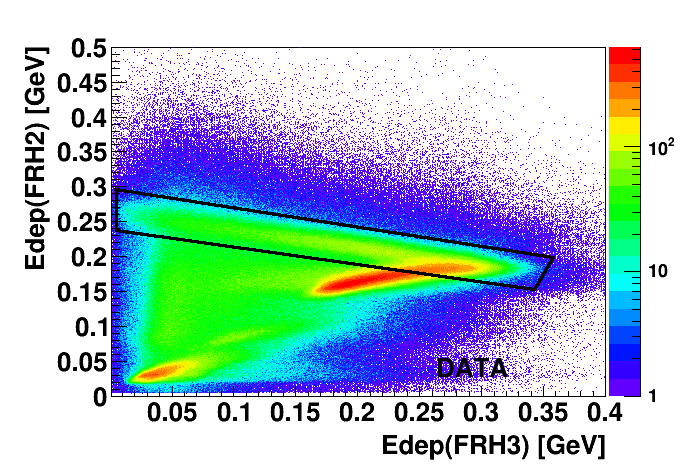} 
\includegraphics[width=6.5cm,height=5.0cm]{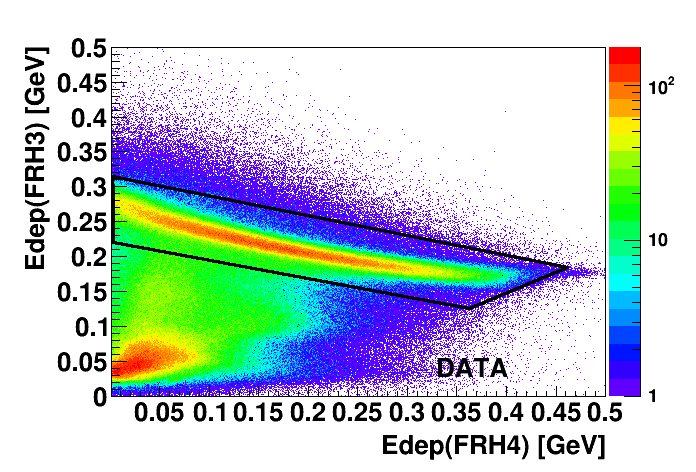}
\caption{(upper panel) Edep(FRH1) vs. Edep(FRH2) spectrum, (left lower panel) Edep(FRH2) vs. Edep(FRH3), (right lower panel) Edep(FRH3) vs. Edep(FRH4). The spectra correspond to events for $\Hea$ stopped in FRH3 or FRH4. Solid line shows graphical cut used in the analysis for the selection of events with $\Hea$ ions.~\label{fig_Edep}}
\end{figure}

Low-energetic $^{3}\hspace{-0.03cm}\mbox{He}$ ions were stopped in the 3rd layer of the Forward Range Hodoscope, while high-energetic in the 4th layer. The helium identification was based on the $\Delta E$--$\Delta E$ method as presented in Fig.~\ref{fig_Edep}. In order to disentangle the $^{3}\hspace{-0.03cm}\mbox{He}$ from other charged particles in FD, a cut in the Edep(FRH1) vs. Edep(FRH2) spectrum was applied (upper panel of Fig.~\ref{fig_Edep}). Next, helium stopped in FRH3 or in FRH4 was selected with cuts in Edep(FRH2) vs. Edep(FRH3) and Edep(FRH3) vs. Edep(FRH4), presented in left and right lower panels of Fig.~\ref{fig_Edep}, respectively.

\begin{figure}[h!]
\centering
\includegraphics[width=6.8cm,height=5.2cm]{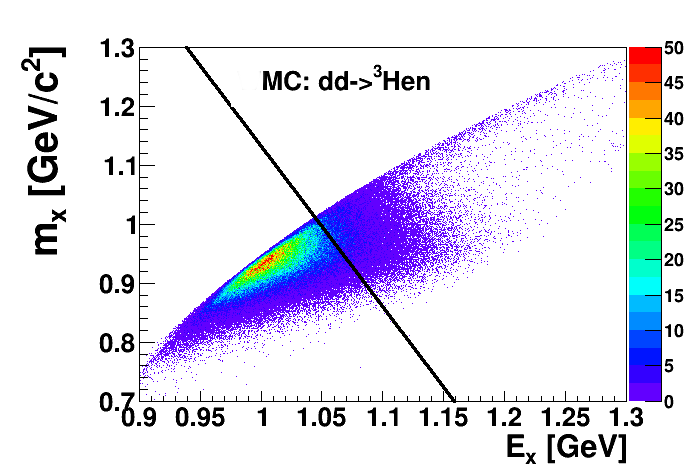}
\includegraphics[width=6.8cm,height=5.2cm]{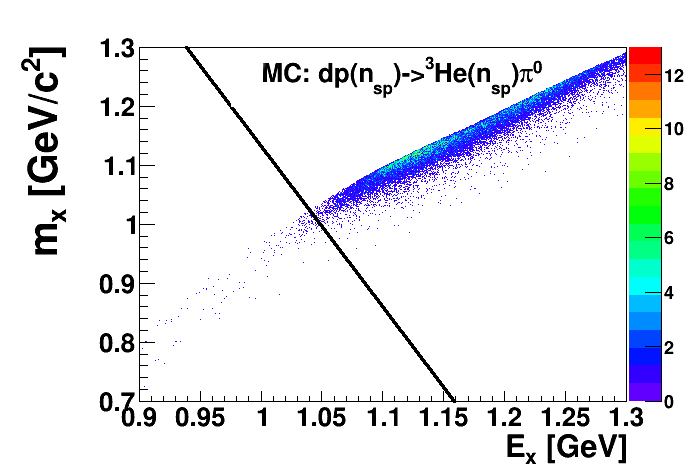}
\includegraphics[width=6.8cm,height=5.2cm]{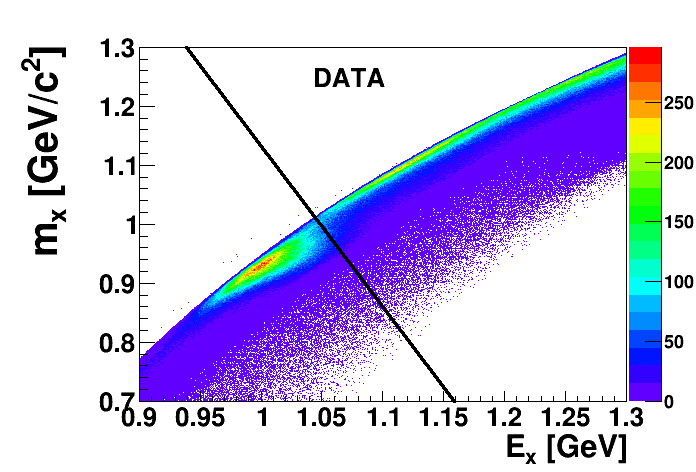} 
\caption{Spectra of missing mass $m_{x}$ vs. missing energy $E_{x}$ for simulation of $\LumReacta$ reaction (left upper panel), simulation of background \mbox{$dp(n_{sp})\rightarrow$ $^{3}\hspace{-0.03cm}\mbox{He} (n_{sp}) \pi^{0}$} reaction (right upper panel) and data (lower panel). Applied cut is marked with black line.~\label{fig_mx}}
\end{figure}

\indent The outgoing neutrons were identified using the missing mass technique. In order to reduce the background originating from the multi-pion reactions like $dd\rightarrow$ $^{3}\hspace{-0.03cm}\mbox{He} n \pi^{0} \pi^{0}$ the number of neutral clusters reconstructed in CD was requested to be less than 2. Then, to reduce background contribution arising from quasi-free \mbox{$dp(n_{sp})\rightarrow$ $^{3}\hspace{-0.03cm}\mbox{He} (n_{sp}) \pi^{0}$}, the cut in missing mass $m_{x}$ vs. missing energy $E_{x}$ spectrum was applied as it is presented in Fig.~\ref{fig_mx}.  

\begin{figure}[h!]
\centering
\includegraphics[width=10.0cm,height=7.0cm]{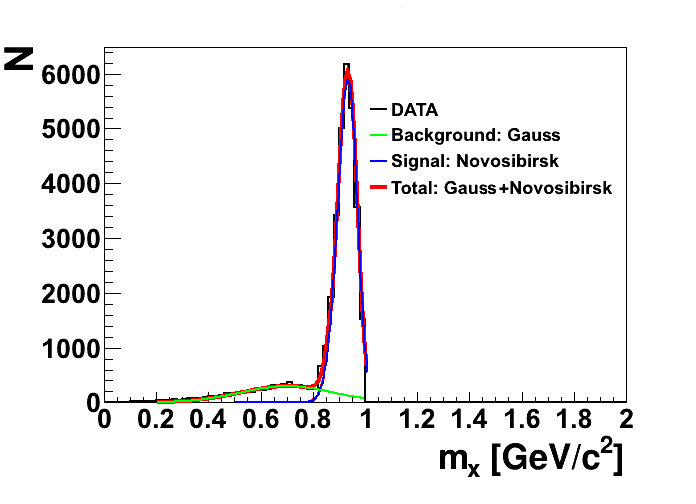}
\caption{The missing mass $m_{x}$ spectrum for $cos\theta^{*}$~$\in$~(0.96,0.98) and $Q$~$\in$~(0,5)~MeV. The red line shows fit to the signal and background while green line shows fit of the Gauss function to the background. Signal peak is marked as a blue line.~\label{fig_mx_nov}}
\end{figure}

\indent Additional, for the high beam momentum region background was subtracted via fitting the signal and background function to the missing mass spectrum for different intervals of $cos\theta^{*}$ and beam momentum. An example of the result for $Q$ $\in$~(0,5)~MeV and $cos\theta^{*}\in$~(0.96,0.98) is presented in Fig.~\ref{fig_mx_nov}. The signal is described with a Novosibirsk function, which is given by formula~(\ref{eq_novv1}) and~(\ref{eq_novv2}) (see Sec.~\ref{SEC_calib}). The background is fitted with a Gauss function. Fit to the signal and background is marked as a red line, while the background alone is marked as a green line.

For the low beam momentum regions, the obtained missing mass spectra were almost background-free, therefore no fitting procedure was necessary.

\vspace{0.5cm}

\noindent \textbf{Luminosity determination}

\noindent In order to calculate the total integrated luminosity, the number of events $N$, the efficiency $\epsilon_{i,j}$, as well as cross section $\frac{d\sigma_{i,j}}{d(cos\theta^{*})}$ was determined for 5 intervals of $cos\theta^{*}$ in the range from 0.88 to 0.98 and 5 intervals of excess energy $Q$ in the range from -70~MeV to 30~MeV. The integrated luminosity was then calculated for each $(i,j)$-th interval in following way:

\begin{equation}
L^{int}_{i,j}=\frac{N_{i,j}}{\epsilon_{i,j}\cdot\frac{d\sigma_{i,j}}{d(cos\theta^{*})}\cdot \Delta(cos\theta^{*})},
\end{equation} 

\noindent where $\Delta(cos\theta^{*})$ is the width of the $cos\theta^{*}$ interval. The overall efficiency including reconstruction efficiency and geometrical acceptance of the detector was determined based on the Monte Carlo simulations and it varies between 50\% and 70\%. 

\begin{figure}[h!]
\centering
\includegraphics[width=12.0cm,height=7.0cm]
{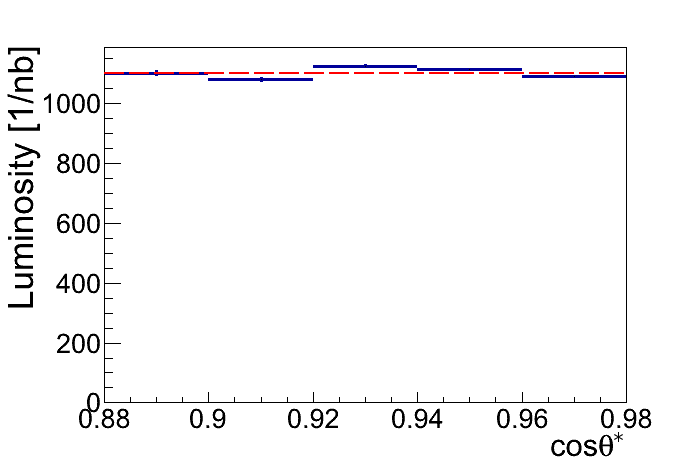}
\vspace{-0.4cm}
\caption{Integrated luminosity as a function of $cos\theta^{*}$.~The statistical uncertainties are marked as vertical bars. The weighted average of integrated luminosity is marked as a dashed red line and is equal to $1102\pm2$~nb$^{-1}$ where only a statistical error is given.~\label{fig_18}}
\end{figure}

\indent The luminosity dependence of $cos\theta^{*}$ for whole excess energy range is presented in Fig.~\ref{fig_18}.

The total integrated luminosity was calculated as a weighted average of the luminosities determined for individual $cos\theta^{*}$ intervals:

\begin{equation}
L^{tot}_{dd\rightarrow ^{3}\hspace{-0.03cm}He n}=\frac{\sum_{i=1}^{5} L_{i} \frac{1}{(\Delta L_{i})^2}}{\sum_{i=1}^{5} \frac{1}{(\Delta L_{i})^2}}, \hspace{0.5cm} \Delta L^{tot}_{dd\rightarrow ^{3}\hspace{-0.03cm}He n}=\left(\sum_{i=1}^{5} \frac{1}{(\Delta L_{i})^2}\right)^{-1/2}.\label{Lum_av}
\end{equation}

\noindent The average integrated luminosity with its statistical uncertainty is equal to
\newline $L^{tot}_{dd\rightarrow ^{3}\hspace{-0.03cm}He n}$ $=(1102\pm2)$~nb$^{-1}$ (see Fig.~\ref{fig_18}).

\section{Dependence on the excess energy -- quasi-free $\mathbold{dd\rightarrow p p n_{sp} n_{sp}}$ reaction analysis~\label{Lum_qf}} 

In order to determine the luminosity dependence on the beam momentum we used the quasi-elastic proton-proton scattering in the deuteron-deuteron collisions: $dd\rightarrow p p n_{sp} n_{sp}$, which, in contrast to $\LumReacta$ reaction, is characterized by the smooth acceptance in the whole momentum range. 
~In this reaction protons from the deuteron beam are scattered on the protons in the deuteron target. We assume that the neutrons are acting only as spectators which means that they do not take part in reactions.

\indent In the case of quasi-free proton-proton scattering the formula for the calculation of the integrated
luminosity can be written in the following form~\cite{Mos_Czyz}:

\begin{equation}
L_{dd\rightarrow p p n_{sp} n_{sp}}=\frac{N_{0} N_{exp}}{2\pi I},
\label{eq:112}
\end{equation}

\noindent where:\\

\begin{center}
$I=\int_{\Delta\Omega(\theta_{lab},\phi_{lab})}\frac{d\sigma}{d\Omega}(\theta^{*},\phi^{*},{p}_{F_{b}}, \theta_{F_{b}}, \phi_{F_{b}},{p}_{F_{t}}, \theta_{F_{t}}, \phi_{F_{t}})$ \\

\vspace{0.3cm}

$\times f({p}_{F_{b}}, \theta_{F_{b}}, \phi_{F_{b}}, {p}_{F_{t}}, \theta_{F_{t}}, \phi_{F_{t}})d{p}_{F_{b}} dcos\theta_{F_{b}} d\phi_{F_{b}} d{p}_{F_{t}} dcos\theta_{F_{t}} d\phi_{F_{t}}d\phi^{*}dcos\theta^{*}$.
\end{center}

\begin{figure}[h!]
\centering
\includegraphics[width=7.0cm,height=5.2cm]{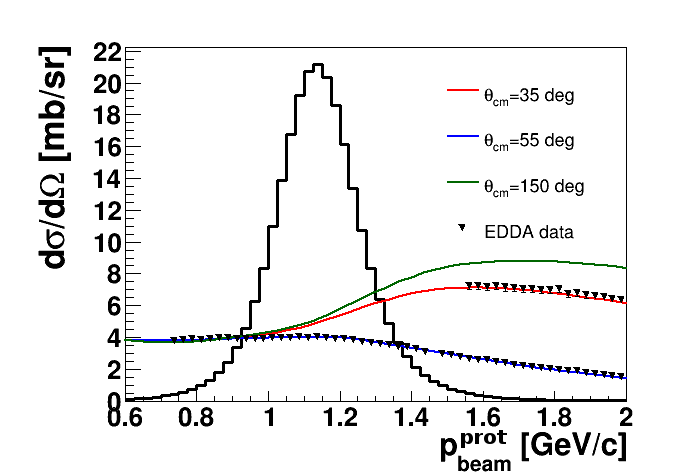}
\hspace{-0.5cm}
\includegraphics[width=7.0cm,height=5.0cm]{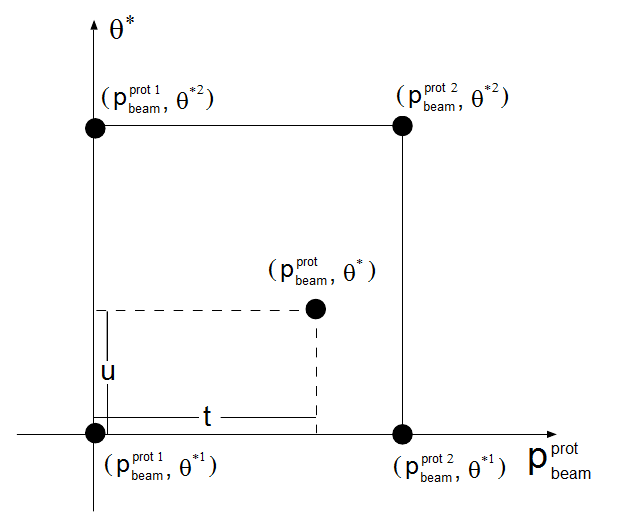} 
\caption{(left) Differential cross sections for proton-proton elastic scattering as a function of the effective beam momentum $p^{prot}_{beam}$ for a three values of the scattering angle $\theta^{*}$ in the CM frame. Black points show EDDA collaboration data~\cite{Albers_EDDA}, while lines denote SAID calculations~\cite{SAID_data}. Distribution of the effective beam momentum for quasi-free $pp\rightarrow p p$ reaction calculated for the deuteron beam momentum range $p_{beam}\in$~(2.127,2.422)~GeV/c is also presented in the figure. (right) Bilinear interpolation of the differential cross section $\frac{d\sigma}{d\Omega}(p^{prot}_{beam},\theta^{*})$. The right figure is adopted from~\cite{Mos_Czyz}.~\label{fig_qf_XS}}
\end{figure}

\noindent The formula is determined based on the fact, that the number of quasi-free scattered protons into the solid angle $\Delta\Omega(\theta_{lab},\phi_{lab})$ is proportional to the integrated luminosity $L$, as well as the inner product of the differential cross section for scattering into the solid angle around $\theta^{*}$ and $\phi^{*}$ expressed in proton-proton CM system: $\frac{d\sigma}{d\Omega}(\theta^{*},\phi^{*},{p}_{F_{b}}, \theta_{F_{b}}, \phi_{F_{b}},{p}_{F_{t}}, \theta_{F_{t}}, \phi_{F_{t}})$ and the probability density of the Fermi momentum distributions: $f({p}_{F_{b}}, \theta_{F_{b}}, \phi_{F_{b}})\cdot f({p}_{F_{t}}, \theta_{F_{t}}, \phi_{F_{t}})$ inside the deuteron beam ($b$) and deuteron target ($t$), respectively. The detailed description of the luminosity calculation for quasi-free reaction one can find in Ref.~\cite{Mos_Czyz}.\\
\indent Due to the complex detection geometry a solid angle corresponding to particular part of the detector cannot be in general expressed in a closed analytical form.~Therefore, the integral in above equation was computed with the Monte Carlo simulation program (its scheme is presented in Appendix~\ref{App_Sim_qf}), containing the geometry of WASA detection system and taking into account detection and reconstruction efficiencies.~The Monte Carlo simulations were carried out for the deuteron beam momentum range $p_{beam}\in$~(2.127,2.422)~GeV/c corresponding to the experimental ramping.~The program first chooses randomly the momentum magnitude of the nucleon inside the deuteron beam and deuteron target, respectively, according to the Fermi distribution~\cite{Lacombe}. The direction of the momentum vector is chosen isotropically. Then, the total proton-proton invariant mass $\sqrt{s_{pp}}$ and the vector of the center-of-mass velocity are determined. Next, the effective proton beam momentum $p^{prot}_{{beam}}$ is calculated in the frame where one of the proton is at rest and momentum of protons is generated isotropically in the proton-proton center-of-mass frame. Further on, the momenta of outgoing particles are transformed to the laboratory frame and are used as an input in the simulation of the detection system response with the GEANT computing package. Detailed description of the next simulation steps is presented in Appendix~\ref{App_Sim_qf}. For each of $N_{0}$ simulated event, we assign a weight corresponding to the differential cross section, which is uniquely determined by the scattering angle and the total proton-proton collision energy~$\sqrt{s_{pp}}$.\\
\indent The factor $N_{0}/2\pi$ in Eq.~(\ref{eq:112}) is a normalization constant.~It results from the fact that the integral is not dimensionless and its units correspond to the units of the cross sections used for the calculations. Therefore, it must be normalized in such a way that the integral over the full solid angle is equal to the total cross section for the elastic scattering averaged over the distribution of the total proton-proton invariant mass $\sqrt{s_{pp}}$ resulting from the Fermi distribution of the target and beam nucleons. In the absence of the Fermi motion it should be simply equal to a total elastic cross section for a given proton beam momentum. A factor $2\pi$ comes from the fact that protons taking part in the scattering are indistinguishable.\\
\indent The differential cross section for quasi free $dd\rightarrow p p n_{sp} n_{sp}$ reaction is a function of the scattering angle $\theta^{*}$ and the total energy in the proton-proton centre-of-mass system $\sqrt{s_{pp}}$ which is dependent on effective proton beam momentum $p^{prot}_{beam}$ seen from the proton in the proton-proton system. In order to calculate it, we have used the cross section values for proton-proton elastic scattering $pp\rightarrow pp$ computed based on the SAID program~\cite{SAID_data} because the EDDA collaboration data base~\cite{Albers_EDDA} is insufficient in the region of interest. The distribution of the effective beam momentum, as well as a comparison of the SAID calculations and the existing differential cross section from the EDDA measurements are shown in Fig.~\ref{fig_qf_XS}. As we can see, the differential cross sections calculated using the SAID programme are in agreement with distributions measured by the EDDA collaboration.

\indent The differential cross section for appropriate $p^{prot}_{beam}$ and $\theta^{*}$ was calculated using bilinear interpolation in the $p^{prot}_{beam}-\theta^{*}$ plane according to the formula:

\begin{equation}
\begin{split}
\frac{d\sigma}{d\Omega}(p^{prot}_{beam},\theta^{*})=(1-t)(1-u)\frac{d\sigma}{d\Omega}(p^{prot 1}_{beam},\theta^{* 1})+
t(1-u)\frac{d\sigma}{d\Omega}(p^{prot 2}_{beam},\theta^{* 1})+\\
tu\frac{d\sigma}{d\Omega}(p^{prot 2}_{beam},\theta^{* 2})+(1-t)u\frac{d\sigma}{d\Omega}(p^{prot 1}_{beam},\theta^{* 2}),
\end{split}
\end{equation}

\noindent where $t$ and $u$ variables are defined in right panel of Fig.~\ref{fig_qf_XS}.

\indent The number of experimental events $N_{exp}$ was determined based on conditions and cuts described in details in reference~\cite{Krzemien_PhD}. The analysis is based on the events selected by the hardware trigger for elastic scattering (Sec.~\ref{Sec_Triggers}) and we carried out primary events selection applying condition of exactly one charged particle in the Forward Detector (FD) and one particle in the Central Detector (CD).

\begin{figure}[h]
\centering
\includegraphics[width=9.0cm,height=6.0cm]{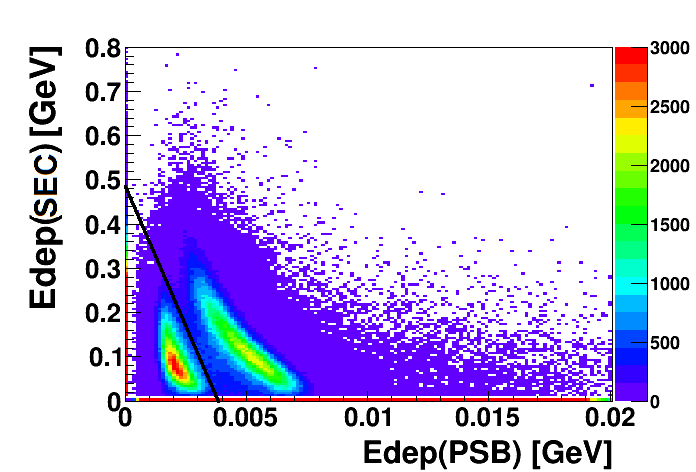} 
\vspace{-0.1cm}
\caption{Experimental spectrum of the energy loss in the Plastic Scintillator Barrel shown as a function of the energy deposited in the Electromagnetic Calorimeter. The applied cut is shown as a black line. Points corresponding to pions are concentrated for Edep(PSB) around 0.002~GeV.~\label{fig_psb}}
\end{figure}

\indent In Ref.~\cite{Krzemien_PhD} we can find detailed studies of the possible background reaction contributions. In case of this analysis the dominating background processes are $dd \rightarrow d \pi^{+} n_{sp} n_{sp}$, $dd\rightarrow d_{b} p_{t} n_{sp}$ and $dd\rightarrow p p_{sp} n n_{sp}$, where the subscripts $sp$, $b$ and $t$ denote the spectators, particles from the beam, and from the target, respectively. In order to reject events corresponding to the charged pions registered in the Central Detector, the cut on the energy deposited in the Electromagnetic Calorimeter (SEC) vs. energy deposited in Plastic Scintillator Barrel (PSB) spectrum was applied and is presented in Fig.~\ref{fig_psb}.

\begin{figure}[h!]
\centering
\includegraphics[width=7.0cm,height=5.0cm]{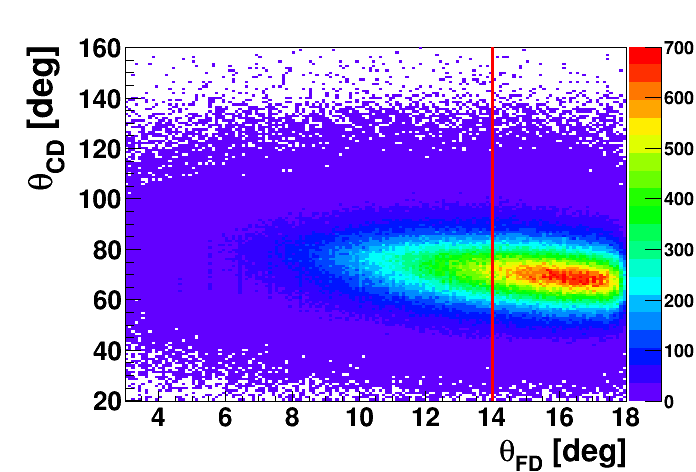}
\hspace{-0.3cm}
\includegraphics[width=7.0cm,height=5.0cm]{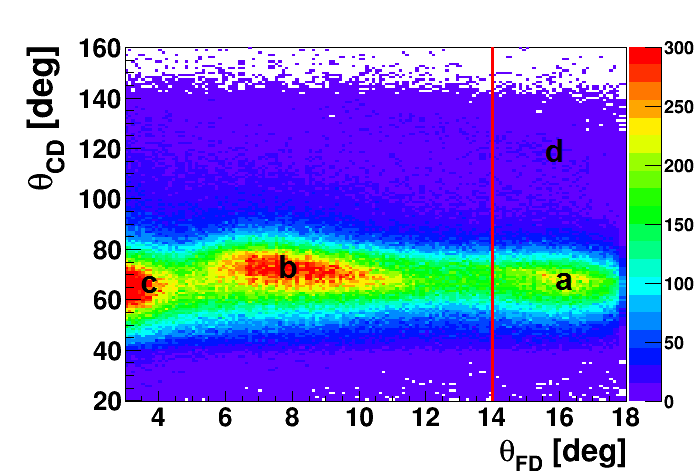} \\
\vspace{-0.1cm}
\caption{Correlations between the polar angles $\theta_{FD}$ and $\theta_{CD}$ of $dd\rightarrow p p n_{sp} n_{sp}$ reaction as simulated (left panel) and obtained in experiment (right panel).~Applied cut is marked with red line. The indicated area correspond to the: a) $dd\rightarrow p p n_{sp} n_{sp}$, b) $dd\rightarrow d_{b} p_{t} n_{sp}$ and $dd\rightarrow p p n_{sp} n_{sp}$, c) $dd\rightarrow p p_{sp} n n_{sp}$, d) $dd\rightarrow p_{b} d_{t} n_{sp}$.~\label{fig_th_copl}}
\end{figure}

\begin{figure}[h!]
\centering
\includegraphics[width=9.0cm,height=6.0cm]{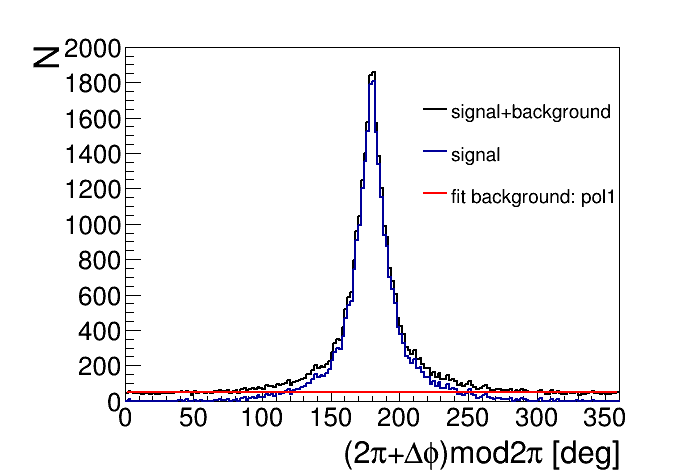}
\caption{Distributions of (2$\pi+\Delta\phi$)mod2$\pi$, where $\Delta\phi=\phi_{CD}-\phi_{FD}$ is the difference of azimuthal angles in Central Detector and Forward Detector.~The example spectrum for one of the $Q$ intervals (black line) with marked fit function (red line) and signal peak after background subtraction (blue line) is presented.~\label{fig_copl}}
\end{figure}

\indent It is not possible to separate quasi-elastic $pp$ scattering from the quasi-elastic $dp$ scattering, however it was investigated that for the forward scattering angles of about $\theta_{FD}$ = 17$^{\circ}$, the $dp$ cross sections are about 20 times smaller than $pp$ cross sections and we take this uncertainty of about 5\% as a systematic error to the final result. The applied cut in polar angle $\theta_{FD}$ is shown in Fig.~\ref{fig_th_copl}. In order to subtract the background coming from $dd\rightarrow p_{b} d_{t} n_{sp}$ reaction, the range $\theta_{CD}\in$~(40,100)$^{\circ}$ was taken into account in further analysis. Additionally, the background was subtracted in $\Delta\phi=\phi_{FD}-\phi_{CD}$ spectrum. In order to symmetrize the background instead of $|\Delta\phi|$ we define (2$\pi+\Delta\phi$)mod2$\pi$. Afterwards, the background was fitted with 1st order polynomial for each of excess energy $Q$ intervals.~The exemplary \mbox{(2$\pi+\Delta\phi$)mod2$\pi$} spectrum is presented in Fig.~\ref{fig_copl}.   
  
\vspace{-0.1cm}   
  
\indent After all cuts and application of conditions described above, the number of experimental events was determined and the luminosity was calculated according to formula (\ref{eq:112}) for each excess energy interval. In the calculations the prescaling factor of the applied experimental trigger equal to 4000, as well as shadowing effect equals 9\% were taken into account.~The latter results from the fact that proton is shadowed by the neutron inside the deuteron which reduces the probability of the quasi-elastic scattering. Unfortunately, there are no experimental results about the shadowing in $dd\rightarrow p p n_{sp} n_{sp}$ collisions. However, we can estimate it based on the probability that a neutron shadows the proton in one deuteron which equals 0.045~\cite{Chiavasa} and assume that shadowing appears independently in deuteron beam and deuteron target. The rough estimation of the probability that the shadowing will not take place in $dd$ reaction \mbox{(1 - 0.045)$^{2}$} gives about 0.91.

\begin{figure}[h!]
\centering
\includegraphics[width=11.0cm,height=7.0cm]{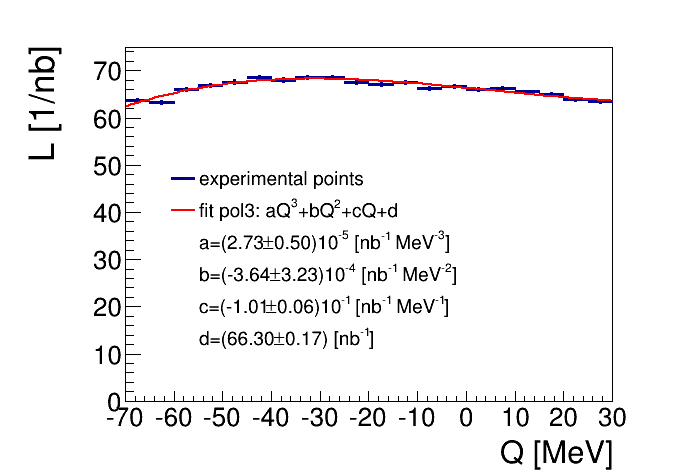}
\caption{Integrated luminosity calculated for experimental data for quasi-free $dd\rightarrow p p n_{sp} n_{sp}$ reaction (blue points) with fitted third degree polynomial function (red line).~\label{lum_norm_fit}}
\end{figure}

\indent The final result is presented in Fig.~\ref{lum_norm_fit}. The statistical uncertainty of each point is about 1\%. The luminosity variation (increase in the excess energy range from about -70~MeV to -40~MeV, and then decrease) is caused by the change of the beam-target overlapping during the acceleration cycle and also by adiabatic beam size shrinking~\cite{Lorentz}. The obtained total integrated luminosity within its statistical uncertainty is equal to $L^{tot}_{dd\rightarrow p p n_{sp} n_{sp}}$ $=(1326\pm2)$~nb$^{-1}$. For further analysis the luminosity was fitted by third degree polynomial $aQ^{3}+bQ^{2}+cQ+d$. The fitted function is marked with the red line in Fig.~\ref{lum_norm_fit}.  

\newpage 

\section{Systematics}

\noindent In case of the $dd\rightarrow$ $^{3}\hspace{-0.03cm}\mbox{He} n$ reaction one source of the systematic error originates from the definition of the cuts used for separation of high-energetic helium in Forward Detector and is equal to about 2\%. Additionally we have also taken into account an uncertainty due to the method used for the background subtraction amounting to 1.6\%. Another source of the luminosity calculation error is connected to normalization to SATURNE experiment and originates from three independent sources: (i) statistical error of the SATURNE data (6.5\%), (ii) normalization uncertainty of the SATURNE data for the $dd\rightarrow$ $^{3}\hspace{-0.03cm}\mbox{He} n$ cross sections (7\%) and (iii) assumption of linear interpolation between SATURNE points used for the estimation of the correction for the parametrized cross section presented in Fig.~\ref{fig_Bizzard} ($<$1.8\%). 

\indent The systematic errors for $dd\rightarrow p p n_{sp} n_{sp}$ analysis resulting from the cuts used for the separation of the quasi-free $pp$ scattering from the background (Fig. \ref{fig_psb} and Fig. \ref{fig_th_copl}) is equal to about 4.1\%. Another contribution to the systematic error comes from the assumption of the potential model of the nucleon bound inside the deuteron and is equal to about 0.8\%. This uncertainty was established as the difference between results determined using the Paris~\cite{Lacombe} and the CDBonn~\cite{CDBonn} potentials. The next source of the systematic error may be attached to the assumption of the bilinear approximation of the cross section shown in Fig.~\ref{fig_qf_XS} (right).~This systematic uncertainty was estimated using assumption in which instead of the interpolation we took the cross section value from the closest data point in the effective proton beam momentum - scattering angle plane. The performed calculations give the difference of about 1.8\%. Additionally we have also taken into account an uncertainty related to the background subtraction in (2$\pi+\Delta\phi$)mod2$\pi$ spectra which does not exceed 0.6\%. The systematic uncertainty includes also contribution connected to the shadowing effect. Until now, we have no theoretical estimation of the possible error of this effect, therefore conservatively we take as an systematic uncertainty half of this effect: 4.5\%. In the systematic error calculation we also take into account the uncertainty 5\% resulting from the background of the quasi-elastic $dp$ scattering. The normalization error includes also normalization uncertainty of the EDDA data (4\%) and the systematic errors for $pp$ elastic scattering cross-sections (2.7\%)~\cite{Albers_EDDA}. The cross section was approximated by the calculation using the SAID procedure. Because, the SAID cross section very well describes EDDA data, we assume the systematic errors of the differential cross section based on EDDA calculations.\\ 
\indent The total integrated luminosity determined based on $dd\rightarrow$ $ ^{3}\hspace{-0.03cm}\mbox{He} n$ and the quasi-free $dd\rightarrow p p n_{sp} n_{sp}$ reactions with statistical, systematic and normalization error are equal to $L^{tot}_{dd\rightarrow ^{3}\hspace{-0.03cm}He n}=(1102\pm2_{stat}\pm28_{syst}\pm107_{norm})$~nb$^{-1}$ and $L^{tot}_{dd\rightarrow p p n_{sp} n_{sp}}=(1326\pm2_{stat}\pm108_{syst}\pm64_{norm})$~nb$^{-1}$, respectively. The systematic and normalization errors were calculated by adding in quadrature the appropriate contributions described above.

\indent To summarize, the luminosity was calculated based on two reactions: $dd\rightarrow$ $ ^{3}\hspace{-0.03cm}\mbox{He} n$ and the quasi-free $dd\rightarrow p p n_{sp} n_{sp}$. The obtained results are consistent, however within large normalization errors.

\chapter{Results and interpretation~\label{Results_interpr}} 

\noindent This chapter presents the determination of the excitation function for $\BcgReacta$ reaction. Moreover, the contribution from the main experimental background in the "Signal Rich" and "Signal Poor" region was investigated. The obtained results are described in sections below.

\section{Excitation function}

\begin{figure}[h!]
\centering
\includegraphics[width=7.0cm,height=4.7cm]{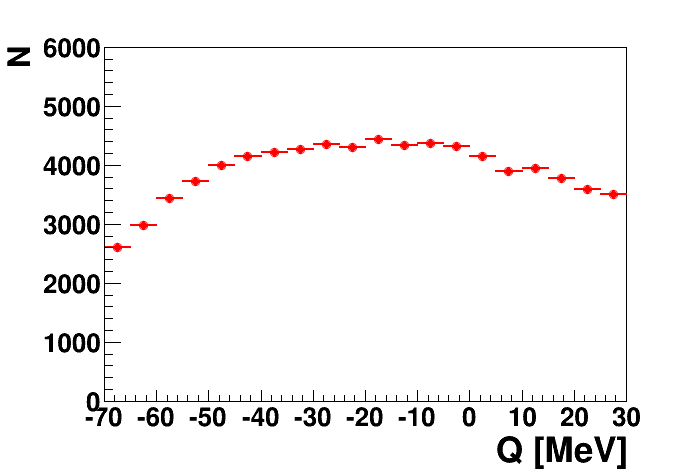}
\hspace{-0.3cm}
\includegraphics[width=7.0cm,height=4.7cm]{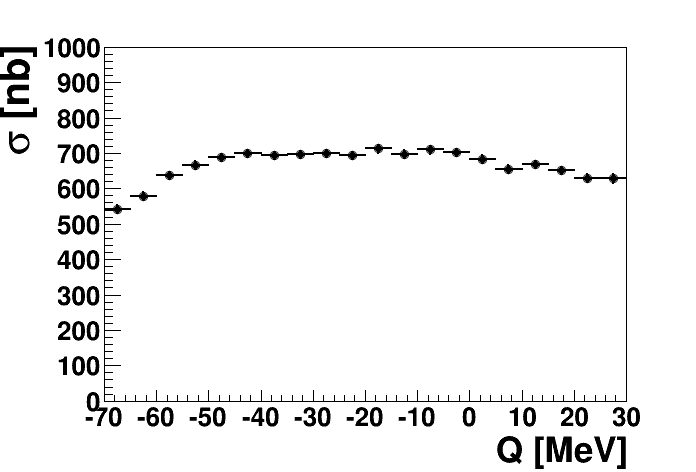}
\vspace{-1.0cm}
\caption{(left) Excitation function for the $dd\rightarrow$ $^{3}\hspace{-0.03cm}\mbox{He} n \pi{}^{0}$ reaction including events selected via applied conditions and cuts. (right) Excitation function obtained by normalizing the events selected in individual excess energy intervals by the corresponding integrated luminosities and efficiencies.~\label{exc_fcn_allcuts}}
\end{figure}

The excitation function for $\BcgReacta$ process was determined for the region where we expect the signal ("Signal Rich" region) after all applied conditions and cuts presented in Chapter~\ref{Analysis}. The excitation curve is shown in left panel of Fig.~\ref{exc_fcn_allcuts}. 

\vspace{-0.3cm}

\noindent The number of events in each excess energy interval was divided by the corresponding integrated luminosity $L(Q)$ determined based on quasi-free proton-proton scattering (Sec.~\ref{Lum_qf}) and corrected for total efficiency (Chapter~\ref{Efficiency}).~The normalized excitation function is presented in the right panel of Fig.~\ref{exc_fcn_allcuts}.


\section{Upper limit of the total cross section}

\begin{figure}[h!]
\centering
\includegraphics[width=9.0cm,height=6.0cm]{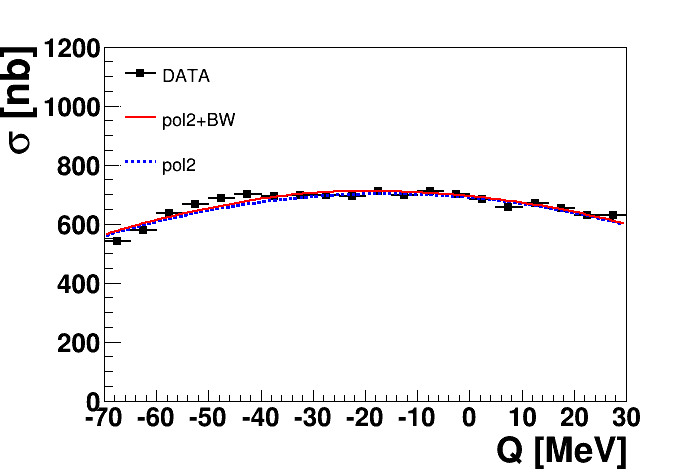}
\caption{Excitation function for the $dd\rightarrow$ $^{3}\hspace{-0.03cm}\mbox{He} n \pi{}^{0}$ reaction obtained by normalizing the events selected in individual excess energy intervals by the corresponding integrated luminosities and corrected for acceptance and effciency. The red solid line represents a fit with second order polynomial combined with a Breit-Wigner function with fixed binding energy and width equal to 30 and 40~MeV, respectively. The blue dotted line shows the second order polynomial corresponding to the background.~\label{ex_fcn_fit}}
\end{figure}

The shape of obtained excitation function can be well described with a quadratic function fit resulting in the $\chi^{2}$ value per degree of freedom of 1.1 (Fig.~\ref{ex_fcn_fit}). The excitation function does not indicate significant sharp enhancement for energies below the $\eta$ production threshold which could be interpreted as a resonance-like structure. Therefore, we can only determine an upper limit for the cross-section for formation of the $^{4}\hspace{-0.03cm}\mbox{He}$-$\eta$ bound state and its decay into the $^{3}\hspace{-0.03cm}\mbox{He} n \pi{}^{0}$ channel. The excitation function was fitted with quadratic function describing the background combined with the Breit-Wigner function which can account for the signal from the bound state:

\begin{equation}
\sigma(Q,\Gamma,B_{s},A)=\frac{A\cdot\Gamma^{2}/4}{\left(Q-B_{s}\right)^{2}+\Gamma^{2}/4}\label{eq:BW},
\end{equation}

\noindent where: 
\noindent $B_{s}$ -binding energy,\\
\noindent $\Gamma$ - width,\\
\noindent $A$ - amplitude.\\

In applied fit the polynomial coefficients and the amplitude $A$ of the Breit-Wigner distribution are treated as free parameters while the binding energy $B_{s}$ and width $\Gamma$ are fixed parameters. The fit was performed for various values of $B_{s}$ and $\Gamma$. The binding energy and the width were varied in the range from 0 to 40 MeV
and from 5 to 50~MeV, respectively. An example of the fit for $\Gamma=40$~MeV and $B_{s}=30$~MeV is presented in Fig.~\ref{ex_fcn_fit}.

\begin{figure}[h!]
\centering
\includegraphics[width=11.0cm,height=7.0cm]{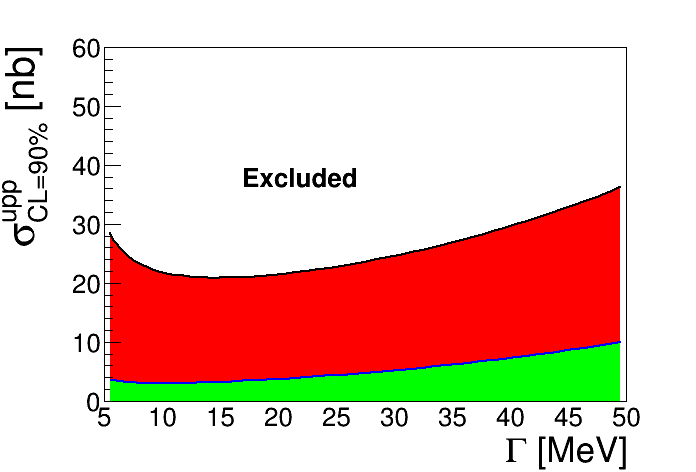}
\caption{Upper limit of the total cross-section for $\MainReact$ reaction as a function of the width of the bound state. The binding energy was set to 30~MeV. The green area denotes the systematic uncertainties.~\label{Result_sigma_upp}}
\end{figure}

The upper limit of the total cross section was calculated as:

\begin{equation}
\sigma^{upp}_{CL=90\%}=k \cdot \sigma_{A},\label{eq:sigma_upp}
\end{equation}

\noindent where $\sigma_{A}$ is uncertainty of the amplitude $A$ obtained from the fit and $k$ is statistical factor equal to 1.64485 corresponding to the confidence level (CL) of 90\%. The example values of the obtained upper limit are given in Table~\ref{tab_upp_lim}. As one can see, the upper limit depends mainly on the width of the bound state while its dependence on the binding energy is only slight. The obtained upper limit as a function of bound state width is presented in Fig.~\ref{Result_sigma_upp} for binding energy 30~MeV. Its value varies between 21 to 36~nb. The green area denotes the systematic errors which contributions are described in details in the next section.

\vspace{-0.3cm}

\begin{table}[h!]
\begin{normalsize}
\begin{center}
\begin{tabular}{|c|c|c||c|c|c|}
\hline
$B_{s}$~[MeV] & $\Gamma$~[MeV] & $\sigma^{upp}_{CL=90\%}$ [nb] &$B_{s}$~[MeV] & $\Gamma$~[MeV] & $\sigma^{upp}_{CL=90\%}$ [nb]\\
\hline
10 &10 &22.05 &30 &10 &21.04\\
10 &20 &21.34 &30 &20 &21.35\\
10 &30 &24.32 &30 &30 &24.32\\
10 &40 &29.26 &30 &40 &29.27\\
10 &50 &36.13 &30 &50 &36.15\\
20 &10 &22.61 &40 &10 &21.55\\
20 &20 &22.72 &40 &20 &20.08\\
20 &30 &27.00 &40 &30 &22.12\\
20 &40 &33.93 &40 &40 &25.81\\
20 &50 &39.24 &40 &50 &30.94\\
\hline 
\end{tabular}
\end{center}
\vspace{-0.8cm}
\begin{center}
\caption{The upper limit of the total cross-section for the $\MainReact$ process determined at CL=90\% for different values of binding energy $B_{s}$ and width $\Gamma$.\label{tab_upp_lim}}
\end{center}
\end{normalsize}
\end{table}

\newpage

\section{Systematics}

Systematic studies were carried out based on Ref.~\cite{BS_WASA}. The variation of the selection conditions by $\pm$10\% gives the systematic error which equals about 6\% (the highest contribution to the error comes from changing the range of the "Signal Rich" region A (see Sec.~\ref{kinem_cuts_Main}). Another, significant contribution to the systematic error of the upper limit is connected to the luminosity determination based on quasi-free $pp$ reaction (Chapter~\ref{Lum_determ}) and equals about 8\% and 5\% which correspond to the systematic and normalization error, respectively. Additional source of systematic errors comes from the assumption of the Fermi momentum distribution of nucleons inside the $\Heb$. The distribution was used in Monte Carlo simulation of the bound state production and decay (Chapter~\ref{Sim_MainReact}). Current analysis was performed with the Fermi momentum distribution based on AV18-TM potential model~\cite{Nogga2}. Another available momentum distribution based on CDB2000-UIX model only slightly changes the acceptance for simultaneous registration of all particles in WASA detector (see Sec.~\ref{Nucl_mom_4He}) providing systematic error of about 1\%. The uncertainty caused by the fit of quadratic or linear function to the background, estimated as $\frac{\sigma_{quad}-\sigma_{lin}}{2}$, changes from about 5\% ($\Gamma$=5~MeV) to 25\% ($\Gamma$=50~MeV). The systematic error is calculated by adding in quadrature all contributions described above and varies from 12\% to 27\% as shown by green area in Fig.~\ref{Result_sigma_upp}.
 
\section{Background studies~\label{Bcg_stud}} 

The excitation function presented in the previous section is dominated by the background which interferes in the $\MainReact$ process. Therefore, understanding of the background processes is crucial for the considered investigations.~We performed the studies of two reactions being the main background contributions:

\begin{enumerate}

\item $\BcgReacta$,

\item $\BcgReactb$.

\end{enumerate}

The simulations of above reactions were carried out for the beam momentum range corresponding to the experimental range $p_{beam}\in$(2.127,2.422) GeV/c.~The first reaction was simulated according to direct production with the uniform distribution over the phase space. The second process proceeds with excitation of an $N^{*}$ resonance which subsequently decays in the nucleon-$\pi^{0}$ pair. The detailed description of the simulations of $\BcgReactb$ reaction can be found in the Appendix~\ref{Sim_Nstar}.

\begin{figure}[h!]
\centering
\includegraphics[width=7.2cm,height=5.0cm]{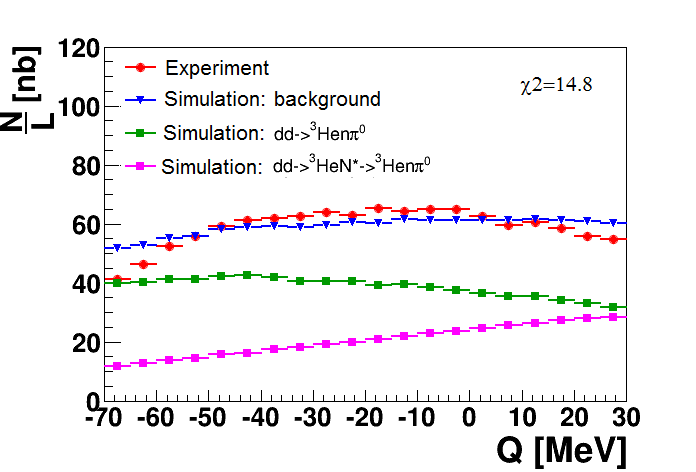}
\hspace{-0.7cm}
\includegraphics[width=7.2cm,height=5.0cm]{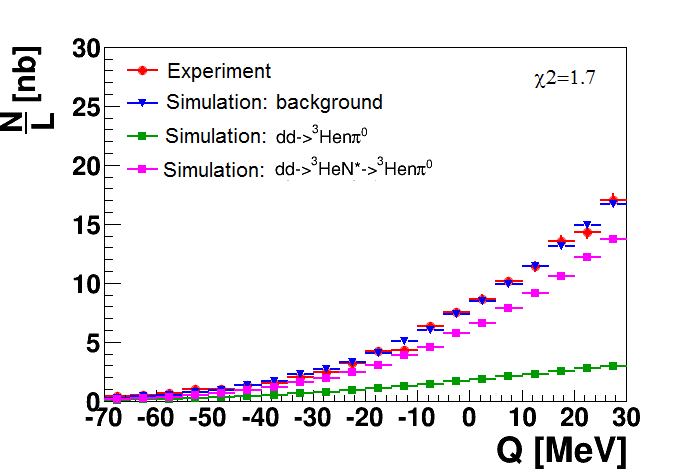}
\caption{Experimental excitation functions (red circles) fitted with two background reactions: $\BcgReacta$ (green squares) and $\BcgReactb$ (magenta squares). A sum of both background contributions is shown as blue triangles. Left and right panels show results for the "Signal Rich" and "Signal Poor" regions, respectively.~\label{fit_plots}}
\end{figure}

We compared the excitation curves obtained from simulations of background processes with the excitation function determined from the experiment. The excitation function for experimental data normalized over luminosity was fitted with a \mbox{$A\cdot WMC_{bcg1}+ B\cdot WMC_{bcg2}$} function where $WMC_{bcg1}$ and $WMC_{bcg2}$ are the excitation functions obtained from the simulations of $\BcgReacta$ and $\BcgReactb$ reactions, respectively. The simulations were carried out using the conditions and cuts described in Chapter~\ref{Analysis}. The result of the fit is presented in Fig.~\ref{fit_plots} for region where we expect the signal from the bound state ($p^{cm}_{^{3}\hspace{-0.05cm}He}\in$~(0.07,0.2)~GeV/c) and for the region poor in signal ($p^{cm}_{^{3}\hspace{-0.05cm}He}\in$~(0.3,0.4)~GeV/c). It is important to stress that the experimental excitation function for the "Signal Rich" region cannot be well described only by the background contributions (see left panel of Fig.~\ref{fit_plots}). It can indicate missing contribution from another background processes, or the influence of wide \mbox{$\BS$} or \mbox{$\Hea$-$N^{*}$} bound state. In contrast, as shown in the right panel of Fig.~\ref{fit_plots}, the experimental excitation function from the "Signal Poor" region is very well described by the background reactions.

\newpage

\chapter{Conclusions and outlook}

This dissertation describes the search for the $\BS$ bound state via the study of the excitation function for the $\BcgReacta$ reaction. It includes detailed description of the performed experiment, the data analysis and obtained results.  

The measurement of $\BcgReacta$ reaction was performed in 2010 with the WASA-at-COSY detection system using the deuteron beam and deuteron pellet target. To reduce the systematic uncertainties of the beam momentum value, the ramped beam technique was used. The experiment was performed for the beam momentum range from 2.127~GeV/c to 2.422~GeV/c corresponding to the excess energy range of $Q\in$~(-70,30)~MeV for the $\BS$ system.

The performed analysis allowed for the determination of the excitation function for $\BcgReacta$ process and the estimation of the upper limit of the cross section for the $\eta$-mesic $\Heb$ formation and decay. Events corresponding to production of the $\eta$-mesic bound state were selected via cuts on the $\Hea$ momentum, neutron and $\pi^{0}$ kinetic energies as well as on the opening angle between $n$-$\pi^{0}$ in the center of mass frame.~The cuts were adjusted based on Monte Carlo simulations of the $\BSbound$ production and decay. The simulations assume that the $\eta$ meson inside bound state is absorbed by the one of neutron and excites it to $N^{*}$ resonance which subsequently decays into $n$-$\pi^{0}$ pair. The $\Hea$ plays here the role of spectator moving with Fermi momentum in the center of mass system. 

The total integrated luminosity in the experiment was determined based on the $dd\rightarrow$ $^{3}\hspace{-0.03cm}\mbox{He} n$ and quasi-free $pp\rightarrow p p$ reactions. It amounts to $L^{tot}_{dd\rightarrow ^{3}\hspace{-0.03cm}He n}=(1102\pm2_{stat}\pm28_{syst}\pm107_{norm})$~nb$^{-1}$ and $L^{tot}_{dd\rightarrow p p n_{sp} n_{sp}}=(1326\pm2_{stat}\pm108_{syst}\pm64_{norm})$~nb$^{-1}$, respectively. The obtained results are consistent within systematics and normalization error.  The excess energy dependence of luminosity used for relative normalization of the $\BcgReacta$ reaction events was determined based on quasi-free $pp\rightarrow p p$ reaction for which the WASA acceptance is a smooth function of the beam momentum.

The obtained excitation function of the $\BcgReacta$ reaction does not reveal the resonance-like structure which could be interpreted as the indication of the $\BS$ bound state with width less than 50~MeV. The fit of the excitation curve with the Breit-Wigner distribution combined with  second order polynomial allows to determine the upper limit of the $\MainReact$ cross section as function of the bound state width and binding energy. The upper limit varies from 21 to 36~nb for the width varying from 5~MeV to 50~MeV. The obtained upper limit is by factor of five larger than the theoretically estimated value of the cross section for the $dd\rightarrow$ $(^{4}\hspace{-0.03cm}\mbox{He}$-$\eta)_{bound}$ $\rightarrow$ $^{3}\hspace{-0.03cm}\mbox{He} p \pi{}^{-}$ reaction~\cite{WycechKrzemien}. Therefore, we can conclude, that the current measurement does not exclude the existence of $\BS$ bound state in considered process. 

The search for $\eta$-mesic $\Heb$ is complex due to huge background contribution. This background largely comprises two processes $\BcgReacta$ and $\BcgReactb$ being direct production and production via $N^{*}$ resonance without formation of $\BSbound$, respectively. The contribution of considered processes to the experimental excitation function was investigated for the regions rich in signal ("Signal Rich") and poor in signal ("Signal Poor") selected in the $\Hea$ momentum. In case of the region rich in signal the combination of the considered background reactions is not sufficient to properly describe the experimental data. This may suggest an influence of some other background process not taken into account, or the sign of the very wide \mbox{$\BS$} or \mbox{$\Hea$-$N^{*}$} bound state. This result is a subject of interpretation of few theoretical groups~\cite{Kelkar_2015,Hirenzaki_2015}. 
 
In 2014 the search for $\eta$-mesic bound states at WASA was extended to \mbox{$\Hea$-$\eta$} system. Based on the new research hypothesis about the mechanism of the decay of $\eta$-mesic nucleus we have elaborated an experimental proposal~\cite{Proposal_new} for the search of the $\Hea$-$\eta$ bound state which was accepted by the COSY Advisory Committee in February 2014 and already in the year 2014 we completed successfully the experimental run. Analysis of the collected data is in progress. The search for $\eta$ and $\eta'$ -mesic nuclei is carried out also by another experimental groups, e.g. at J-PARC~\cite{Fujioka_ActPhysPol2010, Fujioka2} and at GSI~\cite{yoshiki_gsi, fujioka_gsi_fair}.

\appendix

\chapter{The $\mathbold{\eta}$ meson~\label{Etaprop}}

The $\eta$ meson was discovered in 60' by Pevsner~\cite{Pevsner} as a resonance in three pion invariant mass spectrum. Its properties were investigated for many years and are summarized by the Particle Data Group~\cite{PDG}. The main information about the $\eta$ meson is briefly presented in Table~\ref{Tab_eta}.

\begin{table}[h!]
\begin{normalsize}
\begin{center}
\begin{tabular}{l l}\hline
\hline
\vspace{0.2cm}
mass & $547.853\pm0.024$ MeV\\
\vspace{0.2cm}
width &$1.30\pm0.07$ keV  \\
\vspace{0.2cm}
$I^{G}(J^{PC})$ &$0^{+}(0^{-+})$\\
\hline
\hline
\textbf{Decay modes} & \textbf{Branching ratio} \\
\hline\\
\vspace{0.2cm}
\textbf{Charged modes} &  28.10$\pm$0.34 \% \\
\vspace{0.2cm}
$\eta \rightarrow \pi^{+} \pi^{-} \pi^{0}$ & 22.74$\pm$0.28 \% \\
\vspace{0.2cm}
$\eta \rightarrow \pi^{+} \pi^{-} \gamma$ & 4.60$\pm$0.16 \% \\
\vspace{0.2cm}
other modes & 0.76 \%  \\
\vspace{0.2cm}
\textbf{Neutral modes} & 71.91$\pm$0.34 \% \\
\vspace{0.2cm}
$\eta \rightarrow 2\gamma$ & 39.31$\pm$0.20 \% \\
\vspace{0.2cm}
$\eta \rightarrow 3\pi^{0}$ & 32.57$\pm$0.23 \%   \\
\vspace{0.2cm}
other modes & 0.03 \%\\
\hline
\hline
\end{tabular}
\end{center}
\vspace{-0.8cm}
\begin{center}
\caption{$\eta$ meson properties and its main decay modes. The data are taken from~\cite{PDG}.\label{Tab_eta}}
\end{center}
\end{normalsize}
\end{table}

\begin{figure}[h!]
\centering
\includegraphics[width=8.5cm,height=7.5cm]{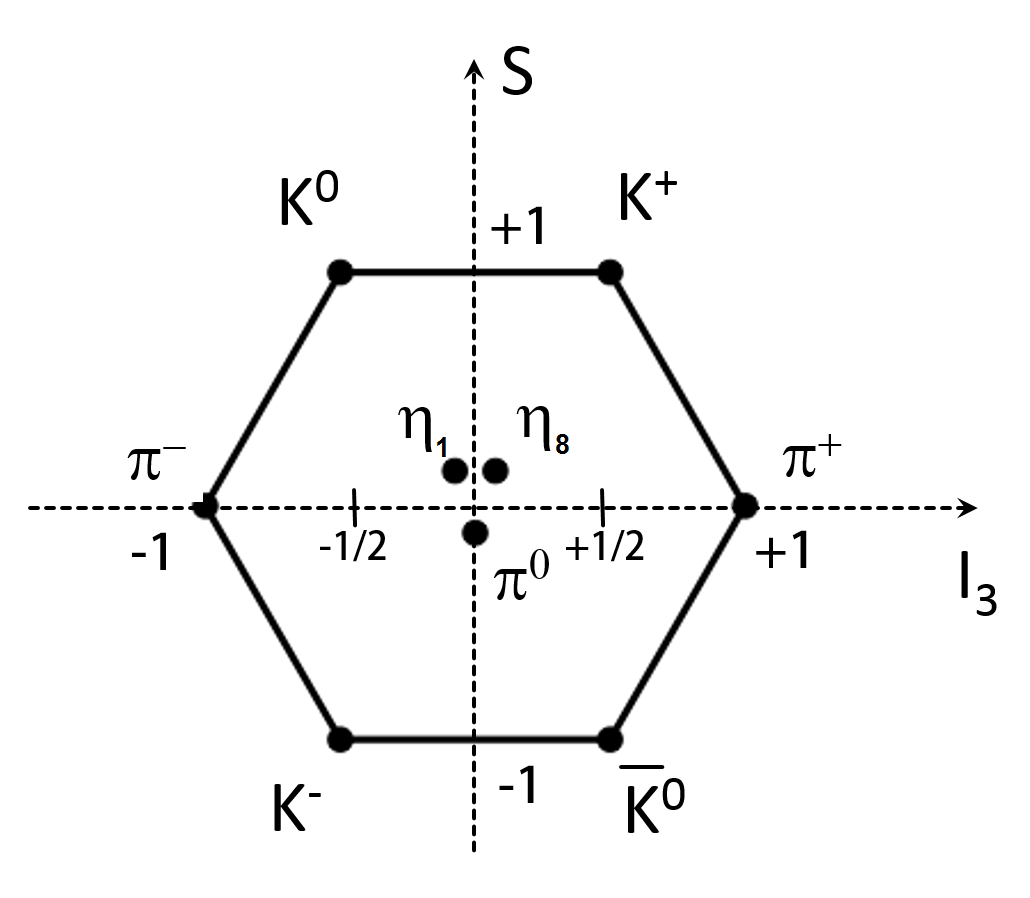}  
\caption{Multiplet of pseudoscalar mesons.~The horizontal axis corresponds to $3^{rd}$ component of the isospin $I_{3}$ while the vertical axis corresponds to strangeness $S$.\label{eta_nonet}}
\end{figure}

\indent $\eta$ is a neutral meson with zero spin $J$ and negative parity $P$. Together with $\eta'$, pions: $\pi^{0}$, $\pi^{+}$, $\pi^{-}$ and kaons: $K^{0}$, $\overline{K}^{0}$, $K^{+}$, $K^{-}$ it is a part of pseudoscalar meson nonet which is schematically presented in Fig.~\ref{eta_nonet}.~The mesons are arranged here according to strangeness $S$ along the $Y$ axis and according to the isospin component $I_{3}$ along the $X$ axis.


\indent In the Gell-Mann quark model mesons consist of quark-antiquark pairs. According to SU(3) flavour symmetry of three lightest quarks $u$, $d$ and $s$ the mesons form a singlet and an octet with the following quark contributions:\\

\noindent $\eta_{1}=\frac{1}{\sqrt{3}}(d\overline{d} + u\overline{u} + s\overline{s})$,\\

\noindent $\eta_{8}=\frac{1}{\sqrt{6}}(d\overline{d} + u\overline{u} - 2s\overline{s})$.\\

\noindent The observed $\eta$ particle is the combination of the $\eta_{1}$ and $\eta_{8}$ states:\\

\noindent $|\eta>=\eta_{8}cos{\theta} - \eta_{1}sin{\theta}$,\\

\noindent where $\theta$ is mixing angle determined experimentally and equals $\theta$ = -15.5$^{\circ}\pm$1.3$^{\circ}$~\cite{Bramon}. According to Bass and Thomas the flavour singlet component $\eta_{1}$ can mix with pure gluonic states~\cite{BassTom,BassTomek} what is important from the point of view of the $\eta$ meson properties inside the nuclear matter.\\
\indent Many strong or electromagnetic decay channels of $\eta$ meson is forbidden by $C$, $P$, $CP$ or $G$ symmetry conservations. The decay into two or four pions is forbidden due to $P$ and $CP$ invariance while charge conjugation does not allow to occur the $\eta \rightarrow \pi^{0}\gamma$, $\eta \rightarrow \pi^{0}\pi^{0}\gamma$ and $\eta \rightarrow \pi^{0}\pi^{0}\pi^{0}\gamma$ processes. \mbox{$G$-parity} conservation forbids the decays into three pions, however they occur with isospin violation and are dominant processes together with second-order electromagnetic $\eta \rightarrow \gamma\gamma$ decay (Table~\ref{Tab_eta}). The investigation of rare $\eta$ symmetry violating decay processes is very important for precise studies of the QCD symmetries.

\chapter{Simulation of $\mathbold{dd\rightarrow p p n_{sp} n_{sp}}$ reaction~\label{App_Sim_qf}} 

\begin{figure}[h!]
\centering
\includegraphics[width=13.0cm,height=7.5cm]{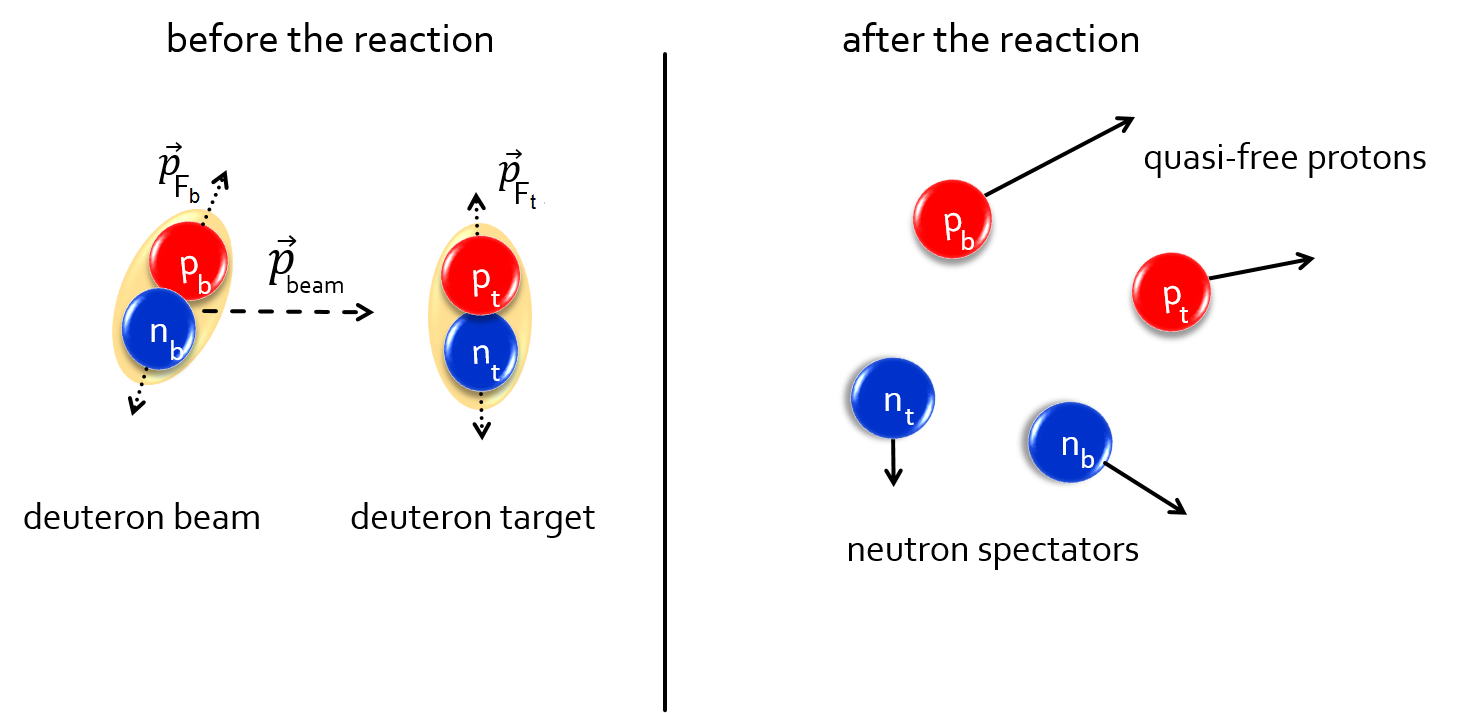}
\caption{Schematic picture of the quasi-free $pp \rightarrow pp$ reaction. Red and blue circles
represent protons and neutrons respectively. The Fermi momentum of the nucleons inside the deuteron beam and deuteron target is presented by the dotted arrows and the deuteron beam momentum by the dashed one.
\label{Scheme_qf}}
\end{figure}

\noindent The quasi-free $pp$ reaction described in Sec.~\ref{Lum_qf} is schematically presented in Fig.~\ref{Scheme_qf} and its simulation is described below.\\

\begin{enumerate}
\item $p_{beam}$ is generated with uniform probability density distribution in the range of $p_{beam}\in$~(2.127,2.422)~GeV/c.

\item The deuteron beam ($b$) as well as the deuteron target ($t$) are considered as a proton-neutron bound systems ($p_{b}$,$n_{b}$) and ($p_{t}$,$n_{t}$), respectively. The neutron $n_{b}$ and $n_{t}$ momentum vectors are distributed isotropically in the spherical coordinates of the deuteron beam and the deuteron target \mbox{(${p}_{F_{b,t}}$, $\theta_{F_{b,t}}$, $\phi_{F_{b,t}}$)} with Fermi momentum distribution of nucleons inside deuteron~\cite{Skurzok_Master} and transformed into Cartesian coordinates. Fermi momentum distributions of proton and neutron bound inside a deuteron derived from two different potential models, namely PARIS~\cite{Lacombe} and CDBonn~\cite{CDBonn} are shown in Fig.~\ref{fig_qfree}. The neutrons four-momenta are calculated based on spectator model assumption ($|\mathbb{P}_{n_{b,t}}|^{2}=m^{2}_{n_{b,t}}$) in the center of mass (CM) frame and transformed using Lorentz transformation into the laboratory frame.

\begin{figure}[h!]
\centering
\includegraphics[width=10.0cm,height=7.5cm]{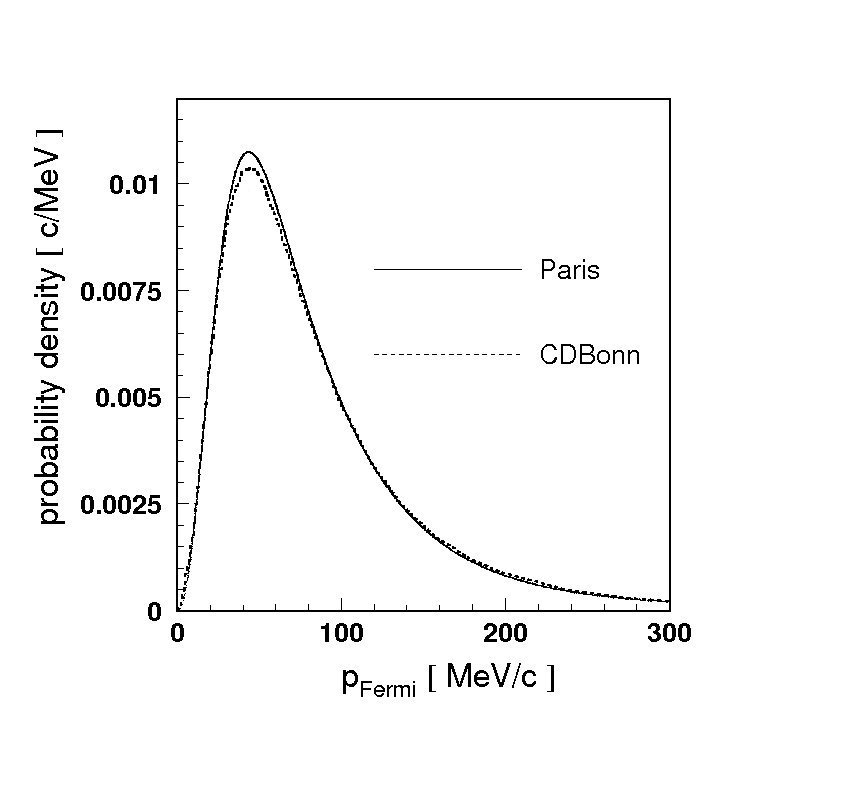}
\vspace{-1.0cm}
\caption{Fermi momentum distribution of nucleons inside the deuteron for PARIS (full line~\cite{Lacombe} and CDBonn (dashed line)~\cite{CDBonn} potentials. The distributions were normalized to unity in the momentum range from 0 to 300~MeV/c.\label{fig_qfree}}
\end{figure}

\item Based on Fermi momentum ${\vec{p}}_{F_{b,t}}$ values, proton $p_{b}$ mass in the deuteron beam and proton $p_{t}$ mass in the deuteron target are calculated according to equation~(\ref{eq:10}).

\begin{equation}
m_{{p_{b,t}}}=\left(m^{2}_{d}+m^{2}_{n}-2m_{d}\sqrt{m^{2}_{n}+|\vec{p}_{F_{b,t}}|^{2}}\right)^{\frac{1}{2}}.
\label{eq:10}
\end{equation}

Four-momenta of both of protons are calculated in the CM systems of deuterons and are transformed into the laboratory (LAB) frame. 

\item The proton-proton invariant mass $\sqrt{s_{pp}}$ is calculated in LAB system according to equation~(\ref{eq:11}):

\begin{equation}
\sqrt{s_{pp}}=\left(E_{p_{b}}+E_{p_{t}}\right)^{2}-\left(\vec{p}_{p_{b}}+\vec{p}_{p_{t}}\right)^{2},
\label{eq:11}
\end{equation}

\noindent where $E_{p_{b}},E_{p_{t}}$ and $\vec{p}_{p_{b}},\vec{p}_{p_{t}}$ are energies and momenta in LAB frame for $p_{b}$ and $p_{t}$, respectively.

\item Based on $\sqrt{s_{pp}}$, $p_{b}$ momentum is calculated in the system where $p_{t}$ is at rest:

\begin{equation}
p^{prot}_{beam}=\sqrt{\left(\frac{s_{pp}-m^{2}_{p_{b}}-m^{2}_{p_{t}}}{2m_{p_{t}}}\right)^{2}-m^{2}_{p_{b}}}.
\label{eq:12}
\end{equation}

\item Four-vectors of protons $p_{b}$ and $p_{t}$  are transformed into the proton-proton CM frame.

\item Scattering between protons is considered.~Scattering angle $\theta^{*}$ as well as azimutal angle $\phi^{*}$ are simulated isotropically. Four-momenta of scattered protons are calculated and transformed to the laboratory system. 

\item WASA Monte Carlo (simulation of the detection system response by GEANT package) is carried out for generated events.  

\end{enumerate}

\chapter{Simulation of $\mathbold{\BcgReactb}$ reaction~\label{Sim_Nstar}}

The simulation of $\BcgReactb$ reaction, being one of the main processes contributing to the background (Sec.~\ref{Bcg_stud}), was carried out for the beam momentum range $p_{beam}\in(2.127,2.422)$~GeV/c corresponding to the experimental ramping. The main assumptions of simulation are schematically described in following points:

\begin{enumerate}

\item The deuteron beam momentum value $p_{beam}$ is generated with uniform probability density distribution in range of $p_{beam}\in(2.127,2.422)$~GeV/c and then the square of invariant mass of the whole system $s_{dd}$ is calculated using Eq.~(\ref{eq_3}) presented in Sec.~\ref{Sim_scheme}.



\item The invariant mass $\sqrt{s_{dd}}$ is distributed randomly according to the distribution presented as follows:


\begin{equation}
\sigma(\sqrt{s_{dd}})=\int^{W_{max}}_{W_{min}} PS(W)\cdot BW\left(\sqrt{s_{dd}}-W-m_{^{3}\hspace{-0.05cm}He},\Gamma_{N^{*}},E_{N^{*}}\right)\cdot dW,
\label{eq_1}
\end{equation}

where:


$\bullet$  $W=\sqrt{s_{dd}}-m_{N^{*}}-m_{^{3}\hspace{-0.05cm}He}$ \\

is the excess energy avaliable in the CM frame with minimum and maximum values equal to $W_{min}=0$ and 
$W_{max}=\sqrt{s_{dd}}-m_{\pi^{0}}-m_{n}-m_{^{3}\hspace{-0.05cm}He}$, respectively; \\

$\bullet$ $PS(W)=\sqrt{W} \left[\sqrt{s_{dd}}+m_{N^{*}}+m_{^{3}\hspace{-0.05cm}He}\right]^{1/2} \left[s_{dd}-(m_{N^{*}}-m_{^{3}\hspace{-0.05cm}He})^{2}\right]^{1/2}/\left(2\sqrt{s_{dd}}^{3}\right)$\\
$=\sqrt{W}\left[2\sqrt{s_{dd}-W}\right]^{1/2} \left[s_{dd}-(\sqrt{s_{dd}}-W-2m_{^{3}\hspace{-0.05cm}He})^{2}\right]^{1/2}/\left(2\sqrt{s_{dd}}^{3}\right)$\\

is a Phase Space factor for 2-body reactions which is proportional to $\sqrt{W}$ near the $\eta$ production threshold and to 1/W above the threshold~\cite{Goshaw};\\

$\bullet$ $BW\left(\sqrt{s_{dd}}-W-m_{^{3}\hspace{-0.05cm}He},\Gamma_{N^{*}},E_{N^{*}}\right)= \frac{\Gamma^{2}_{N^{*}}/4}{(m_{N^{*}}-E_{N^{*}})^{2}+\Gamma^{2}_{N^{*}}/4}$=\\
$\frac{\Gamma^{2}_{N^{*}}/4}{(\sqrt{s_{dd}}-W-m_{^{3}\hspace{-0.05cm}He}-E_{N^{*}})^{2}+\Gamma^{2}_{N^{*}}/4}$ \\

is a Breit Wigner distribution of $N^{*}$ resonance with energy $E_{N^{*}}$=1535~MeV and width $\Gamma_{N^{*}}$=150~MeV. The $BW$ distribution is presented schematically in Fig.~\ref{fig_BW} while $\sigma(\sqrt{s_{dd}})$ distribution in Fig.~\ref{Sigma_sdd}.

\begin{figure}[h!]
\centering
\includegraphics[width=12.0cm,height=8.0cm]{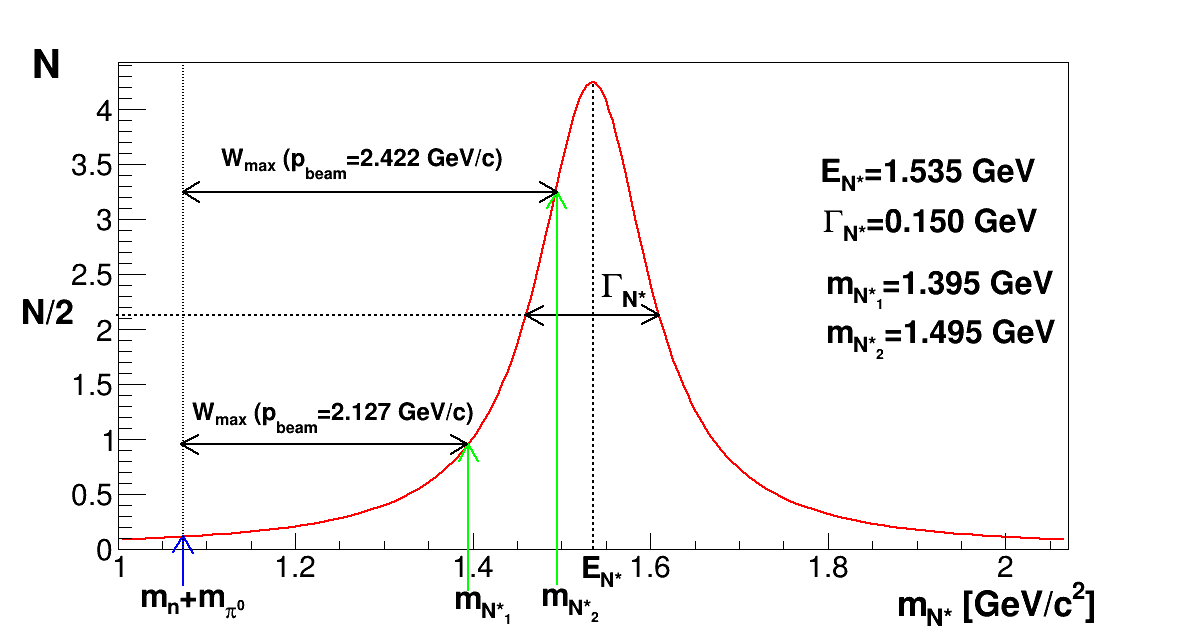}
\caption{Breit Wigner distribution of $N^{*}$ resonance with energy $E_{N^{*}}$=1535~MeV and width $\Gamma_{N^{*}}$=150~MeV. The green arrows show the maximum resonance masses $m_{N^{*}_{1}}$ and $m_{N^{*}_{2}}$ for the beam momentum \mbox{$p_{beam}=2.127$~GeV/c} and $p_{beam}=2.422$~GeV/c, respectively. Blue arrow shows the sum of pion and neutron masses ($m_{\pi^{0}}$ and $m_{n}$) which is the lower limit of the resonance mass in $\BcgReactb$ process. \label{fig_BW}}
\end{figure}


\begin{figure}[h!]
\centering
\includegraphics[width=10.0cm,height=7.0cm]{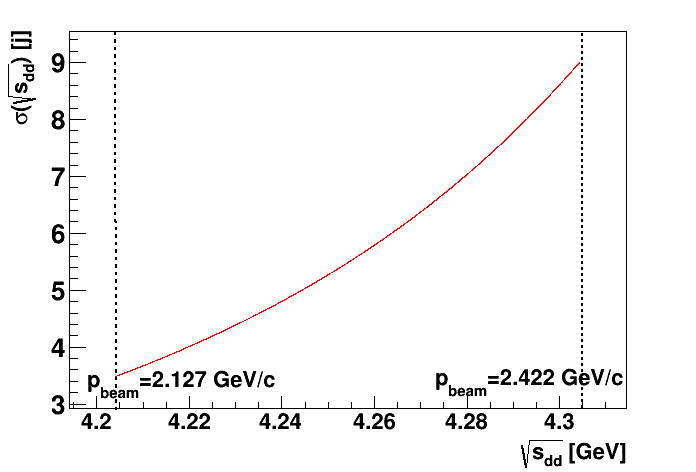}
\caption{$\sigma(\sqrt{s_{dd}})$ distribution determined for the considered beam momentum range. The dashed line denotes the minimal and maximal values of $p_{beam}$ which are equal $2.127$ GeV/c and $2.422$~GeV/c, respectively.\label{Sigma_sdd}}
\end{figure}


\item Excess energy available in the CM frame $W$ is distributed according to the $PS(W)\cdot BW\left(\sqrt{s_{dd}}-W-m_{^{3}\hspace{-0.05cm}He},\Gamma_{N^{*}},E_{N^{*}}\right)$ distribution.

\item The resonance mass $m_{N^{*}}$ is calculated as $m_{N^{*}}=\sqrt{s_{dd}}-W-m_{^{3}\hspace{-0.05cm}He}$ and is limited, because of two conditions: 



$\bullet$ $m_{N^{*}} + m_{^{3}\hspace{-0.05cm}He} \leq \sqrt{s_{dd}}$ (the whole available energy is used to produce $N^{*}$ and $\Hea$),

$\bullet$ $m_{N^{*}} \geq m_{\pi^{0}}+m_{n}$ (resonance mass should be enough to decay into neutron and $\pi^{0}$).








\item The neutron and pion momentum vectors are simulated isotropically in the $N^{*}$ frame in spherical coordinates and transformed into Cartesian coordinates. The absolute value of neutron and pion momenta $\vec{p}^{\,\,**}_{n,\pi^{0}}$ is fixed by equation~(\ref{eq:101}) described in Sec.~\ref{Sim_scheme}.


\item The gamma quanta are simulated isotropically in the $\pi^{0}$ frame in spherical coordinates with momenta \mbox{$\vec{p}^{\,\,***}_{\gamma}=m_{\pi^{0}}/2$}. 

\item The four-momentum vectors of $\Hea$, neutron and gamma quanta are transformed into the center of mass frame and next into laboratory frame by means of Lorentz transformation. 

\item The WMC simulations of the WASA detector response are carried out. 

\end{enumerate}

\clearpage

\addcontentsline{toc}{chapter}{List of Abbreviations}

\chapter*{List of Abbreviations~\label{Abbrev}} \label{Abbrev}

\textbf{WASA} Wide Angle Shower Apparatus\\
\textbf{COSY} Cooler Synchrotron\\
\textbf{PDG} Particle Data Group\\
\textbf{FD} Forward Detector\\
\textbf{CD} Central Detector\\
\textbf{FWC} Forward Window Counter\\
\textbf{FPC} Forward Proportional Chamber\\
\textbf{FTH} Forward Trigger Hodoscope\\
\textbf{FRH} Forward Range Hodoscope\\
\textbf{FRI} Forward Range Intermediate Hodoscope\\
\textbf{FRA} Forward Absorber\\
\textbf{FVH} Forward Veto Hodoscope\\
\textbf{MDC} Mini Drift Chamber\\
\textbf{PSB} Plastic Scintillator Barrel\\
\textbf{SCS} Superconducting Solenoid\\
\textbf{SEC} Scintillator Electromagnetic Calorimeter\\
\textbf{WMC} Wasa Monte Carlo\\
\textbf{CM} Center of Mass System\\
\textbf{LAB} Laboratory System\\
\textbf{DAQ} DATA Acquisition\\
\textbf{FPGA} Field Programmable Gate Array\\
\textbf{QCD} Quantum Chromodynamics\\

\clearpage

\addcontentsline{toc}{chapter}{Acknowledgements}

\chapter*{Acknowledgements} 

\vspace{-0.7cm}

\noindent
\textit{I would like to express my highest gratitude to all the people without whom this thesis would not have been possible.}

\vspace{0.2cm}

\textit{First of all, I would like to express my sincere thanks for my supervisor Prof. Pawe{\l}~Moskal for valuable help, for plenty of hints and suggestions, for understanding and enormous patience. Your vast knowledge and skill was the best assistance in writing this thesis!}

\vspace{0.2cm}

\textit{I would like to thank dr Wojtek Krzemie{\'n}, for his very valuable discussions and for his help in preparation of this thesis.}

\vspace{0.2cm}

\textit{I am very grateful to Prof.~Bogus\l aw Kamys for allowing me preparing this dissertation in the Faculty of Physics, Astronomy and Applied Computer Science of the Jagiellonian University and to Prof. dr James Ritman for a great opportunity to visit the Forschungszentrum J{\"u}lich and work in the Institute f{\"u}r Kernphysik.}

\vspace{0.2cm}

\textit{I would like to express my appreciation to all the \mbox{WASA-at-COSY} members for their help and friendly atmosphere. I am especially grateful to dr Volker Hejny for very useful discussions and advices.}

\vspace{0.2cm}

\textit{I acknowledge support by the Foundation for Polish Science - MPD program, co-financed by the European Union within the European Regional Development Fund and by the Polish National Science Center through grants No.~DEC-2013/11/N/ST2/04152.}

\vspace{0.2cm}

\textit{I also thank all of my colleagues for the scientific contribution and the great time spend together. The highest gratitude I address to Iryna Ozerianska for the great support during my work and understanding all the time!}

\vspace{0.2cm}

\textit{I would like to express my gratitude to my friends, especially to Marcela Batkiewicz who was near me during all years of my studies and encouraged~me.}

\vspace{0.2cm}

\textit{The last, but not least, I want to thank my parents, brother and the rest of my Family for the love, patience and incredible support they provided to me through my entire life.}

\clearpage

\addcontentsline{toc}{chapter}{Bibliography}


\newpage
\thispagestyle{plain}

\end{document}